\documentclass[12pt]{article}
\usepackage{graphicx,amssymb,amsmath,empheq}
\usepackage{authblk}
\usepackage{amsthm}
\usepackage{bm}
\usepackage{empheq}
\usepackage{subfig} % caution: not compatible with revtex4
\usepackage{xcolor}

\usepackage{hyperref}
\hypersetup{colorlinks = true, linkcolor=black, citecolor=black, urlcolor=blue}
\usepackage{url}

\theoremstyle{plain}

\usepackage{tikz}
\usepackage{tikz-cd}
\usetikzlibrary{arrows}
\usetikzlibrary{intersections}
\usetikzlibrary{shapes.geometric}
\usetikzlibrary{decorations.pathmorphing, patterns,shapes}
\usetikzlibrary{decorations.markings}

\tikzset{
  mid arrow/.style={postaction={decorate,decoration={
        markings,
        mark=at position .575 with {\arrow[#1]{stealth}}
      }}},
  near arrow/.style={postaction={decorate,decoration={
        markings,
        mark=at position .275 with {\arrow[#1]{stealth}}
      }}},
   far arrow/.style={postaction={decorate,decoration={
        markings,
        mark=at position .800 with {\arrow[#1]{stealth}}
      }}},
}

\usepackage[nosort]{cite}

\renewcommand{\bar}{\overline}
\renewcommand{\tilde}{\widetilde}

\renewcommand{\Re}{\operatorname{Re}}
\renewcommand{\Im}{\operatorname{Im}}

\newcommand{\dkap}{\delta\kern-1.25pt\varkappa}

\newcommand{\SL}{\operatorname{SL}}

\def\be{\begin{equation}}
\def\ee{\end{equation}}

\title{Exactly solvable
non-unitary time evolution in quantum critical systems I:
Effect of complex spacetime metrics
}

\author[1,2,3]{Xueda Wen}

\affil[1]{\normalsize\it School of Physics, Georgia Institute of Technology, Atlanta, GA 30332, USA}
\affil[2]{\normalsize\it Department of Physics, Harvard University, Cambridge, MA 02138, USA}
\affil[3]{\normalsize\it Department of Physics, University of Colorado, Boulder, CO 80309, USA}

\begin{document}

\maketitle

\begin{abstract}
In this series of works, we study exactly solvable non-unitary time evolutions in one-dimensional quantum critical systems ranging from quantum quenches to time-dependent drivings. In this part I, we are motivated by the recent works  of Kontsevich and Segal \cite{Segal_2021} and Witten\cite{Witten2021_ComplexMetric} on allowable complex spacetime metrics in quantum field theories. In general, such complex spacetime metrics will lead to non-unitary time evolutions. In this work, we study the universal features of
such non-unitary time evolutions based on exactly solvable setups. Various physical quantities including entanglement Hamiltonian and entanglement spectrum, entanglement entropy, and energy density at an arbitrary time can be exactly solved. Due to the damping effect introduced by the complex time, the excitations in the initial state are gradually damped out in time. The non-equilibrium dynamics exhibits universal features that are qualitatively different from the case of real-time evolutions.
For instance, for an infinite system after a global quench, the entanglement entropy of the semi-infinite subsystem will grow logarithmically in time, in contrast to the linear growth in a real-time evolution. Moreover, we study numerically the time-dependent driven quantum critical systems with allowable complex spacetime metrics. It is found that the competition between driving and damping leads to a steady state with an interesting entanglement structure.

\end{abstract}

\newpage

\tableofcontents

\newpage

\section{Introduction}
\label{sec:intro}

\subsection{Introduction and motivation}

Recently in \cite{Segal_2021},  Kontsevich and Segal (KS) studied what is the class of complex spacetime geometries where a generic quantum field theory can be consistently defined, such that the path integral is manifestly convergent on the allowed geometries.
KS's motivation is to develop an alternative to some of the standard axioms of quantum field theories\cite{SW_2000}, 
by postulating that the partition function and the correlation functions extend analytically to a certain domain of complex spacetime metrics.
They studied the \textit{allowable} metrics $g$ by considering the path integral over matter fields, which are taken to be scalars and 
$p$-form fields of all possible ranks with real values.
KS give an explicit description of the allowable metrics $g$.
First one can write the metric in a diagonal form $g_{ij}=\delta_{ij} \lambda_i$ ($\lambda_i$ are complex numbers in general), which can always be done locally, then by requiring the convergence of path integral one can obtain 
\be\label{KS_condition}
\Sigma:=\sum_{i=1}^D |\text{Arg} \, (\lambda_i)|<\pi,
\ee
where $\text{Arg}(z)\in (-\pi,\pi]$ is the argument of $z\in \mathbb C$, and $D$ is the spacetime dimension.
Conversely, if the condition in \eqref{KS_condition} is satisfied, then we say the metric $g$ is allowable.
The condition in \eqref{KS_condition} generalizes the result in an earlier work by Louko and Sorkin in two dimensions \cite{LoukoSorkin1997} to arbitrary dimensions.

As noted by Kontsevich and Segal, the space of allowable metrics is contractible onto the space of Euclidean metrics. 
Note that Euclidean metrics with
\be
ds^2=dt^2+d\vec{x}^{\,2}, 
\ee
are always allowable because $\Sigma=0$. For Lorentz metrics, 
\be\label{L_metric}
ds^2=-dt^2+d\vec{x}^{\,2},
\ee
since $\Sigma=\pi$, they lie on the boundary of the allowable domain of complex metrics (See also the concrete example in \eqref{Complex_metrics} below).

\medskip

Later, Witten proposed to investigate KS's criterion in various interesting examples including quantum gravity \cite{Witten2021_ComplexMetric}, noting that the complex metrics that have proven useful, e.g., complexified black holes, satisfy KS's conditions, while some pathological metrics, e.g., the complex wormholes, do not satisfy KS's conditions. After the proposals in \cite{Segal_2021,Witten2021_ComplexMetric}, there are many recent interests in studying the consequence of allowable complex spacetime metrics in various contexts \cite{2022_Bond,Lehners2021,2022Cosmology,2022_Visser,2203_Loges,2206_Bris,2023_Hertog}.
Note that the allowable complex metrics have also been previously used as a regularization scheme in different contexts including the analysis of gravitational entropy and the real time thermal physics\cite{2016Dong,2021Marolf,2021Dong}.

\medskip

In this work, we will explore the consequence of allowable complex metrics in the context of non-equilibrium dynamics in quantum field theories.
In particular, we consider the following complex spacetime metrics:\cite{Witten2021_ComplexMetric}
\be
\label{Complex_metrics}
ds_{\pm}^2=-(1\mp i\epsilon)^2 dt^2+d\vec{x}^{\,2}, \quad \epsilon>0. 
\ee
One can find that $\Sigma\in (0,\pi)$ and it satisfies KS's criteria in \eqref{KS_condition}.
It is noted that for an allowable metric, $\sqrt {\det g}$ has a positive real part\cite{Segal_2021,Witten2021_ComplexMetric}.
Therefore, the two choices of signs in \eqref{Complex_metrics} differ by the sign of $\sqrt{\det g}$ with $\sqrt {\det g}=\pm i(1\mp i\epsilon)$, where the 
real part of $\sqrt{\det g}$ is always positive. 
As we approach the Lorentzian metric in \eqref{L_metric} by taking $\epsilon\to 0$, $\sqrt {\det g}$ approaches the 
positive or negative imaginary axis, depending on the sign in $\pm i\epsilon$.
These two choices correspond to the time propagation by 
$\exp(-iHt-\epsilon Ht)$ and $\exp(+iHt-\epsilon Ht)$ respectively\cite{Witten2021_ComplexMetric}, where $H$ is the Hamiltonian.
In this work, these two choices will be used in the time evolution of certain initial states $|\psi_0\rangle$ and $\langle \psi_0|$ respectively.
More explicitly, we have
\be
\label{complex_time_general}
|\psi(t)\rangle=e^{-iHt-\epsilon Ht} |\psi_0\rangle, \quad \epsilon>0,
\ee
and the complex conjugate $\langle \psi(t)|=\langle \psi_0| e^{+iHt-\epsilon Ht}$. Apparently, this time evolution is non-unitary, where
the factor $e^{-\epsilon Ht}$ introduces a damping effect which tends to evolve the wavefunction to the ground state of $H$.
In this work, we hope to understand what universal features could appear in such non-unitary time evolutions.

\medskip

Another motivation of this work is to make a connection to the non-unitary time evolution in open quantum systems,
where the non-unitary dynamics is caused by the coupling to the measurement apparatus or more generally the environment.
Some interesting examples include, to name a few, measurement induced phase transitions \cite{2019Skinner,2018_Li,1908_Bao,1908_Jian} and  long-range entangled
states preparation\cite{2112_Nat,2206_Hsieh,2209_Nat,2208_Zhu}.
In particular, it has been demonstrated that the measurement of quantum many-body states are related to non-unitary time evolutions. See, e.g., Ref.\cite{2023Altaman2,2023Altman,2023Jian,2023Jian2,2023Alicea,2024Alicea,2022Granet,2020Chen,2023Schiro,2021_Tang,2022_Milekhin,2023_Silva} for recent works on this topic.

We also noted some recent numerical works on complex time evolution\cite{2024TN,2021_imaginary,2107_Sio} in lattice systems and tensor-network states, 
which are closely related to this work. Their goal is to  provide an efficient numerical way to simulate quantum systems, while in our work we are mainly interested 
in the exactly solvable quantum field theories and universal features in the non-unitary time evolutions.

\medskip
Before we move on to the concrete setup, we want to emphasize that there are other types of non-unitary time evolutions, 
such as the time evolution determined by Lindblad master equations. For free fermion/boson systems, such non-unitary 
time evolutions may be exactly solvable when certain conditions are satisfied \cite{2008_Prosen,2010_Prosen,2013_Horstmann}. See also 
recent progresses in Ref.\cite{2016_Prosen,2018_Rowlands,2019_Shibata,2020_Essler,2021_Essler}.
Although we mainly focus on the time evolution of the form in \eqref{complex_time_general} in this work, 
in the next parts of this series, for different motivations, we will study other types of exactly solvable non-unitary time evolutions.

\subsection{Setup}

Now let us introduce the non-unitary time evolution of a conformal field theory (CFT) after a global/local quantum quench within the complex spacetime metrics.
In general, we consider a given initial state $|\psi_0\rangle$.
Then at $t=0$, we evolve this initial state in \eqref{complex_time_general} with a CFT Hamiltonian:
\footnote{It is emphasized that here $\epsilon$ is not necessarily small -- it is a real number which can be arbitrarily large.}
\be
\label{WF_Quench}
|\psi(t)\rangle=e^{-i H_{\text{CFT}}\, (1-i\epsilon) t }\, |\psi_0\rangle, \quad \epsilon > 0,
\ee
where we fix the sign of $\epsilon$ in \eqref{WF_Quench} to be positive to make sure that it leads to a convergent  path integral.
Note that the complex conjugate of this state is $\langle \psi(t)|=\langle \psi_0| \, e^{ i H_{\text{CFT}}\, (1+i\epsilon) t }$.
If $\epsilon=0$, it reduces to the real time evolution, which is unitary.
For $\epsilon>0$, the time evolution becomes \textit{non-unitary}.

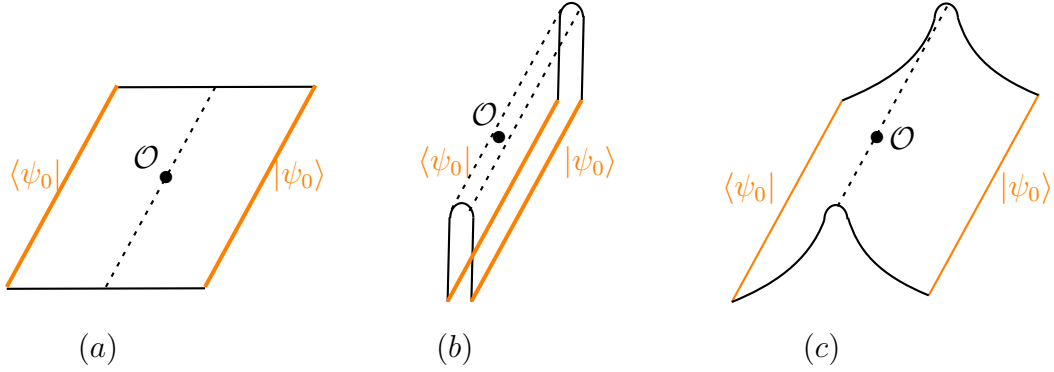
\begin{figure}[tp]
\center
\centering
\begin{tikzpicture}[x=0.75pt,y=0.75pt,yscale=-1,xscale=1]

%% Euclidean

\begin{scope}[xshift=-100pt,yshift=-5pt]
\draw [line width=0.8]    (228,74.27) -- (328,74.03) ;
\draw [line width=0.8]    (172.33,176) -- (272.33,175.75) ;
\draw [line width=0.8]  [dash pattern={on 1.69pt off 2.76pt}]  (272.33-50,130.04+45) -- (328-50,28.32+45) ;

\draw [line width=1.5] [orange]    (272.33-100,130.04+45) -- (328-100,28.32+45) ;
\draw [line width=1.5] [orange]    (272.33+0,130.04+45) -- (328+0,28.32+45) ;

\node at (189.5pt,90pt){$\bullet$};
\node at (182.0pt,82pt){$\mathcal O$};

\node at (240.0pt,89pt){\textcolor{orange}{$|\psi_0\rangle$}};
\node at (140.0pt,89pt){\textcolor{orange}{$\langle \psi_0|$}};

\end{scope}

\draw [line width=0.8]  [orange]  (504,172.67) -- (559.67,70.94) ;
\draw [line width=0.8]  [orange]   (404.67,176) -- (460.33,74.27) ;
\draw [line width=0.8]  [dash pattern={on 1.69pt off 2.76pt}]  (272.33,130.04) -- (328,28.32) ;

%Straight Lines [id:da8047273882856276] 
\draw [line width=0.8]   (261.33,175.75) -- (262.33,133.3) ;
%Straight Lines [id:da36658072828688415] 
\draw [line width=0.8]    (273.33,175.75) -- (274.33,133.3) ;
%Shape: Arc [id:dp20747075589100283] 
\draw  [draw opacity=0][line width=0.8]  (262.66,133.65) .. controls (262.66,133.63) and (262.66,133.6) .. (262.66,133.58) .. controls (262.65,129.45) and (265.26,126.09) .. (268.48,126.09) .. controls (271.62,126.09) and (274.19,129.26) .. (274.33,133.24) -- (268.5,133.58) -- cycle ; \draw  [line width=0.8]  (262.66,133.65) .. controls (262.66,133.63) and (262.66,133.6) .. (262.66,133.58) .. controls (262.65,129.45) and (265.26,126.09) .. (268.48,126.09) .. controls (271.62,126.09) and (274.19,129.26) .. (274.33,133.24) ;

\draw [line width=0.8]   (317,74.03) -- (318,31.57) ;
%Straight Lines [id:da38059229841461995] 
\draw [line width=0.8]    (329,74.03) -- (330,31.57) ;
%Shape: Arc [id:dp8082775311037147] 
\draw  [draw opacity=0][line width=0.8]  (318,32.62) .. controls (318,32.6) and (318,32.58) .. (318,32.56) .. controls (318,28.42) and (320.6,25.07) .. (323.83,25.06) .. controls (326.97,25.06) and (329.54,28.23) .. (329.68,32.21) -- (323.84,32.55) -- cycle ; \draw  [line width=0.8]  (318,32.62) .. controls (318,32.6) and (318,32.58) .. (318,32.56) .. controls (318,28.42) and (320.6,25.07) .. (323.83,25.06) .. controls (326.97,25.06) and (329.54,28.23) .. (329.68,32.21) ;  
%Straight Lines [id:da07935031124740755] 

\draw [line width=1.5]  [orange]  (273.33,175.75) -- (329,74.03) ;
%Straight Lines [id:da19332587232038623] 
\draw [line width=1.5]  [orange]  (261.33,175.75) -- (317,74.03) ;

\node at (250.0pt,79pt){\textcolor{orange}{$|\psi_0\rangle$}};
\node at (195.0pt,79pt){\textcolor{orange}{$\langle \psi_0|$}};

\draw  [draw opacity=0][line width=0.8]  (451.34,134.62) .. controls (451.34,134.6) and (451.34,134.58) .. (451.34,134.55) .. controls (451.33,130.42) and (453.94,127.06) .. (457.16,127.06) .. controls (460.3,127.06) and (462.87,130.23) .. (463.01,134.21) -- (457.18,134.55) -- cycle ; \draw  [line width=0.8]  (451.34,134.62) .. controls (451.34,134.6) and (451.34,134.58) .. (451.34,134.55) .. controls (451.33,130.42) and (453.94,127.06) .. (457.16,127.06) .. controls (460.3,127.06) and (462.87,130.23) .. (463.01,134.21) ;  
%Shape: Arc [id:dp6191645284026938] 
\draw  [draw opacity=0][line width=0.8]  (507,32.89) .. controls (507,32.87) and (507,32.85) .. (507,32.83) .. controls (507,28.69) and (509.6,25.34) .. (512.83,25.34) .. controls (515.97,25.33) and (518.54,28.5) .. (518.68,32.48) -- (512.84,32.82) -- cycle ; \draw  [line width=0.8]  (507,32.89) .. controls (507,32.87) and (507,32.85) .. (507,32.83) .. controls (507,28.69) and (509.6,25.34) .. (512.83,25.34) .. controls (515.97,25.33) and (518.54,28.5) .. (518.68,32.48) ;  
%Curve Lines [id:da858820535723454] 

\draw [line width=0.8]    (463.33,133.57) .. controls (468.33,149.98) and (478.33,162.09) .. (504,172.67) ;
%Curve Lines [id:da7546902182266986] 
\draw [line width=0.8]    (451.34,134.62) .. controls (443.33,156.01) and (425,167.74) .. (404.67,176.01) ;
%Straight Lines [id:da3177649154103529] 
\draw [line width=0.8]  [dash pattern={on 1.69pt off 2.76pt}]  (263.33,129.23) -- (319,27.51) ;

\draw [line width=0.8]    (518.68,32.48) .. controls (523.68,48.9) and (533.68,61) .. (559.34,71.58) ;
%Curve Lines [id:da26258136819199895] 
\draw [line width=0.8]    (507,32.89) .. controls (500.67,54.78) and (480.67,66.01) .. (460.33,74.29) ;

%Straight Lines [id:da3471564125501072] 
\draw [line width=0.8]  [dash pattern={on 1.69pt off 2.76pt}]  (458,127.75) -- (513.67,26.03) ;

\begin{scope}[xshift=170pt]
\node at (245.0pt,89pt){\textcolor{orange}{$|\psi_0\rangle$}};
\node at (140.0pt,89pt){\textcolor{orange}{$\langle \psi_0|$}};
\end{scope}

\node at (215.5pt,70pt){$\bullet$};
\node at (210.0pt,61pt){$\mathcal O$};

\node at (315.5+43pt,70pt){$\bullet$};
\node at (300.0+68pt,70pt){$\mathcal O $};

% Text Node
\draw (85,200) node   {$(a)$};
% Text Node
\draw (265,200) node   {$(b)$};
% Text Node
\draw (450,200) node  {$(c)$};

\end{tikzpicture}
\caption{
Path integral of one-point function with different spacetime metrics (see the main text for details).
(a) Euclidean metric, (b) Lorentz metric, and (c) complex metric. 
}
\label{PI_general}
\end{figure}

\medskip
We will mainly use the path-integral approach to study the complex time evolution in \eqref{WF_Quench}. As an illustration, 
let us consider the one-point function for the local operator $\mathcal O$ in different metrics. 
First, in the Euclidean metric,  to describe $\langle \psi_0| \mathcal O(\tau) |\psi_0\rangle=\langle \psi_0| e^{-H\tau} \mathcal O e^{-H\tau} |\psi_0\rangle$, we need to 
consider a path integral that propagates the initial state $|\psi_0\rangle$ in the imaginary time direction to construct a factor $e^{-H\tau}$,
after which we insert the operator $\mathcal O$ and then propagate the state in the imaginary time direction to construct another factor 
$e^{-H\tau}$, as shown in Fig.\ref{PI_general} (a).
Second, in Lorentz metric, to describe  $\langle \psi_0| \mathcal O(t) |\psi_0\rangle=\langle \psi_0| e^{-i H t} \mathcal O e^{-i H t} |\psi_0\rangle$,
we consider a path integral that propagates the states $|\psi_0\rangle$ forward by a time $t$  to obtain the factor $e^{-iHt}$. 
Then we insert the operator $\mathcal O$ and propagate the state backwards in time to construct the factor $e^{iHt}$, as shown in 
Fig.\ref{PI_general} (b). Third, in  the complex metric, we consider the one point function 
$\langle \psi_0| e^{iHt-\epsilon Ht}  \mathcal O e^{-iHt -\epsilon Ht}|\psi_0\rangle$.
As shown in Fig.\ref{PI_general} (c), the path integral propagates the state $|\Psi\rangle$ both forward and also in the imaginary time direction to
obtain the factor $e^{-iHt -\epsilon Ht}$. After inserting the operator $\mathcal O$, the path integral propagates the state backwards and at the same time in the imaginary time direction to obtain the factor $e^{+iHt -\epsilon Ht}$.

\medskip

Next, for different quantum quenches considered in this work, they correspond to different choices of initial states $|\psi_0\rangle$ in \eqref{WF_Quench}.
The initial states we will consider are the same as those in Ref.\cite{CC2015_global,CC2007_local}, except that now we are interested in the complex time evolution. 

In the global quantum quench, we choose the initial state $|\psi_0\rangle$ as a short-range entangled state, which may be viewed as the ground state of a gapped Hamiltonian. 
In the context of CFT, such initial state can be represented by a regularized conformal boundary state $e^{-\frac{\beta}{4}H_{\text{CFT}}}|B\rangle\rangle$, where
$\beta$ is a positive real number and $|B\rangle\rangle$ is a conformal boundary state.
We study two cases as follows. (1) The total system is of an infinite length defined on $(-\infty,+\infty)$ and the subsystem $A$
is chosen as $A=[0,+\infty)$. (2) The total system is semi-infinite defined on $[0,+\infty)$ and the subsystem $A$ is chosen as a finite interval at the end with  $A=[0,l)$.

In the local quantum quench, we choose the initial state $|\psi_0\rangle$ as the tensor product of ground states of two decoupled CFTs 
defined on $(-\infty,0)$ and $(0,+\infty)$ respectively, i.e., 
$|\psi_0\rangle= e^{-\lambda H_{\text{CFT}}}|G_{L}\rangle\otimes |G_{R}\rangle$, where $|G_{L/R}\rangle$ is the ground state of the CFT on the left/right side and the factor 
$e^{-\lambda H_{\text{CFT}}}$ with $\lambda>0$ plays the role of regularization. At $t=0$, we change the Hamiltonian locally by coupling the two CFTs at their ends at $x=0$, such that the new Hamiltonian $H_{\text{CFT}}$ defined on $(-\infty,+\infty)$ is translationally invariant in space. The subsystem is chosen as $A=[0,+\infty)$.

\medskip

Another setup we will consider in this work is the exactly solvable Floquet CFT\cite{wen2018floquet}, i.e.,  a CFT under a periodic driving,
where one deforms the Hamiltonian density periodically in time.
The minimal setup is based on a two-step driving:
\be
\label{psi_n_real_intro}
|\psi(n)\rangle=\left(e^{-iH_0 T_0}e^{-iH_1 T_1}\right)^n |\psi_0\rangle,
\ee
where $H_0$ and $H_1$ are non-commuting Hamiltonians, and $T_0$ ($T_1$) is the corresponding driving time.
Here the driving Hamiltonians $H_0$ and $H_1$ are obtained from the uniform CFT Hamiltonian by a spatial deformation as
$H_i =\int f_i(x) T_{00}(x) \, dx$,\footnote{More generally, the deformed Hamiltonian has the form 
$H_i=\int f_i(x) T (x) dx+ \int g_i(x)\bar T(x) dx$, where $f_i(x)$ and $g_i(x)$ are smooth real functions that are independent from each other, 
and $T(x)$ and $\bar T(x)$ are holomorphic and anti-holomorphic stress-energy tensors\cite{wen2020periodically}. } 
where $f_i(x)$ is an arbitrary smooth real function, and $T_{00}(x)$ is the Hamiltonian density.
If $f(x)$ is of the simple form $f(x)=a+b\cos\frac{2\pi x}{l}+c\sin\frac{2\pi x}{l}$ where $l$ characterizes the wavelength of deformation, 
one can find the generators of the driving Hamiltonians form
an $\SL(2,\mathbb R)$ algebra\footnote{For a general choice of $f(x)$, the underlying algebra is Virasoro algebra.}. Then the operator evolution in a Floquet CFT is described by a M\"obius transformation, 
which is the underlying reason for the exact solvability of this setup.
The properties of this exactly solvable Floquet CFT as well its generalization have been extensively studied recently\cite{wen2018floquet,
wen2020periodically,
Fan_2020,Wen_2018,Lapierre:2019rwj,fan2020General, lapierre2020geometric,han2020classification,2020Lapierre,2020Andersen,2021Ageev,2021Das,RandomCFT2021,
2022Das_OTOC,2022Bermond,2022Choo,2022Cooling,2023_OBC,2023_StatePrepare,2024Ryu,2023Das,2023_Nozaki_Scrambling, 2024_Krylov,2024_Guo}.
See also Ref.\cite{2021Nozaki,2023Caputa,2023_Geometry, 2023_Briding,2024_Ge,2024Nozaki2,2024Mezei,2023_Brane,2024_Modular} for its holographic dual.
In this work, we will generalize this Floquet CFT setup to a complex time evolution, as follows:
\be
\label{psi_n_intro}
|\psi(n)\rangle=\left(e^{-i(1-i\epsilon)H_0 T_0}e^{-iH_1 T_1}\right)^n |\psi_0\rangle.
\ee
Here we introduce the complex time only in the time evolution with $H_0$.
The motivation is straightforward: On the one hand, by tuning the system to the heating phase of a Floquet CFT\cite{wen2018floquet},
the state will evolve into a highly excited state of $H_0$. On the other hand, the damping factor $e^{-\epsilon H_0 T_0}$ tends to evolve the 
system back to the ground state of $H_0$. This competition between driving and damping may result in a steady state.

\medskip

Before we leave this section, let us introduce several concepts that will be used 
to characterize the non-equilibrium dynamics in the above setups.
One important concept we will study 
 is the so-called entanglement Hamiltonian or modular Hamiltonian, which is related to the reduced density matrix $\rho_A=\text{Tr}_{\bar A}(\rho)$ of a subsystem $A$ as
\be
\label{Def:EH}
\rho_A=e^{-2\pi K_A} \quad \text{or } \quad K_A=-\frac{1}{2\pi}\log \rho_A.
\ee
Here $\bar A$ is the complement of $A$ in the whole system.
In general, the entanglement Hamiltonian $K_A$ is non-local and it is challenging to write down the explicit form of $K_A$.
However, there are a few cases where the analytical form of $K_A$ can be obtained.  
An important example is when the theory is defined on the whole of flat Minkowski space, 
the entanglement Hamiltonian is the boost generator for the half-plane $x_1>0$, i.e., $K_A=\int_{x_1>0} x_1 T_{00}(x) d^{d-1}x$\cite{BW1,BW2}.
Another solvable case is the chiral free fermion system, where the entanglement Hamiltonian can be obtained by using the the resolvent method\cite{2009Casini,2021_Tonni,2019_Reyes}.
In the context of $(1+1)$ dimensional CFTs, it was shown that if the path integral of the reduced density matrix is 
conformally equivalent to an annulus (or cylinder), then 
 the entanglement Hamiltonians can always be written as an integral over the energy-momentum tensor times a local weight, as follows\cite{CardyTonni2016}
\be
\label{EE_temperature}
K_A(t)=\frac{1}{2\pi} \int_a^b \beta_E(x,t)\,  T(x-t) \, dx+ \frac{1}{2\pi} \int_a^b \bar \beta_E(x,t) \, \bar T(x+t) \, dx,
\ee
where $A=[a,\,b]$ is the subsystem.
For the exactly solvable setups as studied in this work, all the entanglement Hamiltonians (for a single interval) are of the form in \eqref{EE_temperature}.
 Here the local weights $\beta_E(x,t)$ and $\bar \beta_E(x,t)\in \mathbb R$ in \eqref{EE_temperature} are called the (inverse) entanglement temperatures. A higher entanglement temperature $1/\beta_E(x)$  (or $1/\bar \beta_E(x)$) indicates there is a stronger entanglement between the region near $x$ and the subsystem 
$\bar A$.

The spectrum of $K_A$, also called entanglement spectrum, 
is useful for characterizing and classifying quantum many-body states\cite{2006Ryu,2008Haldane,2010Pollmann,2012Qi,2014Ludwig}.
The entanglement spectrum determines all the R\'enyi entropies and von Neumann entropy as follows
\be
\label{Renyi_general}
S_A^{(n)}=\frac{1}{1-n} \log \text{Tr}_A (\rho_A^n)=\frac{1}{1-n} \log \left(\text{Tr}_A e^{-2\pi n K_A} \right),
\ee
and
\be
\label{vonNeuman_general}
S_A=-\text{Tr}_A(\rho_A \log \rho_A)=2\pi  \, \text{Tr}_A\left(
K_A e^{-2\pi K_A}
\right).
\ee

\subsection{Summary of results}

Let us first give a brief summary of the main results in this work.
For the above setups of quantum quenches in complex spacetime metrics, we can obtain analytical results for the time evolution of
entanglement Hamiltonian, entanglement spectrum,  entanglement entropy, and energy density at an arbitrary time.

The complex time evolutions of entanglement Hamiltonians $K_A(t)$ and entanglement spectrum for the subsystem $A$
are qualitatively different from those in a real-time evolution.
See, e.g., Eqs.\eqref{KAt_global1}, \eqref{KAtotal_half}, and \eqref{EH_local} for the concrete expressions of $K_A(t)$ 
after different quantum quenches.

 \medskip
The entanglement entropy evolution in a complex spacetime metric and the comparison with the case of Lorentzian metric are summarized as follows.
\begin{enumerate}
\item An infinite system over $(-\infty, +\infty)$ after a global quench, with the ubsystem $A=[0,+\infty)$:
\be
\label{SA_global1_intro}
\left\{
\begin{split}
&\text{Lorentzian metric:}\quad S_A(t)\simeq \frac{\pi c}{3\beta} t,\quad t\to \infty, \\
&\text{Complex metric:}\quad S_A(t)\simeq \frac{c}{6}\log t, \quad t\to \infty.
\end{split}
\right.
\ee
Here $\beta$ is a finite length scale that is introduced in the initial state, which characterizes the correlation length of the initial state,
and $c$ is the central charge of the CFT.
Hereafter, for $t\to \infty$ or long time limit, it means $t$ is much larger than any other finite length scales in the problem.
For example, here in the Lorentzian metric, $t\to \infty$ corresponds to $t\gg \beta$, while in the complex metric it corresponds to 
$t\gg \beta/\epsilon$.

\item A semi-infinite system  over $(0, +\infty)$ after a global quench, with the subsystem $A=[0,\, l]$:
\be
\label{SA_global2_intro}
\left\{
\begin{split}
&\text{Lorentzian metric:}\quad S_A(t)\simeq \frac{\pi c}{3\beta}\,l,\quad t\to \infty, \\
&\text{Complex metric:}\quad S_A(t)\simeq \frac{c}{6} \log l, \quad t\to \infty.
\end{split}
\right.
\ee

\item Local quench. The total system is $(-\infty, +\infty)$, and the subsystem is $A=[0,+\infty)$:

\be
\label{SA_local_intro}
\left\{
\begin{split}
&\text{Lorentzian metric:}\quad S_A(t)\simeq \frac{c}{3}\log t,\quad t\to \infty, \\
&\text{Complex metric:}\quad S_A(t)\simeq \frac{c}{6}\log t, \quad t\to \infty.
\end{split}
\right.
\ee

\end{enumerate}

In both the global and local quenches,
the local energy density $\langle T_{00}(x,t)\rangle$ will decay to zero in time as $1/t^2$. 
See, e.g., the plots in Fig.\ref{LatticeEnergy_global}, Fig.\ref{LatticeEnergy2}, and Fig.\ref{LatticeEnergy_LocalQuench}, as well as the 
concrete expressions of $\langle T_{00}(x,t)\rangle$ in Eqs.\eqref{EnergyGlobal1}, \eqref{T_global2}, \eqref{Tbar_global2} and \eqref{EnergyDensity_Local}, respectively.
One particularly interesting result is for the case of local quench in Sec.\ref{Sec:Local}, where two pulses of energy density are generated after the quench.
One can see clearly how these two pulses of energy density die out as they propagate in space due to the complex spacetime metric, as shown in Fig.\ref{LatticeEnergy_LocalQuench}.

\medskip

For each case introduced above,  we also performed a numerical calculation on the entanglement entropy and energy density evolution based on free fermion lattice models, 
and find a good agreement with the field theory results.

\medskip

Moreover, we study the effect of complex spacetime metrics in a time-dependent driven quantum critical system, more specifically in a Floquet CFT\cite{wen2018floquet}.
Since the complex time evolution introduces a damping effect, there will be a competition between the driving and damping in a driven system.
Based on a numerical study of the Floquet CFT in complex metrics, we find this competition results in a steady state, where the interesting patterns of entanglement
and energy density in space inherit from those of a Floquet CFT in the real time evolution.

\bigskip

The structure of the rest of the paper is organized as follows:
We consider exactly solvable time evolutions in a CFT after different setups of quantum quenches. 
In Sec.\ref{Sec:Global1}, we study an infinite system after a global quench, where the subsystem is chosen as a
semi-infinite system.
In Sec.\ref{Sec:Global2}, we study a semi-infinite system after a global quench, and the subsystem of a finite length is at the end of this semi-infinite system.
In Sec.\ref{Sec:Local}, we study a local quantum quench by joining two CFTs at their ends suddenly.
In Sec.\ref{Sec:FloquetAllowableM}, we consider a periodically driven quantum critical system on a lattice in a complex spacetime metric, 
with the goal of studying the competition between driving and damping. In appendix \ref{Appendix:FreeFermion}, we give some details on the 
complex time evolution in a free fermion lattice model.

\section{Global quench with complex metrics: An infinite system}
\label{Sec:Global1}

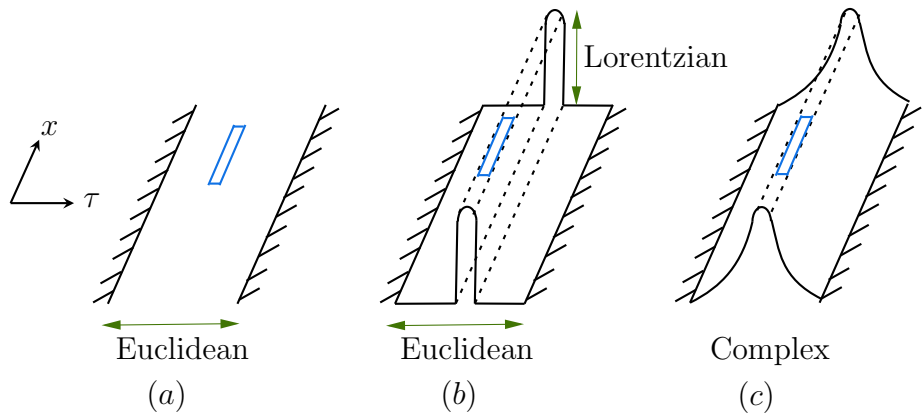
\begin{figure}[h]
\center
\centering
\begin{tikzpicture}[x=0.75pt,y=0.75pt,yscale=-0.8,xscale=0.8]
%uncomment if require: \path (0,235); %set diagram left start at 0, and has height of 235

%Straight Lines [id:da40162350023909277] 
\draw [line width=0.8]  [dash pattern={on 1.69pt off 2.76pt}]  (448.33,135.5) -- (504,10.5) ;
%Straight Lines [id:da9294299589899353] 
\draw [line width=0.8]  [dash pattern={on 1.69pt off 2.76pt}]  (439.33,134.5) -- (495,9.5) ;
%Straight Lines [id:da3513089450426442] 
\draw [line width=0.8]  [dash pattern={on 1.69pt off 2.76pt}]  (260.33,137.17) -- (316,12.17) ;
%Straight Lines [id:da7175995703917796] 
\draw [line width=0.8]    (85.67,68) -- (30,193.5) ;
%Straight Lines [id:da5478532843608803] 
\draw [line width=0.8]    (167.33,68) -- (111.67,193.5) ;
%Straight Lines [id:da9047378354123039] 
\draw [line width=0.8]    (83.67,72) -- (70.67,79.33) ;
%Straight Lines [id:da6330289656480901] 
\draw [line width=0.8]    (79,84) -- (66,91.33) ;
%Straight Lines [id:da6404254092939483] 
\draw [line width=0.8]    (72.67,97) -- (59.67,104.33) ;
%Straight Lines [id:da16671361023818343] 
\draw [line width=0.8]    (68,109) -- (55,116.33) ;
%Straight Lines [id:da6252095957508644] 
\draw [line width=0.8]    (61.67,121) -- (48.67,128.33) ;
%Straight Lines [id:da12872990997127698] 
\draw [line width=0.8]    (57,133) -- (44,140.33) ;
%Straight Lines [id:da39353899760122435] 
\draw [line width=0.8]    (50.67,146) -- (37.67,153.33) ;
%Straight Lines [id:da5133398742699342] 
\draw [line width=0.8]    (46,158) -- (33,165.33) ;
%Straight Lines [id:da24349993083656518] 
\draw [line width=0.8]    (38.33,173.33) -- (25.33,180.67) ;
%Straight Lines [id:da8657234475292466] 
\draw [line width=0.8]    (33.67,185.33) -- (20.67,192.67) ;
%Straight Lines [id:da30818289636382856] 
\draw [line width=0.8]    (175.67,70.33) -- (162.67,77.67) ;
%Straight Lines [id:da3895200240996044] 
\draw [line width=0.8]    (171,82.33) -- (158,89.67) ;
%Straight Lines [id:da5961349613116551] 
\draw [line width=0.8]    (164.67,95.33) -- (151.67,102.67) ;
%Straight Lines [id:da9725219750176017] 
\draw [line width=0.8]    (160,107.33) -- (147,114.67) ;
%Straight Lines [id:da449809324082154] 
\draw [line width=0.8]    (153.67,119.33) -- (140.67,126.67) ;
%Straight Lines [id:da12331305421684091] 
\draw [line width=0.8]    (149,131.33) -- (136,138.67) ;
%Straight Lines [id:da5126924271083251] 
\draw [line width=0.8]    (142.67,144.33) -- (129.67,151.67) ;
%Straight Lines [id:da26594069146634103] 
\draw [line width=0.8]    (138,156.33) -- (125,163.67) ;
%Straight Lines [id:da10137973768913111] 
\draw [line width=0.8]    (130.33,171.67) -- (117.33,179) ;
%Straight Lines [id:da09773917374421348] 
\draw [line width=0.8]    (125.67,183.67) -- (112.67,191) ;
%Straight Lines [id:da5280781148840789] 
\draw [color={rgb, 255:red, 16; green, 114; blue, 229 }  ,draw opacity=1 ][line width=0.8]    (109.67,80.67) -- (93.33,118.67) ;
%Straight Lines [id:da5557864737079223] 
\draw [color={rgb, 255:red, 16; green, 114; blue, 229 }  ,draw opacity=1 ][line width=0.8]    (116.67,80.67) -- (100.33,118.67) ;
%Straight Lines [id:da5295049919664251] 
\draw [color={rgb, 255:red, 16; green, 114; blue, 229 }  ,draw opacity=1 ][line width=0.8]    (108.67,81.67) -- (115.67,81.67) ;
%Straight Lines [id:da22484305846377672] 
\draw [color={rgb, 255:red, 16; green, 114; blue, 229 }  ,draw opacity=1 ][line width=0.8]    (94.33,117.67) -- (101.33,117.67) ;
%Straight Lines [id:da08149398105580108] 
\draw [line width=0.8]    (266.67,68) -- (211,193.5) ;
%Straight Lines [id:da2985119950885793] 
\draw [line width=0.8]    (348.33,68) -- (292.67,193.5) ;
%Straight Lines [id:da15051174471371198] 
\draw [line width=0.8]    (264.67,72) -- (251.67,79.33) ;
%Straight Lines [id:da46415373233829926] 
\draw [line width=0.8]    (260,84) -- (247,91.33) ;
%Straight Lines [id:da2776906895320953] 
\draw [line width=0.8]    (253.67,97) -- (240.67,104.33) ;
%Straight Lines [id:da33045681435812646] 
\draw [line width=0.8]    (249,109) -- (236,116.33) ;
%Straight Lines [id:da8388210095064927] 
\draw [line width=0.8]    (242.67,121) -- (229.67,128.33) ;
%Straight Lines [id:da9349216462006649] 
\draw [line width=0.8]    (238,133) -- (225,140.33) ;
%Straight Lines [id:da5942356881502429] 
\draw [line width=0.8]    (231.67,146) -- (218.67,153.33) ;
%Straight Lines [id:da394879164791631] 
\draw [line width=0.8]    (227,158) -- (214,165.33) ;
%Straight Lines [id:da9917787818111135] 
\draw [line width=0.8]    (219.33,173.33) -- (206.33,180.67) ;
%Straight Lines [id:da11696399135516056] 
\draw [line width=0.8]    (214.67,185.33) -- (201.67,192.67) ;
%Straight Lines [id:da5490804081101984] 
\draw [line width=0.8]    (356.67,70.33) -- (343.67,77.67) ;
%Straight Lines [id:da454655033363909] 
\draw [line width=0.8]    (352,82.33) -- (339,89.67) ;
%Straight Lines [id:da9735725831851769] 
\draw [line width=0.8]    (345.67,95.33) -- (332.67,102.67) ;
%Straight Lines [id:da18622924963713428] 
\draw [line width=0.8]    (341,107.33) -- (328,114.67) ;
%Straight Lines [id:da5209590551869456] 
\draw [line width=0.8]    (334.67,119.33) -- (321.67,126.67) ;
%Straight Lines [id:da477295071379901] 
\draw [line width=0.8]    (330,131.33) -- (317,138.67) ;
%Straight Lines [id:da6665626550700028] 
\draw [line width=0.8]    (323.67,144.33) -- (310.67,151.67) ;
%Straight Lines [id:da5065069404170579] 
\draw [line width=0.8]    (319,156.33) -- (306,163.67) ;
%Straight Lines [id:da9674257529692954] 
\draw [line width=0.8]    (311.33,171.67) -- (298.33,179) ;
%Straight Lines [id:da38883776461934616] 
\draw [line width=0.8]    (306.67,183.67) -- (293.67,191) ;
%Straight Lines [id:da8522873568483882] 
\draw [color={rgb, 255:red, 16; green, 114; blue, 229 }  ,draw opacity=1 ][line width=0.8]    (286.33,75.33) -- (270,113.33) ;
%Straight Lines [id:da6791182041616514] 
\draw [color={rgb, 255:red, 16; green, 114; blue, 229 }  ,draw opacity=1 ][line width=0.8]    (279.33,76.33) -- (286.33,76.33) ;
%Straight Lines [id:da3781050226434217] 
\draw [color={rgb, 255:red, 16; green, 114; blue, 229 }  ,draw opacity=1 ][line width=0.8]    (263,112.33) -- (270,112.33) ;
%Straight Lines [id:da5092330628209587] 
\draw [line width=0.8]    (211,193.5) -- (249.33,193.33) ;
%Straight Lines [id:da8047273882856276] 
\draw [line width=0.8]    (249.33,193.33) -- (250.33,141.17) ;
%Straight Lines [id:da36658072828688415] 
\draw [line width=0.8]    (261.33,193.33) -- (262.33,141.17) ;
%Shape: Arc [id:dp20747075589100283] 
\draw  [draw opacity=0][line width=0.8]  (250.66,141.58) .. controls (250.66,141.56) and (250.66,141.54) .. (250.66,141.51) .. controls (250.65,136.43) and (253.26,132.31) .. (256.48,132.31) .. controls (259.64,132.3) and (262.22,136.24) .. (262.33,141.17) -- (256.5,141.51) -- cycle ; \draw  [line width=0.8]  (250.66,141.58) .. controls (250.66,141.56) and (250.66,141.54) .. (250.66,141.51) .. controls (250.65,136.43) and (253.26,132.31) .. (256.48,132.31) .. controls (259.64,132.3) and (262.22,136.24) .. (262.33,141.17) ;  
%Straight Lines [id:da6687429808940756] 
\draw [line width=0.8]    (261.33,193.33) -- (292.67,193.5) ;
%Straight Lines [id:da4455587713357161] 
\draw [line width=0.8]    (266.67,68.5) -- (305,68.33) ;
%Straight Lines [id:da7051319916457033] 
\draw [line width=0.8]    (305,68.33) -- (306,16.17) ;
%Straight Lines [id:da38059229841461995] 
\draw [line width=0.8]    (317,68.33) -- (318,16.17) ;
%Shape: Arc [id:dp8082775311037147] 
\draw  [draw opacity=0][line width=0.8]  (306,17.44) .. controls (306,17.42) and (306,17.4) .. (306,17.37) .. controls (306,12.29) and (308.6,8.17) .. (311.83,8.17) .. controls (314.99,8.16) and (317.56,12.1) .. (317.68,17.03) -- (311.84,17.37) -- cycle ; \draw  [line width=0.8]  (306,17.44) .. controls (306,17.42) and (306,17.4) .. (306,17.37) .. controls (306,12.29) and (308.6,8.17) .. (311.83,8.17) .. controls (314.99,8.16) and (317.56,12.1) .. (317.68,17.03) ;  
%Straight Lines [id:da5190201178621242] 
\draw [line width=0.8]    (317,68.33) -- (348.33,68.5) ;
%Straight Lines [id:da07935031124740755] 
\draw [line width=0.8]  [dash pattern={on 1.69pt off 2.76pt}]  (261.33,193.33) -- (317,68.33) ;
%Straight Lines [id:da19332587232038623] 
\draw [line width=0.8]  [dash pattern={on 1.69pt off 2.76pt}]  (249.33,193.33) -- (305,68.33) ;
%Straight Lines [id:da9841686782454616] 
\draw [line width=0.8]    (452.67,67.33) -- (397,192.83) ;
%Straight Lines [id:da6541797487679921] 
\draw [line width=0.8]    (534.33,67.33) -- (478.67,192.83) ;
%Straight Lines [id:da6791273531862074] 
\draw [line width=0.8]    (450.67,71.33) -- (437.67,78.67) ;
%Straight Lines [id:da5158993877072018] 
\draw [line width=0.8]    (446,83.33) -- (433,90.67) ;
%Straight Lines [id:da6016309330029606] 
\draw [line width=0.8]    (439.67,96.33) -- (426.67,103.67) ;
%Straight Lines [id:da28443255495356934] 
\draw [line width=0.8]    (435,108.33) -- (422,115.67) ;
%Straight Lines [id:da0014053964013563958] 
\draw [line width=0.8]    (428.67,120.33) -- (415.67,127.67) ;
%Straight Lines [id:da8562265815398841] 
\draw [line width=0.8]    (424,132.33) -- (411,139.67) ;
%Straight Lines [id:da5314431759216951] 
\draw [line width=0.8]    (417.67,145.33) -- (404.67,152.67) ;
%Straight Lines [id:da8251614804306637] 
\draw [line width=0.8]    (413,157.33) -- (400,164.67) ;
%Straight Lines [id:da14169058247284216] 
\draw [line width=0.8]    (405.33,172.67) -- (392.33,180) ;
%Straight Lines [id:da06359750588420232] 
\draw [line width=0.8]    (400.67,184.67) -- (387.67,192) ;
%Straight Lines [id:da16467296496406825] 
\draw [line width=0.8]    (542.67,69.67) -- (529.67,77) ;
%Straight Lines [id:da9280695490185826] 
\draw [line width=0.8]    (538,81.67) -- (525,89) ;
%Straight Lines [id:da7159396568913782] 
\draw [line width=0.8]    (531.67,94.67) -- (518.67,102) ;
%Straight Lines [id:da07360335126251005] 
\draw [line width=0.8]    (527,106.67) -- (514,114) ;
%Straight Lines [id:da5074635121646071] 
\draw [line width=0.8]    (520.67,118.67) -- (507.67,126) ;
%Straight Lines [id:da7777832695335362] 
\draw [line width=0.8]    (516,130.67) -- (503,138) ;
%Straight Lines [id:da3040577139944898] 
\draw [line width=0.8]    (509.67,143.67) -- (496.67,151) ;
%Straight Lines [id:da9898183741989007] 
\draw [line width=0.8]    (505,155.67) -- (492,163) ;
%Straight Lines [id:da26861781714194377] 
\draw [line width=0.8]    (497.33,171) -- (484.33,178.33) ;
%Straight Lines [id:da9669465014119923] 
\draw [line width=0.8]    (492.67,183) -- (479.67,190.33) ;
%Straight Lines [id:da060584888232735445] 
\draw [color={rgb, 255:red, 16; green, 114; blue, 229 }  ,draw opacity=1 ][line width=0.8]    (467.33,74.67) -- (451,112.67) ;
%Straight Lines [id:da002102267203868746] 
\draw [color={rgb, 255:red, 16; green, 114; blue, 229 }  ,draw opacity=1 ][line width=0.8]    (474.33,74.67) -- (458,112.67) ;
%Straight Lines [id:da3899158062623955] 
\draw [color={rgb, 255:red, 16; green, 114; blue, 229 }  ,draw opacity=1 ][line width=0.8]    (466.33,75.67) -- (473.33,75.67) ;
%Straight Lines [id:da4865305677421562] 
\draw [color={rgb, 255:red, 16; green, 114; blue, 229 }  ,draw opacity=1 ][line width=0.8]    (452,111.67) -- (459,111.67) ;
%Shape: Arc [id:dp9899556766380025] 
\draw  [draw opacity=0][line width=0.8]  (436.34,141.77) .. controls (436.34,141.75) and (436.34,141.73) .. (436.34,141.71) .. controls (436.33,136.63) and (438.94,132.5) .. (442.16,132.5) .. controls (445.32,132.49) and (447.9,136.43) .. (448.01,141.36) -- (442.18,141.7) -- cycle ; \draw  [line width=0.8]  (436.34,141.77) .. controls (436.34,141.75) and (436.34,141.73) .. (436.34,141.71) .. controls (436.33,136.63) and (438.94,132.5) .. (442.16,132.5) .. controls (445.32,132.49) and (447.9,136.43) .. (448.01,141.36) ;  
%Shape: Arc [id:dp6191645284026938] 
\draw  [draw opacity=0][line width=0.8]  (492,16.77) .. controls (492,16.75) and (492,16.73) .. (492,16.71) .. controls (492,11.63) and (494.6,7.5) .. (497.83,7.5) .. controls (500.99,7.49) and (503.56,11.43) .. (503.68,16.36) -- (497.84,16.7) -- cycle ; 
\draw  [line width=0.8]  (492,16.77) .. controls (492,16.75) and (492,16.73) .. (492,16.71) .. controls (492,11.63) and (494.6,7.5) .. (497.83,7.5) .. controls (500.99,7.49) and (503.56,11.43) .. (503.68,16.36) ;  
%Curve Lines [id:da858820535723454] 
\draw [line width=0.8]    (448.33,140.5) .. controls (453.33,160.67) and (454,177.33) .. (479.67,190.33) ;
%Curve Lines [id:da7546902182266986] 
\draw [line width=0.8]    (436.34,141.77) .. controls (430,168.67) and (417.33,182.67) .. (397,192.83) ;
%Curve Lines [id:da3512144857873969] 
\draw [line width=0.8]    (504,16.17) .. controls (509,36.33) and (509.67,53) .. (535.33,66) ;
%Curve Lines [id:da020131033634765205] 
\draw [line width=0.8]    (492,17.44) .. controls (485.67,44.33) and (473,58.33) .. (452.67,68.5) ;
%Straight Lines [id:da3177649154103529] 
\draw [line width=0.8]  [dash pattern={on 1.69pt off 2.76pt}]  (251.33,136.17) -- (307,11.17) ;
%Straight Lines [id:da7782128573548234] 
\draw [color={rgb, 255:red, 16; green, 114; blue, 229 }  ,draw opacity=1 ][line width=0.8]    (279.33,75.33) -- (263,113.33) ;
%Straight Lines [id:da913219793364192] 
\draw [color={rgb, 255:red, 65; green, 117; blue, 5 }  ,draw opacity=1 ]   (28,206.98) -- (111,206.02) ;
\draw [shift={(113,206)}, rotate = 179.34] [fill={rgb, 255:red, 65; green, 117; blue, 5 }  ,fill opacity=1 ][line width=0.08]  [draw opacity=0] (12,-3) -- (0,0) -- (12,3) -- cycle    ;
\draw [shift={(26,207)}, rotate = 359.34] [fill={rgb, 255:red, 65; green, 117; blue, 5 }  ,fill opacity=1 ][line width=0.08]  [draw opacity=0] (12,-3) -- (0,0) -- (12,3) -- cycle    ;
%Straight Lines [id:da6463153074874814] 
\draw [color={rgb, 255:red, 65; green, 117; blue, 5 }  ,draw opacity=1 ]   (207.33,207.31) -- (290.33,206.36) ;
\draw [shift={(292.33,206.33)}, rotate = 179.34] [fill={rgb, 255:red, 65; green, 117; blue, 5 }  ,fill opacity=1 ][line width=0.08]  [draw opacity=0] (12,-3) -- (0,0) -- (12,3) -- cycle    ;
\draw [shift={(205.33,207.33)}, rotate = 359.34] [fill={rgb, 255:red, 65; green, 117; blue, 5 }  ,fill opacity=1 ][line width=0.08]  [draw opacity=0] (12,-3) -- (0,0) -- (12,3) -- cycle    ;
%Straight Lines [id:da46232905668850754] 
\draw [color={rgb, 255:red, 65; green, 117; blue, 5 }  ,draw opacity=1 ]   (325.34,66.33) -- (325.66,10) ;
\draw [shift={(325.67,8)}, rotate = 90.32] [fill={rgb, 255:red, 65; green, 117; blue, 5 }  ,fill opacity=1 ][line width=0.08]  [draw opacity=0] (12,-3) -- (0,0) -- (12,3) -- cycle    ;
\draw [shift={(325.33,68.33)}, rotate = 270.32] [fill={rgb, 255:red, 65; green, 117; blue, 5 }  ,fill opacity=1 ][line width=0.08]  [draw opacity=0] (12,-3) -- (0,0) -- (12,3) -- cycle    ;

% Text Node
% Text Node
\draw (33.33,212.73) node [anchor=north west][inner sep=0.75pt]  [font=\normalsize]  {Euclidean};
% Text Node
\draw (212.67,213.07) node [anchor=north west][inner sep=0.75pt]  [font=\normalsize]  {Euclidean};
% Text Node
\draw (328.33,28.73) node [anchor=north west][inner sep=0.75pt]  [font=\normalsize]  {Lorentzian};
% Text Node
\draw (408,212.73) node [anchor=north west][inner sep=0.75pt]  [font=\normalsize]  {Complex};
% Text Node
\draw (52.67,240.07) node [anchor=north west][inner sep=0.75pt]  [font=\normalsize]  {$( a)$};
% Text Node
\draw (238,240.07) node [anchor=north west][inner sep=0.75pt]  [font=\normalsize]  {$( b)$};
% Text Node
\draw (425.33,240.73) node [anchor=north west][inner sep=0.75pt]  [font=\normalsize]  {$( c)$};

%%% x-t axis
     \begin{scope}[xshift=-145pt]
\draw [line width=0.8] [>=stealth,->]   (162,130) -- (202,130) ;
\draw [line width=0.8] [>=stealth,->]   (162,130) -- (180,90) ;
\node at (140pt,62pt){$x$};
\node at (160pt,96pt){$\tau$};
\end{scope}
%%% x-t axis above

\end{tikzpicture}
\caption{Path integral of the reduced density matrix $\rho_A$ after a global quench in a CFT with
(a) Euclidean (b) Partial Euclidean and partial Lorentz, and (c) Complex spacetime metrics.
The fields living on the upper and lower edges of the brunch cut (in blue) correspond to the rows and columns of $\rho_A$.
}
\label{Global}
\end{figure}

For a CFT after a global quantum quench, here we consider the setup introduced in Ref.\cite{CC2015_global}.
We start from an initial state of the following form
\be
|\psi_0\rangle=e^{-\frac{\beta}{4} H_{\text{CFT}}} |B\rangle\rangle,
\label{psi0_global}
\ee
where $|B\rangle\rangle$ is a conformal boundary state. Noting that there is no scale in the conformal boundary state, $|B\rangle\rangle$ is not normalizable itself and has zero real-space entanglement \cite{2015Miyaji}.
By introducing the regularization factor $e^{-\frac{\beta}{4} H_{\text{CFT}}}$ in \eqref{psi0_global}, $|\psi_0\rangle$ is normalizable and has a finite real-space entanglement.
Physically, $\beta$ characterizes the correlation length of the initial state.

In the Lorentz metric, the state after a quantum quench is $|\psi(t)\rangle=e^{-i H_{\text{CFT}}t } |\psi_0\rangle$. The time dependent density matrix can be written as
\be
\rho(t)=e^{-i H_{\text{CFT}}t} e^{-\frac{\beta}{4}H_{\text{CFT}}} |B\rangle\rangle \langle \langle B| e^{-\frac{\beta}{4}H_{\text{CFT}}} e^{i H_{\text{CFT}}t}.
\ee
Then the path integral of 
$\rho_A=\text{Tr}_{\bar A} \,\rho(t)$ for subsystem $A=[x_1,\,x_2]$ can be represented in Fig.\ref{Global} (b) by sewing together the degrees of freedom in region $\bar A$. Here the width of the strip in Fig.\ref{Global} (b) is $\beta/2$ which is introduced by the regularization factor in \eqref{psi0_global}, and the branch cut (blue lines) corresponds to the subsystem $A$.\footnote{For readers who are not familiar with the path integral interpretation of reduced density matrix, see the recent review \cite{Witten2018_RMP,2023_Casini}.}
To study $\rho_A$, it is convenient to take the analytical continuation $\tau=it$ first, by writing $\rho(t)$ as
\be
\rho(\tau)=e^{-(\frac{\beta}{4}+\tau)H_{\text{CFT}}} |B\rangle\rangle \langle \langle B| e^{-(\frac{\beta}{4}-\tau)H_{\text{CFT}}}.
\ee
The path integral of $\rho_A(\tau)$ corresponds to the configuration in the Euclidean space in Fig.\ref{Global} (a).
Then one can evaluate $\rho_A(t)$ or equivalently the entanglement Hamiltonian $K_A(t)$ by taking $\tau\to it$ in the final step \cite{CC2015_global,CardyTonni2016}.

\medskip
Now we are interested in the complex time evolution in \eqref{WF_Quench}, and the density matrix $\rho(t)$ can be written as
\be
\label{rho_t_complex}
\rho(t)=e^{-i H_{\text{CFT}} t}e^{-\left(\frac{\beta}{4}+\epsilon t\right) H_{\text{CFT}}}|B\rangle\rangle\langle\langle B| e^{-\left(\frac{\beta}{4}+\epsilon t\right) H_{\text{CFT}}}
e^{i H_{\text{CFT}} t}.
\ee
The path integral of $\rho_A(t)=\text{Tr}_{\bar A} \rho(t)$ corresponds to the configuration in Fig.\ref{Global} (c). To evaluate $\rho_A(t)$ as well as other physical quantities, 
we define the two-time density matrix $\rho(t_1,t_2)$ as 
\be
\rho(t_1,t_2)=e^{-i H_{\text{CFT}} t_2}e^{-\left(\frac{\beta}{4}+\epsilon t_1\right) H_{\text{CFT}}}|B\rangle\rangle\langle\langle B| e^{-\left(\frac{\beta}{4}+\epsilon t_1\right) H_{\text{CFT}}}
e^{i H_{\text{CFT}} t_2}.
\ee
Then by taking $\tau_2=it_2$ and $\tau_1=t_1$, we have
\be
\label{rho_tau1tau2}
\rho(\tau_1,\tau_2)=e^{-\left(\frac{\beta}{4}+\epsilon \tau_1+\tau_2\right) H_{\text{CFT}}}|B\rangle\rangle\langle\langle B| e^{-\left(\frac{\beta}{4}+\epsilon \tau_1-\tau_2\right) H_{\text{CFT}}},
\ee
The path integral of $\rho(\tau_1,\tau_2)$ as well as the reduced density matrix $\rho_A(\tau_1,\tau_2)=\text{Tr}_{\bar A} \rho(\tau_1,\tau_2)$
is defined in the Euclidean space in Fig.\ref{Global} (a).
In the final step, by considering the following analytical continuation
\be
\tau_1\to t\quad \text{and } \quad \tau_2\to it,
\ee 
one can obtain the time evolution of $\rho_A(t)=\text{Tr}_{\bar A} \, \rho(t)$ and the corresponding entanglement Hamiltonian $K_A(t)$, where $\rho(t)$ is defined in \eqref{rho_t_complex}.

\subsection{Entanglement Hamiltonian and entanglement spectrum evolution}
\label{Sec:EH_global1}

\begin{figure}
\centering
\begin{tikzpicture}[x=0.75pt,y=0.75pt,yscale=-1,xscale=1]
%uncomment if require: \path (0,235); %set diagram left start at 0, and has height of 235

%Shape: Ellipse [id:dp012446082530921054] 
\draw  [color={rgb, 255:red, 0; green, 0; blue, 0 }  ,draw opacity=1 ][line width=0.75]  (335.3,136.71) .. controls (328.39,136.81) and (322.56,121.22) .. (322.28,101.89) .. controls (322,82.56) and (327.39,66.81) .. (334.3,66.71) .. controls (341.21,66.61) and (347.04,82.2) .. (347.32,101.53) .. controls (347.6,120.86) and (342.22,136.61) .. (335.3,136.71) -- cycle ;
%Straight Lines [id:da7734540903331133] 
\draw [color={rgb, 255:red, 0; green, 0; blue, 0 }  ,draw opacity=1 ][line width=0.75]    (334.3,66.71) -- (457.3,66.71) ;
%Straight Lines [id:da6128329957930577] 
\draw [color={rgb, 255:red, 0; green, 0; blue, 0 }  ,draw opacity=1 ][line width=0.75]    (335.3,136.71) -- (460.3,136.71) ;
%Shape: Ellipse [id:dp8597876963858283] 
\draw  [color={rgb, 255:red, 0; green, 0; blue, 0 }  ,draw opacity=1 ][line width=0.75]  (458.3,136.71) .. controls (451.39,136.81) and (445.56,121.22) .. (445.28,101.89) .. controls (445,82.56) and (450.39,66.81) .. (457.3,66.71) .. controls (464.21,66.61) and (470.04,82.2) .. (470.32,101.53) .. controls (470.6,120.86) and (465.22,136.61) .. (458.3,136.71) -- cycle ;
%Straight Lines [id:da47739200755107314] 
\draw [color={rgb, 255:red, 0; green, 0; blue, 0 }  ,draw opacity=1 ][line width=0.75]    (89.3,70.71) -- (239.3,70.71) ;
%Straight Lines [id:da9698760738489608] 
\draw [color={rgb, 255:red, 0; green, 0; blue, 0 }  ,draw opacity=1 ][line width=0.75]    (90.3,140.71) -- (240.3,140.71) ;
%Straight Lines [id:da18939175151330934] 
\draw    (98.3,59.71) -- (89.3,70.71) ;
%Straight Lines [id:da06944178967320291] 
\draw    (109.3,59.71) -- (100.3,70.71) ;
%Straight Lines [id:da7448565153489306] 
\draw    (119.3,59.71) -- (110.3,70.71) ;
%Straight Lines [id:da9519827428067682] 
\draw    (129.3,59.71) -- (120.3,70.71) ;
%Straight Lines [id:da3776595275870759] 
\draw    (139.3,59.71) -- (130.3,70.71) ;
%Straight Lines [id:da6638110766009964] 
\draw    (150.3,59.71) -- (141.3,70.71) ;
%Straight Lines [id:da3886425809450218] 
\draw    (160.3,59.71) -- (151.3,70.71) ;
%Straight Lines [id:da9818385370392111] 
\draw    (170.3,59.71) -- (161.3,70.71) ;
%Straight Lines [id:da9185002173609925] 
\draw    (178.3,59.71) -- (169.3,70.71) ;
%Straight Lines [id:da7477416967423337] 
\draw    (189.3,59.71) -- (180.3,70.71) ;
%Straight Lines [id:da6080274156353704] 
\draw    (199.3,59.71) -- (190.3,70.71) ;
%Straight Lines [id:da40327570124328593] 
\draw    (209.3,59.71) -- (200.3,70.71) ;
%Straight Lines [id:da18694783147397998] 
\draw    (219.3,59.71) -- (210.3,70.71) ;
%Straight Lines [id:da5108752474177245] 
\draw    (230.3,59.71) -- (221.3,70.71) ;
%Straight Lines [id:da22667012655188246] 
\draw    (240.3,59.71) -- (231.3,70.71) ;
%Straight Lines [id:da27238953577608416] 
\draw    (97.3,140.71) -- (88.3,151.71) ;
%Straight Lines [id:da3504034621132074] 
\draw    (108.3,140.71) -- (99.3,151.71) ;
%Straight Lines [id:da9293624454930537] 
\draw    (118.3,140.71) -- (109.3,151.71) ;
%Straight Lines [id:da9895373576764644] 
\draw    (128.3,140.71) -- (119.3,151.71) ;
%Straight Lines [id:da09419375015533626] 
\draw    (138.3,140.71) -- (129.3,151.71) ;
%Straight Lines [id:da6942110288482801] 
\draw    (149.3,140.71) -- (140.3,151.71) ;
%Straight Lines [id:da15679204681522818] 
\draw    (159.3,140.71) -- (150.3,151.71) ;
%Straight Lines [id:da1540639766418559] 
\draw    (169.3,140.71) -- (160.3,151.71) ;
%Straight Lines [id:da832361968629939] 
\draw    (177.3,140.71) -- (168.3,151.71) ;
%Straight Lines [id:da21713192602410047] 
\draw    (188.3,140.71) -- (179.3,151.71) ;
%Straight Lines [id:da6748623508624015] 
\draw    (198.3,140.71) -- (189.3,151.71) ;
%Straight Lines [id:da5651849566678426] 
\draw    (208.3,140.71) -- (199.3,151.71) ;
%Straight Lines [id:da18039882620302228] 
\draw    (218.3,140.71) -- (209.3,151.71) ;
%Straight Lines [id:da7001373936241664] 
\draw    (229.3,140.71) -- (220.3,151.71) ;
%Straight Lines [id:da6137911420841692] 
\draw    (239.3,140.71) -- (230.3,151.71) ;
%Shape: Arc [id:dp1341284100509258] 
\draw  [draw opacity=0] (164.95,100.69) .. controls (163.72,102.68) and (161.51,104) .. (159,104) .. controls (155.13,104) and (152,100.87) .. (152,97) .. controls (152,93.13) and (155.13,90) .. (159,90) .. controls (161.2,90) and (163.17,91.02) .. (164.45,92.61) -- (159,97) -- cycle ; \draw   (164.95,100.69) .. controls (163.72,102.68) and (161.51,104) .. (159,104) .. controls (155.13,104) and (152,100.87) .. (152,97) .. controls (152,93.13) and (155.13,90) .. (159,90) .. controls (161.2,90) and (163.17,91.02) .. (164.45,92.61) ;  
%Straight Lines [id:da9434842873808061] 
\draw [color={rgb, 255:red, 74; green, 144; blue, 226 }  ,draw opacity=1 ][line width=1.5]    (164.45,92.61) -- (239.45,92.61) ;
%Straight Lines [id:da29925487470050016] 
\draw [color={rgb, 255:red, 74; green, 144; blue, 226 }  ,draw opacity=1 ][line width=1.5]    (164.95,100.69) -- (239.95,100.69) ;
%Curve Lines [id:da6833932055919283] 
\draw [color={rgb, 255:red, 74; green, 144; blue, 226 }  ,draw opacity=1 ][line width=1.5]    (345,79.5) .. controls (388,73) and (457,83.5) .. (469,92.5) ;
%Curve Lines [id:da8616106169039495] 
\draw [color={rgb, 255:red, 74; green, 144; blue, 226 }  ,draw opacity=1 ][line width=1.5]    (346.32,85.53) .. controls (389.32,79.03) and (458.32,89.53) .. (470.32,98.53) ;
%Straight Lines [id:da31306338595342365] 
\draw    (265,102.5) -- (293,102.5) ;
\draw [shift={(296,102.5)}, rotate = 180] [fill={rgb, 255:red, 0; green, 0; blue, 0 }  ][line width=0.08]  [draw opacity=0] (8.93,-4.29) -- (0,0) -- (8.93,4.29) -- cycle    ;

\small
\draw (178,162) node  {$|B\rangle\rangle$};
\draw (138,95) node  {$|a\rangle\rangle$};

\draw (162,115) node  {$0+i\tau_2$};

\draw (48,99) node  {$\frac{\beta}{2}+2\epsilon \tau_1$};

\draw (280,88) node  {$w=f(z)$};

\begin{scope}[xshift=30pt]
\draw (140pt,40pt)--(155pt,40pt);
\draw (140pt,40pt)--(140pt,25pt);
\node at (147pt,30pt){$z$};
\end{scope}

\begin{scope}[xshift=190pt]
\draw (140pt,40pt)--(155pt,40pt);
\draw (140pt,40pt)--(140pt,25pt);
\node at (147pt,30pt){$w$};
\end{scope}

\draw [line width=0.5] [>=stealth,<->]   (82,70) -- (82,141) ;

\begin{scope}[xshift=200pt,yshift=-19pt]
\draw [line width=0.5] [>=stealth,<->]   (68,170) -- (195,170) ;
\end{scope}

\draw (400,155) node  {$W$};
\draw (485,122) node  {$|a\rangle\rangle$};
\draw (310,122) node  {$|B\rangle\rangle$};

\begin{scope}[xshift=80pt,yshift=20pt]
\draw [line width=0.5] [>=stealth,->]   (282,95) -- (282,75) ;
\draw [line width=0.5] [>=stealth,->]   (282,95) -- (300,95) ;

\draw (285,68) node  {$v$};
\draw (308,92) node  {$u$};
\end{scope}

\end{tikzpicture}
\caption{The path integral of $\rho_A=\text{Tr}_{\bar A} \rho(\tau_1,\tau_2)$ for $A=[0,+\infty)$ in the Euclidean space.
The width of the strip is $\beta/2+2\epsilon \tau_1$, with the two boundaries located along $\Im z=-\beta/4-\epsilon \tau_1$ 
and $\beta/4+\epsilon \tau_1$ respectively. The branch cut corresponding to subsystem $A$ is along $C=\{i\tau_2+x, \, x\ge 0\}$.
A small disk of radius $\epsilon_0$ is removed at the entanglement point $z_0=0+i\tau_2$, with a conformal boundary condition
$|a\rangle\rangle$ imposed along the boundary of this removed disk.
After a conformal mapping $w=f(z)$, the strip (left) is mapped to a cylinder (right) of length $W$ in the $\Re w$ direction.
The circumference of the cylinder along the $\Im w$ direction is $2\pi$.
}
\label{Rho_A_Global1}
\end{figure}
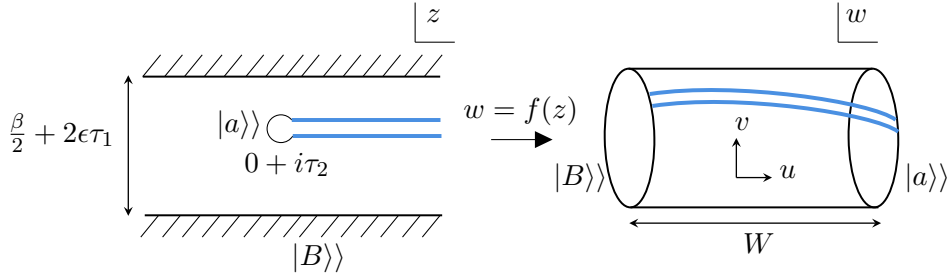

Let us start from $\rho_A(\tau_1,\tau_2)$ in the Euclidean space, as shown in Fig.\ref{Rho_A_Global1}.
Here $\rho_A(\tau_1,\tau_2)$ is obtained from $\rho(\tau_1,\tau_2)$ in \eqref{rho_tau1tau2} by sewing together the degrees of freedom in $\bar A$,
and the branch cut along $C=\{i\tau_2+x, \, x\ge 0\}$ corresponds to the subsystem $A=[0,+\infty)$.

Following the method in \cite{CardyTonni2016}, we consider the conformal mapping
\be
\label{ConformalMap_Global1}
w=f(z)=\log\left(
\frac{\sinh[\pi(z-i\tau_2)/(\beta+4\epsilon \tau_1)]}
{\cosh[\pi(z+i\tau_2)/(\beta+4\epsilon \tau_1)]}
\right),
\ee 
which maps $\rho_A(\tau_1,\tau_2)$ in $z$-plane to a $w$-cylinder, where $w=u+iv$, as shown in Fig.\ref{Rho_A_Global1}.
The entanglement Hamiltonian $K_A$ after the above conformal mapping $w=f(z)$ can be considered as the generator of translation along  the cylinder in the $v$-direction, i.e.,
\be
K_A=-2\pi \int T_{vv} \, d u,
\ee
where the integral is along a constant $v$, and $T_{vv}$ is the Hamiltonian density in Euclidean signature. One can further write $K_A$ in terms of the holomorphic (anti-holomorphic) component of the stress-energy tensor 
$T$ ($\bar T$) as
\be
K_A=\int_{f(C)}T(w)\, dw+\int_{\bar{f(C)} } \bar T(\bar w) \, d\bar w,
\ee
where we have considered the convention $T_{vv}=-(T+\bar T)/2\pi$. By mapping back to the original $z$-plane, the entanglement Hamiltonian becomes
\be
\label{KA_general}
K_A=\int_C \frac{T(z)}{f'(z)} dz+\int_{\bar C} \frac{\bar T(\bar z)}{\bar{f'(z)}} d\bar z.
\ee
Here we have ignored the Schwartzian derivative term, since it will be canceled in the calculation of entanglement entropy by introducing the normalization of $\rho_A$.
As a remark, the above procedure of studying the entanglement Hamiltonian is not limited to the setup of a global quantum quench in Fig.\ref{Rho_A_Global1}.
We can apply the same procedure to other setups in Sec.\ref{Sec:Global2} and Sec.\ref{Sec:Local} later.

Based on \eqref{ConformalMap_Global1} and \eqref{KA_general}, we can obtain the concrete form of 
entanglement Hamiltonian as follows:
\be
\begin{split}
K_A=&\frac{\beta+4\epsilon \tau_1}{\pi}\int_0^\infty \frac{\sinh[\pi x/(\beta+4\epsilon \tau_1)]\cdot \cosh[\pi(x+2i\tau_2)/(\beta+4\epsilon \tau_1)]}{\cosh[2\pi i\tau_2/(\beta+4\epsilon \tau_1)]} T(x+i\tau_2) \,dx\\
&+\frac{\beta+4\epsilon \tau_1}{\pi}\int_0^\infty \frac{\sinh[\pi x/(\beta+4\epsilon \tau_1)]\cdot \cosh[\pi(x-2i\tau_2)/(\beta+4\epsilon \tau_1)]}{\cosh[2\pi i\tau_2/(\beta+4\epsilon \tau_1)]} \bar T(x-i\tau_2) \,dx.
\end{split}
\ee
By taking the analytical continuation $\tau_1\to t$ and $\tau_2\to it$, the entanglement Hamiltonian becomes
\be
\label{KAt_global1}
\begin{split}
K_A(t)=&\frac{\beta+4\epsilon t}{\pi}\int_0^\infty \frac{\sinh[\pi x/(\beta+4\epsilon t)]\cdot \cosh[\pi(x-2t)/(\beta+4\epsilon t)]}{\cosh[2\pi t/(\beta+4\epsilon t)]} T(x-t) \,dx\\
&+\frac{\beta+4\epsilon t}{\pi}\int_0^\infty \frac{\sinh[\pi x/(\beta+4\epsilon t)]\cdot \cosh[\pi(x+2t)/(\beta+4\epsilon t)]}{\cosh[2\pi t/(\beta+4\epsilon t)]} \bar T(x+t) \,dx.
\end{split}
\ee
This entanglement Hamiltonian fully characterizes the property of the subsystem $A$ under a complex time evolution. There are several remarks in order:

First, by setting $\epsilon=0$, the entanglement Hamiltonian $K_A(t)$ reduces to the case in a real-time evolution \cite{CardyTonni2016}.

Second, for the complex time evolution with $\epsilon >0$,  let us consider the  (inverse) entanglement temperature defined in \eqref{EE_temperature}.
For a general $x\gg \beta$, in the long time limit with $t\gg x$ and $t\gg \beta/\epsilon$, one can find
\be
\beta_E(x,t)\simeq \bar \beta_E(x,t)\simeq  2\pi x, \quad \epsilon>0,
\ee
which corresponds to the result in the ground state. 
This is due to the damping effect introduced by the complex time evolution.
As a comparison, in the real time evolution with $x\gg \beta$ and $t\gg x$, we have
\be
\beta_E(x,t)\simeq \beta, \quad \bar\beta_E(x,t)\to \infty, \quad \epsilon=0,
\ee
where the asymmetry can be understood based on the quasi-particle picture, i.e., the entanglement entropy in $A$ is mainly contributed by the right-moving quasi-particles 
emitted from the initial state \cite{CardyTonni2016}.

To have a more straightforward understanding of the time-dependent entanglement Hamiltonian, 
it is helpful to study its spectrum, i.e., the entanglement spectrum, as follows.

\bigskip

\textit{-- Time evolution of entanglement spectrum:}

\bigskip

The spacing of the entanglement spectrum as a function of time $t$ depends on the length $W(t)$ of the cylinder in Fig.\ref{Rho_A_Global1} after an 
analytical continuation $\tau_1\to t$ and $\tau_2\to it$:
\be
\label{ES_general}
E_i(t)-E_j(t)=\frac{\pi (\Delta_i-\Delta_j)}{W(t)},
\ee
where $\Delta_i$ are the scaling dimensions of boundary operators.
The length $W(t)$ can be obtained as follows. First, by defining
\be
\label{Chiral_W_global1}
\mathcal W=f(i\tau_2+\infty)-f(i\tau_2+\epsilon_0),
\ee
the length of the cylinder in Fig.\ref{Rho_A_Global1} has the form
\be
\label{W_global1}
W=\Re \mathcal W=\frac{1}{2}(\mathcal W+\bar {\mathcal W}).
\ee
From \eqref{Chiral_W_global1}, it is straightforward to obtain that 
\be
\mathcal W=\log\left(\frac{\cosh[\pi(\epsilon_0+2i\tau_2)/(\beta+4\epsilon \tau_1)]}{\sinh[\pi\epsilon_0/(\beta+4\epsilon \tau_1)]}\right)-i\frac{2\pi \tau_2}{\beta+4\epsilon \tau_1}.
\ee
Then based on \eqref{W_global1}, after an analytical continuation $\tau_1\to t$ and $\tau_2\to it$, one can obtain
\be
\label{W:Global1}
W(t)=\frac{1}{2} \log\left(\frac{ \cosh[2\pi \epsilon_0/(\beta+4\epsilon t)]+\cosh[ 4\pi t/(\beta+4\epsilon t)]}{2\sinh^2[\pi\epsilon_0/(\beta+4\epsilon t)]}\right), \quad \epsilon\ge 0.
\ee
Now let us consider the long time evolution limit $t\to \infty$ or more precisely $t\gg \beta/\epsilon$. The scaling behavior of $W(t)$ depends on whether
the time evolution is real  ($\epsilon=0$) or complex ($\epsilon> 0$) as follows
\be
\label{W:Global1_longtime}
\left\{
\begin{split}
&W(t\to \infty)\sim \frac{2\pi t}{\beta}, \quad \text{real time evolution }(\epsilon=0),\\
&W(t\to \infty)\sim \log t, \quad \text{complex time evolution (}\epsilon> 0).
\end{split}
\right.
\ee
Then based on the relation in \eqref{ES_general}, one can find that in the long time limit $t\to \infty$, 
\be
\label{Global1_ES}
\left\{
\begin{split}
&\text{Spacing of entanglement spectrum}\sim \frac{1}{t}, \quad \text{real time evolution }(\epsilon=0),\\
&\text{Spacing of entanglement spectrum}\sim \frac{1}{\log t}, \quad \text{complex time evolution }(\epsilon> 0).
\end{split}
\right.
\ee
This feature is closely related to the entanglement entropy evolution as discussed in the next subsection.

\subsection{Entanglement entropy evolution}

As seen from \eqref{Renyi_general} and \eqref{vonNeuman_general}, the R\'enyi entropy and von Neumann entropy is determined by the entanglement Hamiltonian $K_A$
and its spectrum. In the complex time evolution of entanglement entropy, the method as used in \cite{CardyTonni2016} still works, which we do not repeat here.
One can find  the $n$-th R\'enyi entropy and von Neumann entropy are related to $W(t)$ in \eqref{W:Global1} as:
\be
\label{SA_general}
S_A^{(n)}(t)\simeq \frac{c}{12}\left(1+\frac{1}{n}\right)W(t)-g_a-g_b, \quad S_A(t)\simeq \frac{c}{6}W(t)-g_a-g_b,
\ee
where $g_a$ and $g_b$ are boundary entropies \cite{1991AL}.
Here we are mainly interested in the leading term which is proportional to $W(t)$, the expression of which has already been given in \eqref{W:Global1}.
Note that the relation in \eqref{SA_general} can be applied to other cases if  the reduced density matrix 
$\rho_A$ can be conformally mapped to a cylinder \cite{CardyTonni2016}.

\begin{figure}[htp]
\center
\centering
\includegraphics[width=3.25in]{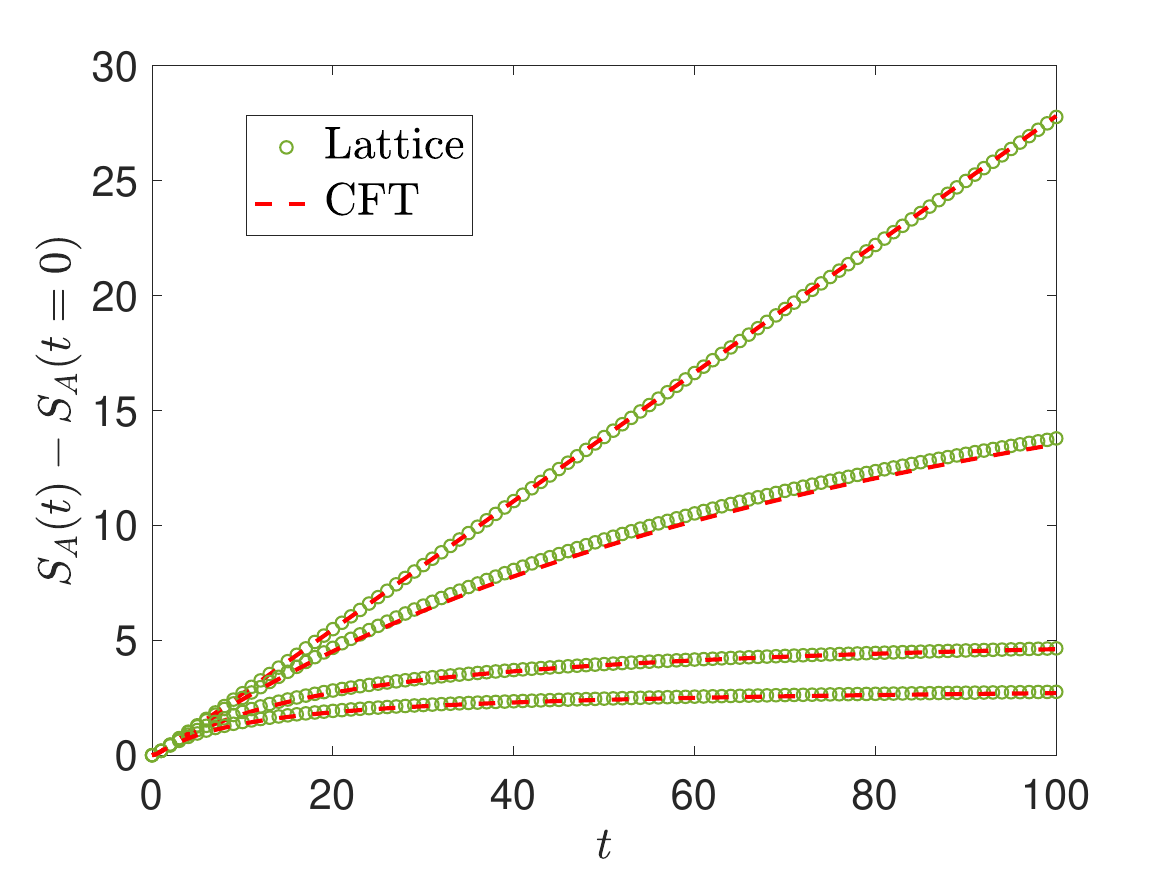}
\caption{
Comparison of complex time evolution of entanglement entropy $S_A(t)$ after a global quench
in lattice system and CFT calculations.
The lattice system is defined on $[0, L]=[0, 800]$, and the subsystem is $A=[0,400]$.
The mass term in the lattice model is set as $m=1/2$ in \eqref{gap_Hamiltonian}.
From top to bottom, we choose $\epsilon$ (which characterizes the imaginary part of the complex time) as 
$\epsilon=0$ (Lorentz metric), $0.01$, $0.05$, and $0.1$
The CFT result is plotted according to \eqref{SA_general} and \eqref{W:Global1}, and the fitting parameters are 
$\beta=3.75$ and $\epsilon_0=0.1$. 
}
\label{LatticeEE}
\end{figure}

As shown in Fig.\ref{LatticeEE}, we compare the CFT results in \eqref{SA_general} and \eqref{W:Global1} 
and a numerical calculation based on the free fermion lattice system (see appendix \ref{Appendix:FreeFermion} for more details), 
and one can find the agreement is remarkable. 
In the real time evolution ($\epsilon=0$), one can find the entanglement entropy grows linearly in time\cite{CC2015_global}.
In the complex time evolution($\epsilon>0$), the entanglement entropy evolution is suppressed in time. 
As we increase $\epsilon$, since the damping effect is stronger, 
one can find $S_A(t)$ is more suppressed, as expected. 

In the long time limit, based on the behavior of $W(t)$ in \eqref{W:Global1_longtime}, one can find
\be
\label{SA:Global1_longtime}
\left\{
\begin{split}
&S_A(t\to \infty)\simeq \frac{c}{3}\cdot \frac{\pi t}{\beta}, \quad \text{real time evolution }(\epsilon=0),\\
&S_A(t\to \infty)\simeq \frac{c}{6}\log t, \quad \text{complex time evolution }(\epsilon> 0).
\end{split}
\right.
\ee
The complex time evolution of entanglement entropy is qualitatively different from that in a real time evolution.

\subsection{Energy density evolution}
\label{Sec:EnergyGlobal1}

Now let us consider the local energy density evolution, i.e., $\langle T_{00}(x,t)\rangle$.
In the real time evolution, it was known that the quench dynamics considered here conserves the energy, with 
$\langle T_{00}(x,t)\rangle=\frac{\pi c}{6\beta^2}$ \cite{2016_CC_review}. Now, because of the damping effect in a complex time evolution, it is expected 
the energy density will decay in time.

More concretely, let us consider the one-point function $\langle T(z)\rangle$, which corresponds to inserting a holomorphic stress-tensor energy operator $T$ 
at $z=x+i\tau_2$ in the strip configuration in Fig.\ref{Rho_A_Global1} (with the branch cut removed now).
By considering the conformal mapping 
\be
\label{conformalMap_global1}
w=f(z)=e^{\frac{2\pi}{\beta+4\epsilon \tau_1}\cdot z},
\ee
which maps the $z$-strip to the $w$ upper-half-plane (UHP), as follows:
\be
\begin{tikzpicture}[x=0.75pt,y=0.75pt,yscale=-1,xscale=1]
\draw [color={rgb, 255:red, 0; green, 0; blue, 0 }  ,draw opacity=1 ][line width=0.75]    (89.3,70.71) -- (239.3,70.71) ;
%Straight Lines [id:da9698760738489608] 
\draw [color={rgb, 255:red, 0; green, 0; blue, 0 }  ,draw opacity=1 ][line width=0.75]    (90.3,140.71) -- (240.3,140.71) ;
%Straight Lines [id:da18939175151330934] 
\draw    (98.3,59.71) -- (89.3,70.71) ;
%Straight Lines [id:da06944178967320291] 
\draw    (109.3,59.71) -- (100.3,70.71) ;
%Straight Lines [id:da7448565153489306] 
\draw    (119.3,59.71) -- (110.3,70.71) ;
%Straight Lines [id:da9519827428067682] 
\draw    (129.3,59.71) -- (120.3,70.71) ;
%Straight Lines [id:da3776595275870759] 
\draw    (139.3,59.71) -- (130.3,70.71) ;
%Straight Lines [id:da6638110766009964] 
\draw    (150.3,59.71) -- (141.3,70.71) ;
%Straight Lines [id:da3886425809450218] 
\draw    (160.3,59.71) -- (151.3,70.71) ;
%Straight Lines [id:da9818385370392111] 
\draw    (170.3,59.71) -- (161.3,70.71) ;
%Straight Lines [id:da9185002173609925] 
\draw    (178.3,59.71) -- (169.3,70.71) ;
%Straight Lines [id:da7477416967423337] 
\draw    (189.3,59.71) -- (180.3,70.71) ;
%Straight Lines [id:da6080274156353704] 
\draw    (199.3,59.71) -- (190.3,70.71) ;
%Straight Lines [id:da40327570124328593] 
\draw    (209.3,59.71) -- (200.3,70.71) ;
%Straight Lines [id:da18694783147397998] 
\draw    (219.3,59.71) -- (210.3,70.71) ;
%Straight Lines [id:da5108752474177245] 
\draw    (230.3,59.71) -- (221.3,70.71) ;
%Straight Lines [id:da22667012655188246] 
\draw    (240.3,59.71) -- (231.3,70.71) ;
%Straight Lines [id:da27238953577608416] 
\draw    (97.3,140.71) -- (88.3,151.71) ;
%Straight Lines [id:da3504034621132074] 
\draw    (108.3,140.71) -- (99.3,151.71) ;
%Straight Lines [id:da9293624454930537] 
\draw    (118.3,140.71) -- (109.3,151.71) ;
%Straight Lines [id:da9895373576764644] 
\draw    (128.3,140.71) -- (119.3,151.71) ;
%Straight Lines [id:da09419375015533626] 
\draw    (138.3,140.71) -- (129.3,151.71) ;
%Straight Lines [id:da6942110288482801] 
\draw    (149.3,140.71) -- (140.3,151.71) ;
%Straight Lines [id:da15679204681522818] 
\draw    (159.3,140.71) -- (150.3,151.71) ;
%Straight Lines [id:da1540639766418559] 
\draw    (169.3,140.71) -- (160.3,151.71) ;
%Straight Lines [id:da832361968629939] 
\draw    (177.3,140.71) -- (168.3,151.71) ;
%Straight Lines [id:da21713192602410047] 
\draw    (188.3,140.71) -- (179.3,151.71) ;
%Straight Lines [id:da6748623508624015] 
\draw    (198.3,140.71) -- (189.3,151.71) ;
%Straight Lines [id:da5651849566678426] 
\draw    (208.3,140.71) -- (199.3,151.71) ;
%Straight Lines [id:da18039882620302228] 
\draw    (218.3,140.71) -- (209.3,151.71) ;
%Straight Lines [id:da7001373936241664] 
\draw    (229.3,140.71) -- (220.3,151.71) ;
%Straight Lines [id:da6137911420841692] 
\draw    (239.3,140.71) -- (230.3,151.71) ;

\draw    (265,102.5) -- (293,102.5) ;
\draw [shift={(296,102.5)}, rotate = 180] [fill={rgb, 255:red, 0; green, 0; blue, 0 }  ][line width=0.08]  [draw opacity=0] (8.93,-4.29) -- (0,0) -- (8.93,4.29) -- cycle    ;

\begin{scope}[xshift=170pt,yshift=-10pt]
\draw    (97.3,140.71) -- (88.3,151.71) ;
%Straight Lines [id:da3504034621132074] 
\draw    (108.3,140.71) -- (99.3,151.71) ;
%Straight Lines [id:da9293624454930537] 
\draw    (118.3,140.71) -- (109.3,151.71) ;
%Straight Lines [id:da9895373576764644] 
\draw    (128.3,140.71) -- (119.3,151.71) ;
%Straight Lines [id:da09419375015533626] 
\draw    (138.3,140.71) -- (129.3,151.71) ;
%Straight Lines [id:da6942110288482801] 
\draw    (149.3,140.71) -- (140.3,151.71) ;
%Straight Lines [id:da15679204681522818] 
\draw    (159.3,140.71) -- (150.3,151.71) ;
%Straight Lines [id:da1540639766418559] 
\draw    (169.3,140.71) -- (160.3,151.71) ;
%Straight Lines [id:da832361968629939] 
\draw    (177.3,140.71) -- (168.3,151.71) ;
%Straight Lines [id:da21713192602410047] 
\draw    (188.3,140.71) -- (179.3,151.71) ;
%Straight Lines [id:da6748623508624015] 
\draw    (198.3,140.71) -- (189.3,151.71) ;
%Straight Lines [id:da5651849566678426] 
\draw    (208.3,140.71) -- (199.3,151.71) ;
%Straight Lines [id:da18039882620302228] 
\draw    (218.3,140.71) -- (209.3,151.71) ;
%Straight Lines [id:da7001373936241664] 
\draw    (229.3,140.71) -- (220.3,151.71) ;
%Straight Lines [id:da6137911420841692] 
\draw    (239.3,140.71) -- (230.3,151.71) ;

\draw [color={rgb, 255:red, 0; green, 0; blue, 0 }  ,draw opacity=1 ][line width=0.75]    (90.3,140.71) -- (240.3,140.71) ;

\small
\draw (158,99) node  {$\bullet$};
\draw (172,99) node  {$w$};
\end{scope}

\small
%\draw (178,162) node  {$|B\rangle\rangle$};
\draw (138,99) node  {$\bullet$};
\draw (152,99) node  {$z$};

\draw (48,99) node  {$\frac{\beta}{2}+2\epsilon \tau_1$};

\draw (280,88) node  {$w=f(z)$};

%\begin{scope}[xshift=30pt]
%\draw (140pt,40pt)--(155pt,40pt);
%\draw (140pt,40pt)--(140pt,25pt);
%\node at (147pt,30pt){$z$};
%\end{scope}

\draw [line width=0.5] [>=stealth,<->]   (82,70) -- (82,141) ;

\end{tikzpicture}
\ee
The stress-energy tensor transforms as
\be\label{T_transform3}
T_{\text{strip}}(z)=\left(\frac{dz}{dw}\right)^{-2} 
T_{\text{UHP}}(w)- \frac{c}{12}\{w,z\},
\ee
where $\{w,\,z\}$ is the Schwarzian derivative defined by
\be
\label{Sch_def2}
\{w,z\}:=\frac{d^3w/dz^3}{dw/dz}-\frac{3}{2}\left(
\frac{d^2w/dz^2}{dw/dz}
\right)^2.
\ee
Since $\langle T_{\text{UHP}}(w)\rangle=0$, we have
$
\langle T_{\text{strip}}(z)\rangle=-\frac{c}{12} \{w,\,z\}$,
based on which and \eqref{conformalMap_global1} one can obtain (after an analytical continuation $\tau_1\to t$ and $\tau_2\to it$)
\be
\langle T_{\text{strip}}(x, t)\rangle=\frac{c}{24}\cdot \frac{\pi^2}{\left(\frac{\beta}{2}+2\epsilon t\right)^2}.
\ee
One can obtain the same expression for $\langle \bar T_{\text{strip}}(x, t)\rangle$.
Therefore, the local energy density evolves in time as
\be
\label{EnergyGlobal1}
\langle
T_{00}(x,t)\rangle
=\frac{1}{2\pi}\big[\langle T(x, t)\rangle+\langle \bar T(x, t)\rangle\big]
= \frac{\pi c}{24(\frac{\beta}{2}+ 2\epsilon t)^2}.
\ee
For $\epsilon=0$, we reproduce the real time evolution result in \cite{2016_CC_review,1986_BCN}.
For $\epsilon> 0$, the energy density decays in time because of the complex time.
As $t\to +\infty$, the energy density decays to the ground state value in time as $1/t^2$, which is consistent with our analysis in the 
entanglement Hamiltonian evolution in Sec.\ref{Sec:EH_global1}.

\medskip
A comparison of the energy density evolution in lattice and CFT calculations for different values of $\epsilon$ can be found in Fig.\ref{LatticeEnergy_global}, 
where one can find a remarkable agreement. See appendix \ref{Appendix:FreeFermion} for more details on the lattice calculations.

\begin{figure}[h]
\center
\centering
\includegraphics[width=3.25in]{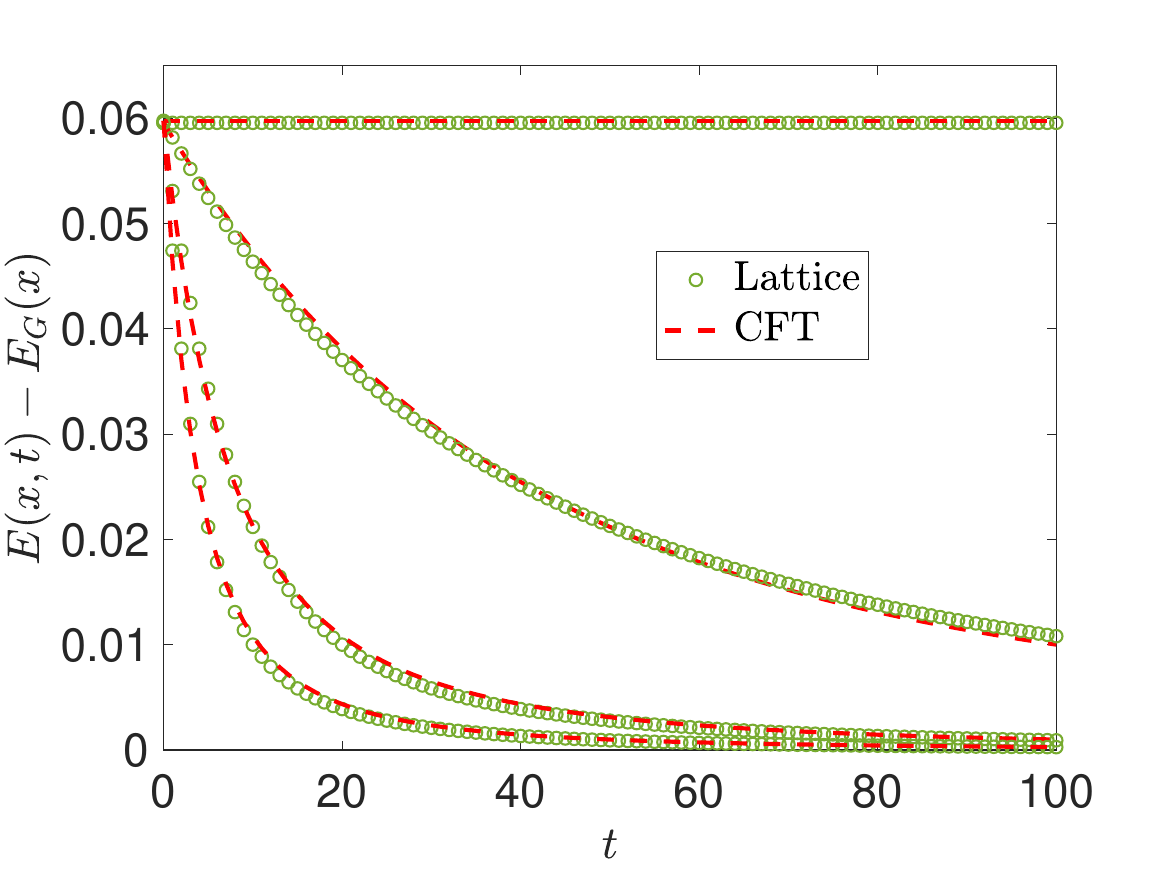}
\caption{
Comparison of energy density evolution $E(x,t)-E_G(x)$ after a global quench in complex spacetime metric
in a free fermion lattice system (green circles) and in a CFT calculation (red dashed lines). Here $E_G(x)$ is the energy density in the ground state of a CFT.
The lattice system is defined on $[0, L]=[0, 800]$. The mass term used in the initial state of the lattice model is set as $m=1/2$ in \eqref{gap_Hamiltonian}.
The energy density is calculated by considering the average $\frac{1}{l}\int_{L/2-l/2}^{L/2+l/2} E(x,t)dx$ with $l=100$ in the lattice calculation. From top to bottom, we have $\epsilon=0$, $0.01$, $0.05$, and $0.1$.
The fitting parameter in CFT is $\beta=2.96$.
}
\label{LatticeEnergy_global}
\end{figure}

\section{Global quench with complex metrics: A semi-infinite system}
\label{Sec:Global2}

Now we consider a semi-infinite system in $[0,+\infty)$ after a global quantum quench, where the subsystem $A=[0,\,l]$ 
is at the end of this semi-infinite system.
In the real-time evolution, it is known that the entanglement  entropy evolution as well as other quantities depends on the length scale $l$\cite{2016_CC_review}.
For example, the entanglement entropy $S_A(t)$ grows linearly in time if $t<l$, and saturates if $t>l$. 
Also, as studied in Ref.\cite{2018WenRyuLudwig,2020Zhu}, for $t>l$, the entanglement Hamiltonian and entanglement spectrum for subsystem
$A=[0,l]$ approaches those in a thermal ensemble at a finite temperature exponentially fast in time.
Here we will generalize this setup to the case of a complex-time evolution.

\subsection{Entanglement Hamiltonian and entanglement spectrum evolution}

Let us first consider the path integral of $\rho_A(\tau_1,\tau_2)=\text{Tr}\rho(\tau_1,\tau_2)$ in 
Euclidean spacetime. Here $\rho(\tau_1,\tau_2)$ has the same definition in \eqref{rho_tau1tau2}
except that now the initial state is defined over a semi-infinite system. Pictorially, the path integral of $\rho_A(\tau_1,\tau_2)$ corresponds to
a half strip in $z$-plane as follows:
\be
\label{rhoA_semi_PI}
\begin{tikzpicture}[x=0.75pt,y=0.75pt,yscale=-1,xscale=1]
%uncomment if require: \path (0,235); %set diagram left start at 0, and has height of 235

%Straight Lines [id:da47739200755107314] 
\draw [color={rgb, 255:red, 0; green, 0; blue, 0 }  ,draw opacity=1 ][line width=0.75]    (100.3,39.71) -- (250.3,39.71) ;
%Straight Lines [id:da9576962348624866] 
\draw [color={rgb, 255:red, 0; green, 0; blue, 0 }  ,draw opacity=1 ][line width=0.75]    (100.3,100.71) -- (250.3,100.71) ;
%Straight Lines [id:da6513516693134163] 
\draw    (100,101) -- (100.3,39.71) ;
%Straight Lines [id:da4291444325429079] 
\draw    (110,40) -- (120,30) ;
%Straight Lines [id:da4412399050425261] 
\draw    (100.3,39.71) -- (110.3,29.71) ;
%Straight Lines [id:da8280452883882754] 
\draw    (120,40) -- (130,30) ;
%Straight Lines [id:da489097453810567] 
\draw    (140,40) -- (150,30) ;
%Straight Lines [id:da5621273261100278] 
\draw    (130.3,39.71) -- (140.3,29.71) ;
%Straight Lines [id:da8973268558599854] 
\draw    (150,40) -- (160,30) ;
%Straight Lines [id:da2639243602786099] 
\draw    (169,40) -- (179,30) ;
%Straight Lines [id:da029491409870951002] 
\draw    (159.3,39.71) -- (169.3,29.71) ;
%Straight Lines [id:da8250753267903806] 
\draw    (179,40) -- (189,30) ;
%Straight Lines [id:da1699849714179208] 
\draw    (199,40) -- (209,30) ;
%Straight Lines [id:da8018672244108169] 
\draw    (189.3,39.71) -- (199.3,29.71) ;
%Straight Lines [id:da8767522443510545] 
\draw    (209,40) -- (219,30) ;
%Straight Lines [id:da1114480538106386] 
\draw    (230,40) -- (240,30) ;
%Straight Lines [id:da12017522029189143] 
\draw    (220.3,39.71) -- (230.3,29.71) ;
%Straight Lines [id:da04039368558741241] 
\draw    (240,40) -- (250,30) ;
%Straight Lines [id:da9625396970722556] 
\draw    (111,111) -- (121,101) ;
%Straight Lines [id:da5085043040654217] 
\draw    (101.3,110.71) -- (111.3,100.71) ;
%Straight Lines [id:da3425532762483626] 
\draw    (121,111) -- (131,101) ;
%Straight Lines [id:da9432466523201901] 
\draw    (141,111) -- (151,101) ;
%Straight Lines [id:da6982977056783359] 
\draw    (131.3,110.71) -- (141.3,100.71) ;
%Straight Lines [id:da008195578884815036] 
\draw    (151,111) -- (161,101) ;
%Straight Lines [id:da5656594237121133] 
\draw    (170,111) -- (180,101) ;
%Straight Lines [id:da05750756058683448] 
\draw    (160.3,110.71) -- (170.3,100.71) ;
%Straight Lines [id:da5718263710831482] 
\draw    (180,111) -- (190,101) ;
%Straight Lines [id:da8856232255178303] 
\draw    (200,111) -- (210,101) ;
%Straight Lines [id:da937019625783976] 
\draw    (190.3,110.71) -- (200.3,100.71) ;
%Straight Lines [id:da9726233525968004] 
\draw    (210,111) -- (220,101) ;
%Straight Lines [id:da9649125016123422] 
\draw    (231,111) -- (241,101) ;
%Straight Lines [id:da35755517695199324] 
\draw    (221.3,110.71) -- (231.3,100.71) ;
%Straight Lines [id:da19656988048432267] 
\draw    (241,111) -- (251,101) ;
%Straight Lines [id:da4774246560920056] 
\draw    (90,111) -- (100,101) ;
%Straight Lines [id:da3973171567437922] 
\draw    (90.3,100.71) -- (100.3,90.71) ;
%Straight Lines [id:da6785772395277093] 
\draw    (90,90) -- (100,80) ;
%Straight Lines [id:da9711843788227461] 
\draw    (90.3,79.71) -- (100.3,69.71) ;
%Straight Lines [id:da8999930357491253] 
\draw    (90,71) -- (100,61) ;
%Straight Lines [id:da8478501864882666] 
\draw    (90.3,60.71) -- (100.3,50.71) ;
%Straight Lines [id:da6714568002918281] 
\draw    (90,50) -- (100,40) ;
%Shape: Arc [id:dp33132549052073923] 
\draw  [draw opacity=0] (161.67,56.79) .. controls (162.92,54.84) and (165.09,53.52) .. (167.59,53.47) .. controls (171.56,53.39) and (174.85,56.55) .. (174.93,60.52) .. controls (175.01,64.49) and (171.85,67.78) .. (167.88,67.86) .. controls (165.17,67.92) and (162.79,66.47) .. (161.51,64.28) -- (167.73,60.67) -- cycle ; \draw   (161.67,56.79) .. controls (162.92,54.84) and (165.09,53.52) .. (167.59,53.47) .. controls (171.56,53.39) and (174.85,56.55) .. (174.93,60.52) .. controls (175.01,64.49) and (171.85,67.78) .. (167.88,67.86) .. controls (165.17,67.92) and (162.79,66.47) .. (161.51,64.28) ;  
%Straight Lines [id:da5567565992405998] 
\draw [color={rgb, 255:red, 74; green, 144; blue, 226 }  ,draw opacity=1 ][line width=1.5]    (100,56) -- (163,56) ;
%Straight Lines [id:da8599214410394936] 
\draw [color={rgb, 255:red, 74; green, 144; blue, 226 }  ,draw opacity=1 ][line width=1.5]    (100,64) -- (162,64) ;

\small

\draw (48,69) node  {$\frac{\beta}{2}+2\epsilon \tau_1$};

\draw (170,74) node  {$l+i \tau_2$};

\draw [line width=0.5] [>=stealth,<->]   (82,40) -- (82,103) ;

\draw [line width=0.5] [>=stealth,->]   (282,95) -- (282,75) ;
\draw [line width=0.5] [>=stealth,->]   (282,95) -- (300,95) ;

\draw (285,68) node  {$\tau$};
\draw (308,92) node  {$x$};

\end{tikzpicture}
\ee
There are three boundaries here: the first two boundaries which correspond to the initial state are along $\Im z=-\beta/4-\epsilon \tau_1$ and $\Im z=\beta/4+\epsilon \tau_1$ respectively. The third boundary along $\Re z=0$ corresponds to the physical boundary of the semi-infinite system.
The branch cut corresponding to the subsystem $A$ is along $C=\{i\tau_2+x, \, 0\le x\le l\}$. To introduce a UV cutoff,  a small disk of radius $\epsilon_0$ is removed at the entangling point at $z=l+i\tau_2$.

Next, we map $\rho_A(\tau_1,\tau_2)$ in \eqref{rhoA_semi_PI} to a $w$-cylinder (see the right plot in Fig.\ref{Rho_A_Global1}) by using the following two-step conformal mapping:
\be
\left\{
\begin{split}
\xi(z)=&\sinh\left(\frac{2\pi z}{\beta+4\epsilon \tau_1}\right),\\
w(\xi)=&-\log \left(\frac{1+\bar \xi_0}{1+\xi_0}\cdot \frac{\xi-\xi_0}{\xi+\bar \xi_0}\right),
\end{split}
\right.
\ee
where $\xi=\xi(z)$ and $\xi_0=\xi(z_0)$ with $z_0=l+i\tau_2$. Then the entanglement Hamiltonian $K_A$ 
corresponds to the generator of translation in the direction of $v=\Im w$ in the $w$-cylinder.
Based on \eqref{KA_general}, one can obtain 
the entanglement Hamiltonian for $A=[0,l]$ at an arbitrary time $t$ as follows (after an analytical continuation $\tau_1\to t$ and $\tau_2\to it$):
\begin{equation}\label{KAtotal_half}
\begin{split}
K_A(t)
=&\frac{\beta+4\epsilon t}{\pi}\int_{l}^{0}
\frac
{
\sinh\left[\frac{\pi (x-l)}{\beta+4\epsilon t}\right]
\cosh\left[\frac{\pi\left(x-2t+l\right)}{\beta+4\epsilon t}\right]
\sinh\left[\frac{\pi(x+l)}{\beta+4\epsilon t}\right]
\cosh\left[\frac{\pi(x-2t-l)}{\beta+4\epsilon t}\right]
}
{\cosh\left(\frac{2\pi}{\beta+4\epsilon t}t\right)\sinh\left(\frac{2\pi}{\beta+4\epsilon t} l\right)
\cosh\left[
\frac{2\pi}{\beta+4\epsilon t}\left(x-t\right)
\right]}
T(x,t)dx\\
\rule{0pt}{.8cm}
&+
\frac{\beta+4\epsilon t}{\pi}\int_{l}^{0}
\frac
{
\sinh\left[\frac{\pi (x-l) }{\beta+4\epsilon t}\right]
\cosh\left[\frac{\pi \left(x+2t+l\right)}{\beta+4\epsilon t}\right]
\sinh\left[\frac{\pi (x+l)}{\beta+4\epsilon t}\right]
\cosh\left[\frac{\pi (x+2t-l)}{\beta+4\epsilon t}\right]
}
{\cosh\left(\frac{2\pi t}{\beta+4\epsilon t}\right)\sinh\left(\frac{2\pi l}{\beta+4\epsilon t}\right)
\cosh\left[
\frac{2\pi \left(x+t\right)}{\beta+4\epsilon t}
\right]}
\overline{T}(x,t)dx.
\end{split}
\end{equation}
For $\epsilon=0$, one can find that $K_A(t)$ reduces to the result in a real-time evolution in \cite{2018WenRyuLudwig}.
One main difference between the real time and complex time evolutions can be observed in the long time limit. In the real time evolution with $t\gg \beta$ and $t>l$, 
the entanglement temperature as defined in \eqref{EE_temperature} is
\be
\beta_E(x,t)\simeq \bar \beta_E(x,t)\simeq \beta, \quad \epsilon=0.
\ee
That is, $\rho_A(t)$ for $A=[0,l]$ is the same as that in a thermal ensemble at a finite temperature $\beta^{-1}$ \cite{2018WenRyuLudwig}.
In the complex time evolution with $t\gg l/\epsilon$, however, the entanglement temperature will reach the following steady value:
\be
\beta_E(x,t)\simeq \bar \beta_E(x,t)\simeq  2\pi \cdot \frac{l^2-x^2}{2l}, \quad \epsilon>0.
\ee
This is nothing but the result for a subsystem $A=[0,l]$ in the ground state of a semi-infinite system in $[0,+\infty)$.
It is due to the damping effect introduced by the complex time.
To have a more intuitive understanding of the time-dependent entanglement Hamiltonian, 
let us take a closer look at its spectrum.

\bigskip

\textit{-- Time evolution of entanglement spectrum:}

\bigskip

By using the same procedure in Sec.\ref{Sec:EH_global1}, one can find the spacing of entanglement spectrum is described by \eqref{ES_general}, 
where $W(t)$ has the expression
\be
\label{W_semi_infinite}
W(t)=\log\left\{
\frac{2\sinh\big[\frac{2\pi}{\beta+4\epsilon t}(l-\frac{\epsilon_0}{2})\big]\cdot \cosh\big(\frac{2\pi t}{\beta+4\epsilon t}\big)}
{\sinh\big(\frac{2\pi}{\beta+4\epsilon t}\cdot\frac{\epsilon_0}{2}\big)\cdot\sqrt{
2\cosh\big(\frac{2\pi}{\beta+4\epsilon t}\cdot 2l\big)+2\cosh\big(\frac{2\pi}{\beta+4\epsilon t}\cdot 2t\big)
}}
\right\},\quad \epsilon\ge 0.
\ee

In the real time evolution ($\epsilon=0$) with $t\gg \beta$, one can find $W(t)$ as\cite{2018WenRyuLudwig}:
\be
\label{Wt_global2_real}
W(t)\simeq \left\{
\begin{split}
\frac{2\pi}{\beta} t, \quad t<l,\\
\frac{2\pi}{\beta} l, \quad t>l.
\end{split}
\right.
\ee
That is, the spacing of entanglement spectrum first decays as $1/t$ for $t<l$, and then saturates at a value which is proportional to $l/\beta$.

In the complex time evolution ($\epsilon>0$), the behavior of $W(t)$ is complicated for a general time $t$, 
but it has very simple features in certain limits:
\be
\label{Wt_global2_complex}
W(t)\simeq \left\{
\begin{split}
\log t, \quad \beta/\epsilon\ll t < l,\\
\log\left(\frac{2 l}{\epsilon_0}\right), \quad \epsilon t\gg l.
\end{split}
\right.
\ee
That is, if $\beta/\epsilon  \ll t <l$, the spacing of entanglement spectrum will decay as $1/\log t$ in time.
 For $\epsilon t\gg l$, the spacing of entanglement spectrum will approach the ground state value.

 As a remark, for certain choices of $\beta$ and $\epsilon$, the condition $\beta/\epsilon \ll l$ in \eqref{Wt_global2_complex} may be not
 satisfied. In this case, $W(t)$ does not have a simple scaling behavior for $t<l$.

\subsection{Entanglement entropy evolution}
\label{Sec:Semi_EE}

\begin{figure}[h]
\center
\centering
\includegraphics[width=3.25in]{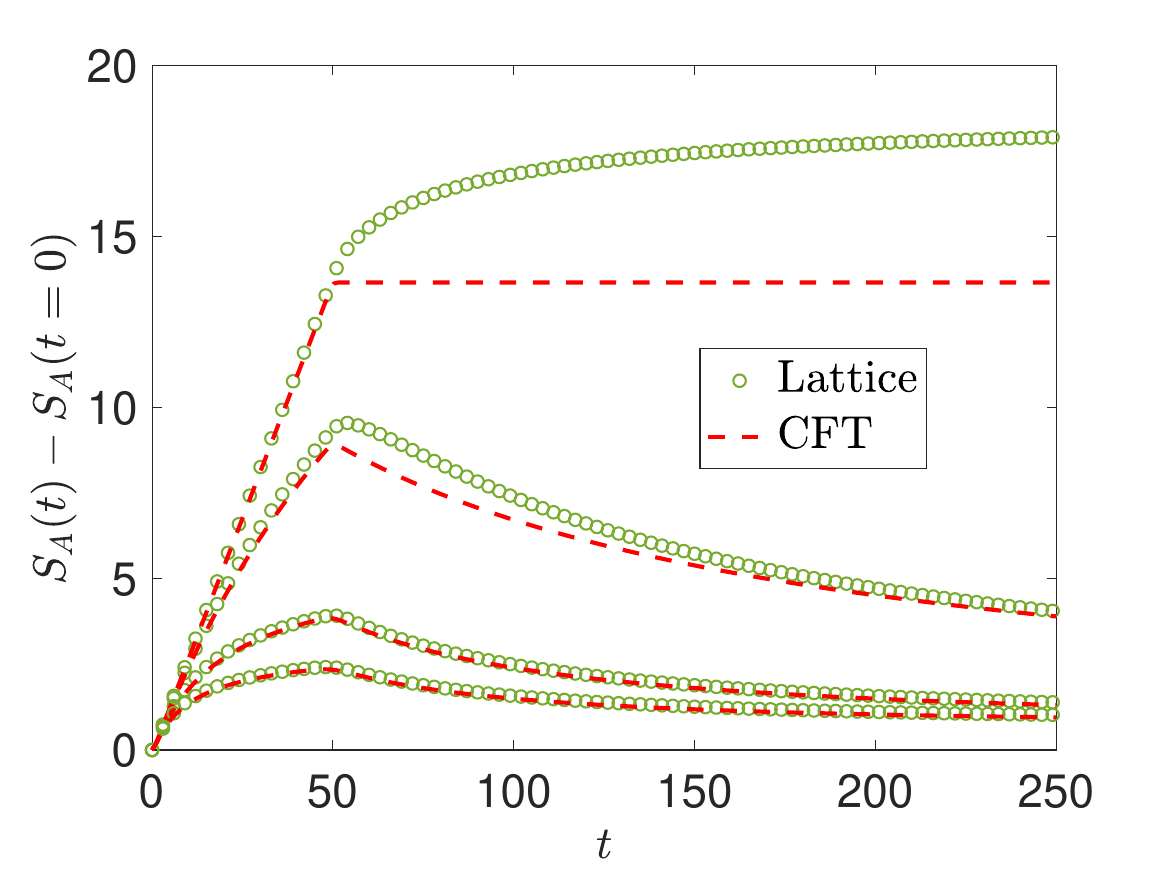}
\includegraphics[width=3.25in]{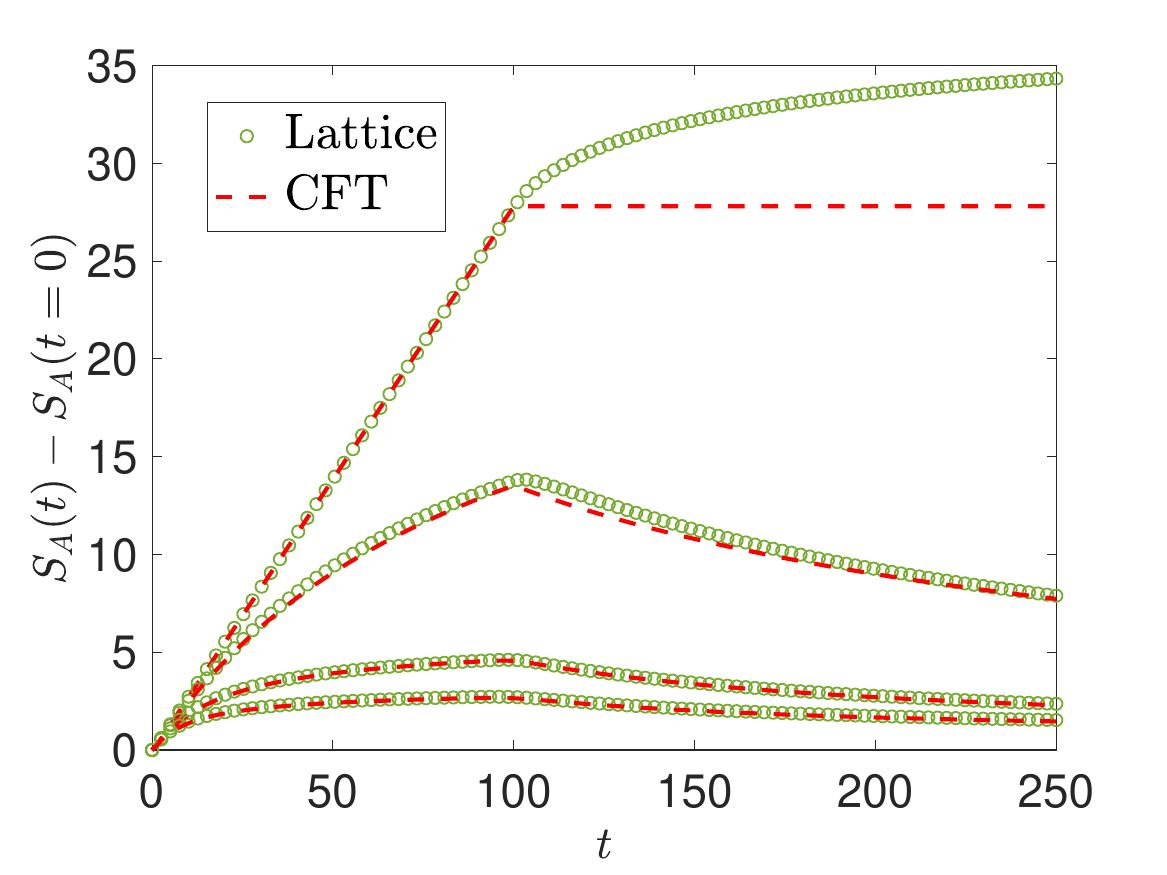}
\caption{
Comparison of complex time evolution of entanglement entropy $S_A(t)$ after a global quench
in lattice system and CFT calculations.
Now the subsystem $[0,l]$ is at the end of the total system.
The parameters are the same as those in Fig.\ref{LatticeEE}  but with $A=[0,50]$ (left plot) and $A=[0,100]$ (right plot).
In each plot, from top to bottom, we have $\epsilon=0$, $0.01$, $0.05$, and $0.1$.
}
\label{LatticeEE2}
\end{figure}

Since the reduced density matrix in \eqref{rhoA_semi_PI} is conformally equivalent to a cylinder, 
the R\'enyi and von-Neumann entropies are related to $W(t)$ through the same equation in \eqref{SA_general}.
Together with \eqref{Wt_global2_real}, one can find that in a real time evolution, $S_A(t)\simeq \frac{\pi c}{3\beta}t$ for $t<l$, and 
$S_A(t)\simeq \frac{\pi c}{3\beta} l$ for $t>l$, which can be understood based on the quasi-particle picture\cite{CC2015_global}. 

\medskip

The feature of $S_A(t)$ becomes qualitatively different in the complex time evolution, as seen in Fig.\ref{LatticeEE2}.
More concretely, one can find that:

\begin{enumerate}

\item For $t<l$, the entanglement entropy evolution is slower than a linear growth. In particular, if $\beta/\epsilon\ll t<l$, one has
\be
\label{EE_logt_case1}
S_A(t)\simeq \frac{c}{6}\cdot \log t.
\ee
Similar to the discussion near \eqref{Wt_global2_complex}, it is possible that $\beta/\epsilon \ll l$ is not satisfied for certain choices of $\beta$ and $\epsilon$.
In this case, $S_A(t)$ does not have a simple form of time dependence as in \eqref{EE_logt_case1}. See, e.g., an exact plot of $S_A(t)$ in Fig.\ref{LatticeEE2}.

\item
For $ t\gg l/\epsilon$, the entanglement entropy reaches a steady value
\be
\label{SA_ground}
S_A(t)\simeq \frac{c}{6}\cdot \log\left(\frac{2l}{\epsilon_0}\right),
\ee
which is the result for the entanglement entropy of $A=[0,l]$ in the ground state of a semi-infinite CFT on $[0,\infty)$.
This is because the excitations injected at $t=0$ will die out due to the
damping effect caused by the complex time. Then in the long time limit $t\gg l/\epsilon$, the state will decay to the ground state of $H_{\text{CFT}}$.

\end{enumerate}

\subsection{Energy density evolution}

The analysis of energy density evolution is similar to that in Sec.\ref{Sec:EnergyGlobal1}.
Now, in the presence of a physical boundary at $x=0$, there will be a modification of the energy density due to Casimir effect.

We consider the 1-point function $\langle T(z)\rangle$ by inserting a holomorphic stress-energy tensor operator 
$T$ at $z=x+i\tau_2$ in the half-strip in \eqref{rhoA_semi_PI} (with the branch cut removed). Then, with 
the conformal mapping
\be
w=f(z)=\sinh\left(\frac{2\pi z}{\beta+4\epsilon \tau_1}\right),
\ee
one can map the half-strip in \eqref{rhoA_semi_PI} to a right-half-plane (RHP) in $w$-coordinate.
Since $\langle T_{\text{RHP}}(w)\rangle=\langle \bar T_{\text{RHP}}(\bar w)\rangle=0$, one can find  
$\langle T_{\text{strip}}(z)\rangle=-\frac{c}{12} \{w,\,z\}$, 
and similarly for $\bar T_{\text{strip}}(\bar z)$,
where $\{w,\,z\}$ is the Schwarzian derivative defined in \eqref{Sch_def2}. One can obtain
\be
\label{T_global2}
\langle T_{\text{strip}}(x,t)\rangle =- \frac{c}{12}\{w,z\}=\frac{\pi^2 c}{6}\cdot \left(\frac{1}{\beta+4\epsilon t}\right)^2\cdot \left[
3 \tanh^2\frac{2\pi (x-t)}{\beta+4\epsilon t}-2
\right].
\ee
Similarly, we have
\be
\label{Tbar_global2}
\langle \bar T_{\text{strip}}(x,t)\rangle =- \frac{c}{12}\{w,z\}=\frac{\pi^2 c}{6}\cdot \left(\frac{1 }{\beta+4\epsilon t}\right)^2\cdot \left[
3 \tanh^2\frac{2\pi (x+t)}{\beta+4\epsilon t}-2
\right].
\ee
One can find that the energy density is not uniform in space.
Before we move on to the time evolution, let us first consider the stress-energy density in the initial state by setting $t=0$.

\subsubsection{Casimir energy density of the gapped initial state with boundary}

\begin{figure}[htp]
\center
\centering
\includegraphics[width=3.05in]{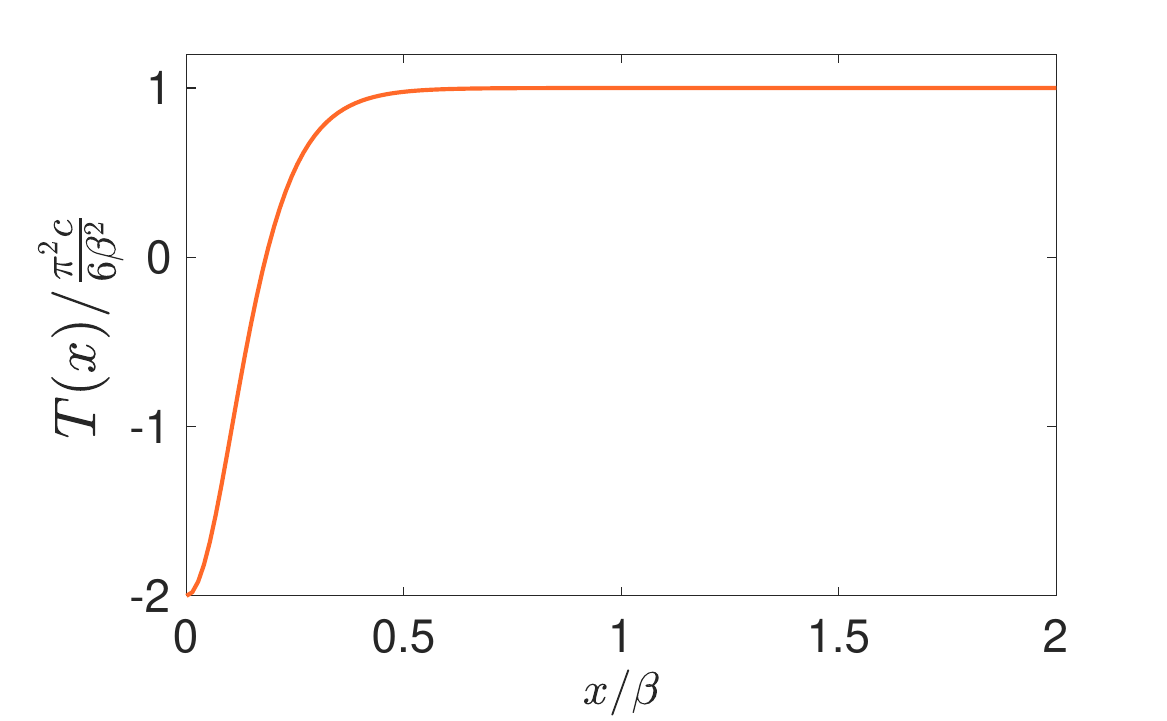}
\caption{
Stress-energy density $\langle T(x)\rangle$ in the initial state $|\psi_0\rangle=e^{-\frac{\beta}{4} H_{\text{CFT}}} |B\rangle\rangle$ for a semi-infinite system in $[0,+\infty)$, where the boundary is 
imposed at $x=0$. For $x>\beta$, $\langle T(x)\rangle$ approaches $\frac{\pi^2 c}{6\beta^2}$, which is the expectation value in an infinite system of finite temperature $\beta$. The negative value near $x=0$ can be understood based on the Casimir effect.
}
\label{Casimir}
\end{figure}

From \eqref{T_global2} and \eqref{Tbar_global2}, the stress-energy density in the initial state is:
\be
\label{TTbar}
\langle T(x)\rangle=\langle \bar T(x) \rangle=
\frac{\pi^2 c}{6\beta^2}\left[
3 \tanh^2\left(\frac{2\pi x}{\beta}\right)-2\right].
\ee
The stress-energy density quickly approaches the value $+\frac{\pi^2 c}{6\beta}$ in an infinite system as $x>\beta$.
However, as we approach the boundary at $x=0$, the energy density becomes negative, which is similar to the feature of Casimir effect.
Indeed, this negative value near the boundary can be understood based on the following physical picture. The initial state $|\psi_0\rangle=e^{-\frac{\beta}{4} H_{\text{CFT}}} |B\rangle\rangle$ has a correlation length $\xi\sim \mathcal O (\beta)$. Near the boundary, the system is approximately a CFT living in a finite interval $[0,  \xi]$, which gives rise to
the negative energy density.\footnote{One way to understand the CFT behavior in this short-range entangled initial state is to consider the entanglement Hamiltonian \eqref{KAtotal_half} for subsystem $[0,l]$ with $l\ll \beta$ at $t=0$. One can find the (inverse) entanglement temperature is $\beta_E(x)=\bar \beta_E(x)\simeq 2\pi (l^2-x^2)/2l$, which corresponds to the CFT result.
} 
Certainly, this argument is not rigorous, since it is apparent from Fig.\ref{Casimir} that the negative energy density near the boundary is not uniform.

As a remark, the above point of view was previously considered in Ref.\cite{2017Ryu}  in the study of entanglement Hamiltonian/spectrum of a gapped system, 
where the entanglement Hamiltonian/spectrum can be considered as that of a CFT living in a length scale of the 
correlation length $\xi$ with appropriate boundary conditions.

\begin{figure}[h]
\center
\centering
\includegraphics[width=3.25in]{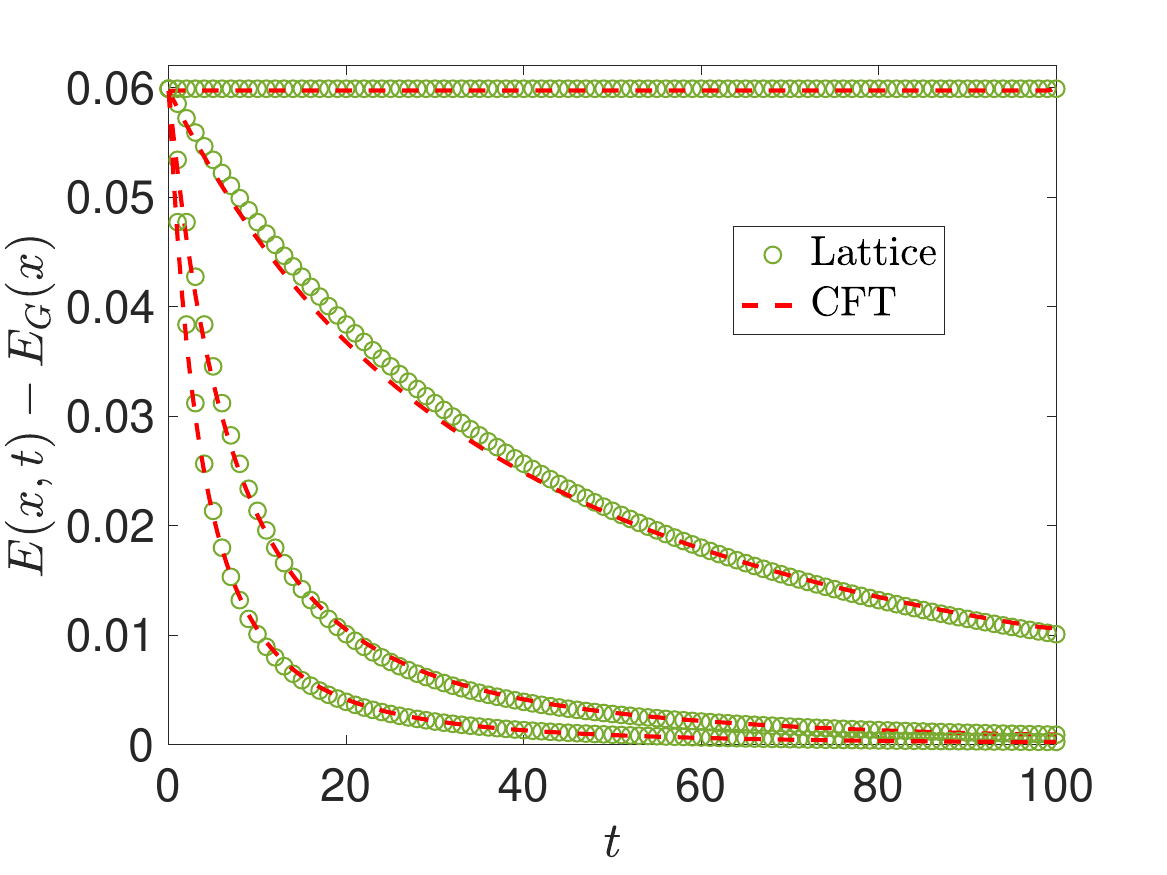}
\caption{
Comparison of complex time evolution of energy density in lattice and CFT calculations.
In the lattice calculation, the energy density is calculated by considering the average $\frac{1}{l}\int_{0}^{l} E(x,t)dx$ in $[0, l]$ with $l=100$ at the end of the system 
of length $L=800$.  From top to bottom, we have $\epsilon=0$, $0.01$, $0.05$, and $0.1$. 
The fitting parameter in CFT is $\beta=2.94$.
}
\label{LatticeEnergy2}
\end{figure}

\subsubsection{Time evolution}

Now let us consider the complex-time evolution of energy density $T_{00}(x,t)=\frac{1}{2\pi}[T(x,t)+\bar{T}(x,t)]$, 
where $T(x,t)$ and $\bar T(x,t)$ are given in \eqref{T_global2} and \eqref{Tbar_global2} respectively. One can find that $T_{00}(x,t)$ decays to the ground state value, which is zero,
 in the long time limit $t\to \infty$.
 
 For a general time $t$,  we plot the averaged energy density in a subregion $[0,l]$ at the end of the system, as shown in Fig.\ref{LatticeEnergy2}, where
 the lattice model results and the CFT results agree very well\footnote{Note that the Casimir effect which happens in a short length scale $\xi\sim \mathcal O(\beta)$ near the boundary is hard to observe in a lattice system, where the energy density fluctuates near the boundary of the lattice. This is why we consider an averaged energy density in the boundary region.}. One can find that as we increase $\epsilon$ in the complex time evolution, the (averaged) energy density 
 decays faster in time, as expected.

\section{Local quenches with complex metrics}
\label{Sec:Local}

For local quantum quenches in a CFT, 
there are various setups where the time evolutions are exactly solvable. See, e.g., Refs.\cite{CC2007_local,2009CC,2013Nozaki,2014Nozaki,2014He}.
In this work, we will consider the setup proposed in Ref.\cite{CC2007_local,2009CC}, which we will briefly review as follows.

\begin{figure}[htp]
\center
\centering
\begin{tikzpicture}[x=0.75pt,y=0.75pt,yscale=-0.8,xscale=0.8]

%Straight Lines [id:da40162350023909277] 
\draw [line width=0.8]  [dash pattern={on 1.69pt off 2.76pt}]  (448.33,157.5) -- (504,55.78) ;
%Straight Lines [id:da9294299589899353] 
\draw [line width=0.8]  [dash pattern={on 1.69pt off 2.76pt}]  (439.33,156.69) -- (495,54.96) ;
%Straight Lines [id:da3513089450426442] 
\draw [line width=0.8]  [dash pattern={on 1.69pt off 2.76pt}]  (272.33,158.04) -- (328,56.32) ;
%Straight Lines [id:da5280781148840789] 
\draw [color={rgb, 255:red, 16; green, 114; blue, 229 }  ,draw opacity=1 ][line width=0.8]    (124.5,127.39) -- (108.17,158.32) ;
%Straight Lines [id:da5557864737079223] 
\draw [color={rgb, 255:red, 16; green, 114; blue, 229 }  ,draw opacity=1 ][line width=0.8]    (131.5,127.39) -- (115.17,158.32) ;
%Straight Lines [id:da5295049919664251] 
\draw [color={rgb, 255:red, 16; green, 114; blue, 229 }  ,draw opacity=1 ][line width=0.8]    (123.5,128.21) -- (130.5,128.21) ;
%Straight Lines [id:da22484305846377672] 
\draw [color={rgb, 255:red, 16; green, 114; blue, 229 }  ,draw opacity=1 ][line width=0.8]    (109.17,157.5) -- (116.17,157.5) ;
%Straight Lines [id:da8522873568483882] 
\draw [color={rgb, 255:red, 16; green, 114; blue, 229 }  ,draw opacity=1 ][line width=0.8]    (298.33,107.72) -- (282,138.65) ;
%Straight Lines [id:da6791182041616514] 
\draw [color={rgb, 255:red, 16; green, 114; blue, 229 }  ,draw opacity=1 ][line width=0.8]    (291.33,108.54) -- (298.33,108.54) ;
%Straight Lines [id:da3781050226434217] 
\draw [color={rgb, 255:red, 16; green, 114; blue, 229 }  ,draw opacity=1 ][line width=0.8]    (275,137.84) -- (282,137.84) ;
%Straight Lines [id:da5092330628209587] 
\draw [line width=0.8]    (201.67,204) -- (261.33,203.75) ;
%Straight Lines [id:da8047273882856276] 
\draw [line width=0.8]    (261.33,203.75) -- (262.33,161.3) ;
%Straight Lines [id:da36658072828688415] 
\draw [line width=0.8]    (273.33,203.75) -- (274.33,161.3) ;
%Shape: Arc [id:dp20747075589100283] 
\draw  [draw opacity=0][line width=0.8]  (262.66,161.65) .. controls (262.66,161.63) and (262.66,161.6) .. (262.66,161.58) .. controls (262.65,157.45) and (265.26,154.09) .. (268.48,154.09) .. controls (271.62,154.09) and (274.19,157.26) .. (274.33,161.24) -- (268.5,161.58) -- cycle ; \draw  [line width=0.8]  (262.66,161.65) .. controls (262.66,161.63) and (262.66,161.6) .. (262.66,161.58) .. controls (262.65,157.45) and (265.26,154.09) .. (268.48,154.09) .. controls (271.62,154.09) and (274.19,157.26) .. (274.33,161.24) ;  
%Straight Lines [id:da6687429808940756] 
\draw [line width=0.8]    (273.33,203.75) -- (330.33,203.33) ;
%Straight Lines [id:da7051319916457033] 
\draw [line width=0.8]    (317,102.03) -- (318,59.57) ;
%Straight Lines [id:da38059229841461995] 
\draw [line width=0.8]    (329,102.03) -- (330,59.57) ;
%Shape: Arc [id:dp8082775311037147] 
\draw  [draw opacity=0][line width=0.8]  (318,60.62) .. controls (318,60.6) and (318,60.58) .. (318,60.56) .. controls (318,56.42) and (320.6,53.07) .. (323.83,53.06) .. controls (326.97,53.06) and (329.54,56.23) .. (329.68,60.21) -- (323.84,60.55) -- cycle ; \draw  [line width=0.8]  (318,60.62) .. controls (318,60.6) and (318,60.58) .. (318,60.56) .. controls (318,56.42) and (320.6,53.07) .. (323.83,53.06) .. controls (326.97,53.06) and (329.54,56.23) .. (329.68,60.21) ;  
%Straight Lines [id:da07935031124740755] 
\draw [line width=0.8]  [dash pattern={on 1.69pt off 2.76pt}]  (273.33,203.75) -- (329,102.03) ;
%Straight Lines [id:da19332587232038623] 
\draw [line width=0.8]  [dash pattern={on 1.69pt off 2.76pt}]  (261.33,203.75) -- (317,102.03) ;
%Straight Lines [id:da060584888232735445] 
\draw [color={rgb, 255:red, 16; green, 114; blue, 229 }  ,draw opacity=1 ][line width=0.8]    (467.33,108) -- (451,138.92) ;
%Straight Lines [id:da002102267203868746] 
\draw [color={rgb, 255:red, 16; green, 114; blue, 229 }  ,draw opacity=1 ][line width=0.8]    (474.33,108) -- (458,138.92) ;
%Straight Lines [id:da3899158062623955] 
\draw [color={rgb, 255:red, 16; green, 114; blue, 229 }  ,draw opacity=1 ][line width=0.8]    (466.33,108.81) -- (473.33,108.81) ;
%Straight Lines [id:da4865305677421562] 
\draw [color={rgb, 255:red, 16; green, 114; blue, 229 }  ,draw opacity=1 ][line width=0.8]    (452,138.11) -- (459,138.11) ;
%Shape: Arc [id:dp9899556766380025] 
\draw  [draw opacity=0][line width=0.8]  (436.34,162.62) .. controls (436.34,162.6) and (436.34,162.58) .. (436.34,162.55) .. controls (436.33,158.42) and (438.94,155.06) .. (442.16,155.06) .. controls (445.3,155.06) and (447.87,158.23) .. (448.01,162.21) -- (442.18,162.55) -- cycle ; \draw  [line width=0.8]  (436.34,162.62) .. controls (436.34,162.6) and (436.34,162.58) .. (436.34,162.55) .. controls (436.33,158.42) and (438.94,155.06) .. (442.16,155.06) .. controls (445.3,155.06) and (447.87,158.23) .. (448.01,162.21) ;  
%Shape: Arc [id:dp6191645284026938] 
\draw  [draw opacity=0][line width=0.8]  (492,60.89) .. controls (492,60.87) and (492,60.85) .. (492,60.83) .. controls (492,56.69) and (494.6,53.34) .. (497.83,53.34) .. controls (500.97,53.33) and (503.54,56.5) .. (503.68,60.48) -- (497.84,60.82) -- cycle ; \draw  [line width=0.8]  (492,60.89) .. controls (492,60.87) and (492,60.85) .. (492,60.83) .. controls (492,56.69) and (494.6,53.34) .. (497.83,53.34) .. controls (500.97,53.33) and (503.54,56.5) .. (503.68,60.48) ;  
%Curve Lines [id:da858820535723454] 
\draw [line width=0.8]    (448.33,161.57) .. controls (453.33,177.98) and (463.33,190.09) .. (489,200.67) ;
%Curve Lines [id:da7546902182266986] 
\draw [line width=0.8]    (436.34,162.61) .. controls (430,184.49) and (410,195.73) .. (389.67,204) ;
%Straight Lines [id:da3177649154103529] 
\draw [line width=0.8]  [dash pattern={on 1.69pt off 2.76pt}]  (263.33,157.23) -- (319,55.51) ;
%Straight Lines [id:da7782128573548234] 
\draw [color={rgb, 255:red, 16; green, 114; blue, 229 }  ,draw opacity=1 ][line width=0.8]    (291.33,107.72) -- (275,138.65) ;
%Straight Lines [id:da913219793364192] 

\draw [color={rgb, 255:red, 65; green, 117; blue, 5 }  ,draw opacity=1 ]   (203.33,215.6) -- (327,215.34) ;
\draw [shift={(329,215.33)}, rotate = 179.88] [fill={rgb, 255:red, 65; green, 117; blue, 5 }  ,fill opacity=1 ][line width=0.08]  [draw opacity=0] (12,-3) -- (0,0) -- (12,3) -- cycle    ;
\draw [shift={(201.33,215.6)}, rotate = 359.88] [fill={rgb, 255:red, 65; green, 117; blue, 5 }  ,fill opacity=1 ][line width=0.08]  [draw opacity=0] (12,-3) -- (0,0) -- (12,3) -- cycle    ;
%Straight Lines [id:da46232905668850754] 
\draw [color={rgb, 255:red, 65; green, 117; blue, 5 }  ,draw opacity=1 ]   (337.35,100.03) -- (337.65,54.93) ;
\draw [shift={(337.67,52.93)}, rotate = 90.39] [fill={rgb, 255:red, 65; green, 117; blue, 5 }  ,fill opacity=1 ][line width=0.08]  [draw opacity=0] (12,-3) -- (0,0) -- (12,3) -- cycle    ;
\draw [shift={(337.33,102.03)}, rotate = 270.39] [fill={rgb, 255:red, 65; green, 117; blue, 5 }  ,fill opacity=1 ][line width=0.08]  [draw opacity=0] (12,-3) -- (0,0) -- (12,3) -- cycle    ;
%Straight Lines [id:da5527334482586331] 
\draw [line width=0.8]    (23.33,204.43) -- (142.67,204.3) ;
%Straight Lines [id:da7981863848630008] 
\draw [line width=0.8]    (79,102.3) -- (198.33,102.16) ;
%Straight Lines [id:da025095408021616028] 
\draw [line width=0.8]    (102,138.11) -- (98,145.16) ;
%Straight Lines [id:da6015045682716055] 
\draw [line width=0.8]    (60.33,138.11) -- (102,138.11) ;
%Straight Lines [id:da10366333987533016] 
\draw [line width=0.8]    (56.33,145.16) -- (98,145.16) ;
%Straight Lines [id:da717226550045954] 
\draw [line width=0.8]    (102,138.11) -- (90.86,144.54) ;
%Straight Lines [id:da9559098737297576] 
\draw [line width=0.8]    (91.43,138.11) -- (80.29,144.54) ;
%Straight Lines [id:da9366968260852183] 
\draw [line width=0.8]    (81.17,138.11) -- (70.02,144.54) ;
%Straight Lines [id:da7966075596254152] 
\draw [line width=0.8]    (71.14,137.87) -- (60,144.31) ;
%Straight Lines [id:da355425816613952] 
\draw [line width=0.8]    (142.33,138.51) -- (138.33,145.57) ;
%Straight Lines [id:da23123540839693535] 
\draw [line width=0.8]    (142.33,138.51) -- (184,138.51) ;
%Straight Lines [id:da44235963861542515] 
\draw [line width=0.8]    (138.33,145.57) -- (180,145.57) ;
%Straight Lines [id:da3902891803651789] 
\draw [line width=0.8]    (184,138.51) -- (172.86,144.95) ;
%Straight Lines [id:da32436690854552586] 
\draw [line width=0.8]    (173.43,138.51) -- (162.29,144.95) ;
%Straight Lines [id:da0986831589917413] 
\draw [line width=0.8]    (163.17,138.51) -- (152.02,144.95) ;
%Straight Lines [id:da8673942460479555] 
\draw [line width=0.8]    (153.14,138.28) -- (142,144.71) ;
%Straight Lines [id:da06901657400831818] 
\draw [line width=0.8]    (267,138.11) -- (263,145.16) ;
%Straight Lines [id:da5126895039454781] 
\draw [line width=0.8]    (225.33,138.11) -- (267,138.11) ;
%Straight Lines [id:da7192378545264756] 
\draw [line width=0.8]    (221.33,145.16) -- (263,145.16) ;
%Straight Lines [id:da7872263494281585] 
\draw [line width=0.8]    (267,138.11) -- (255.86,144.54) ;
%Straight Lines [id:da6277861715898219] 
\draw [line width=0.8]    (256.43,138.11) -- (245.29,144.54) ;
%Straight Lines [id:da5916220578655295] 
\draw [line width=0.8]    (246.17,138.11) -- (235.02,144.54) ;
%Straight Lines [id:da8006257977853254] 
\draw [line width=0.8]    (236.14,137.87) -- (225,144.31) ;
%Straight Lines [id:da8741101996139856] 
\draw [line width=0.8]    (315,137.7) -- (311,144.75) ;
%Straight Lines [id:da6982030931451286] 
\draw [line width=0.8]    (315,137.7) -- (356.67,137.7) ;
%Straight Lines [id:da6370499491382945] 
\draw [line width=0.8]    (311,144.75) -- (352.67,144.75) ;
%Straight Lines [id:da3290520641490535] 
\draw [line width=0.8]    (356.67,137.7) -- (345.52,144.13) ;
%Straight Lines [id:da36966160381874125] 
\draw [line width=0.8]    (346.1,137.7) -- (334.95,144.13) ;
%Straight Lines [id:da956072719267569] 
\draw [line width=0.8]    (335.83,137.7) -- (324.69,144.13) ;
%Straight Lines [id:da3711813831177341] 
\draw [line width=0.8]    (325.81,137.47) -- (314.67,143.9) ;
%Straight Lines [id:da5966772911184187] 
\draw [line width=0.8]    (440,137.56) -- (436,144.62) ;
%Straight Lines [id:da6360891046548492] 
\draw [line width=0.8]    (398.33,137.56) -- (440,137.56) ;
%Straight Lines [id:da24565456194876] 
\draw [line width=0.8]    (394.33,144.62) -- (436,144.62) ;
%Straight Lines [id:da21536944128056812] 
\draw [line width=0.8]    (440,137.56) -- (428.86,144) ;
%Straight Lines [id:da8432521895989455] 
\draw [line width=0.8]    (429.43,137.56) -- (418.29,144) ;
%Straight Lines [id:da8074438105490102] 
\draw [line width=0.8]    (419.17,137.56) -- (408.02,144) ;
%Straight Lines [id:da08081819883156949] 
\draw [line width=0.8]    (409.14,137.33) -- (398,143.76) ;
%Straight Lines [id:da3701571347713508] 
\draw [line width=0.8]    (487,136.89) -- (483,143.94) ;
%Straight Lines [id:da12112449087127342] 
\draw [line width=0.8]    (487,136.89) -- (528.67,136.89) ;
%Straight Lines [id:da4736482742794026] 
\draw [line width=0.8]    (483,143.94) -- (524.67,143.94) ;
%Straight Lines [id:da3484069912517549] 
\draw [line width=0.8]    (528.67,136.89) -- (517.52,143.32) ;
%Straight Lines [id:da6662785963981501] 
\draw [line width=0.8]    (518.1,136.89) -- (506.95,143.32) ;
%Straight Lines [id:da5008252557680369] 
\draw [line width=0.8]    (507.83,136.89) -- (496.69,143.32) ;
%Straight Lines [id:da7455601840499434] 
\draw [line width=0.8]    (497.81,136.65) -- (486.67,143.09) ;
%Straight Lines [id:da43915840702892706] 
\draw [color={rgb, 255:red, 65; green, 117; blue, 5 }  ,draw opacity=1 ]   (24,215.68) -- (139,215.34) ;
\draw [shift={(141,215.33)}, rotate = 179.83] [fill={rgb, 255:red, 65; green, 117; blue, 5 }  ,fill opacity=1 ][line width=0.08]  [draw opacity=0] (12,-3) -- (0,0) -- (12,3) -- cycle    ;
\draw [shift={(22,215.69)}, rotate = 359.83] [fill={rgb, 255:red, 65; green, 117; blue, 5 }  ,fill opacity=1 ][line width=0.05]  [draw opacity=0] (12,-3) -- (0,0) -- (12,3) -- cycle    ;
%Curve Lines [id:da8146996885933107] 
\draw [line width=0.8]    (503.68,60.48) .. controls (508.68,76.9) and (518.68,89) .. (544.34,99.58) ;
%Curve Lines [id:da26258136819199895] 
\draw [line width=0.8]    (492,60.89) .. controls (485.67,82.78) and (465.67,94.01) .. (445.33,102.29) ;
%Straight Lines [id:da010437562767824926] 
\draw [line width=0.8]    (257.33,102.27) -- (317,102.03) ;
%Straight Lines [id:da21076735610220276] 
\draw [line width=0.8]    (329,102.03) -- (386,101.61) ;

% Text Node
\draw (43.33,220.41) node [anchor=north west][inner sep=0.75pt]  [font=\normalsize]  {Euclidean};
% Text Node
\draw (226.67,220.86) node [anchor=north west][inner sep=0.75pt]  [font=\normalsize]  {Euclidean};
% Text Node
\draw (340.33,68.48) node [anchor=north west][inner sep=0.75pt]  [font=\normalsize]  {Lorentzian};
% Text Node
\draw (408.67,219.86) node [anchor=north west][inner sep=0.75pt]  [font=\normalsize]  {Complex};
% Text Node
\draw (64.67,251.84) node [anchor=north west][inner sep=0.75pt]  [font=\normalsize]  {$( a)$};
% Text Node
\draw (248,250.21) node [anchor=north west][inner sep=0.75pt]  [font=\normalsize]  {$( b)$};
% Text Node
\draw (425.33,250.75) node [anchor=north west][inner sep=0.75pt]  [font=\normalsize]  {$( c)$};

%%% x-t axis
     \begin{scope}[xshift=-145pt]
\draw [line width=0.8] [>=stealth,->]   (162,130) -- (202,130) ;
\draw [line width=0.8] [>=stealth,->]   (162,130) -- (180,90) ;
\node at (140pt,62pt){$x$};
\node at (160pt,96pt){$\tau$};
\end{scope}
%%% x-t axis above

\end{tikzpicture}
\caption{Path integral representation of 
the reduced density matrix $\rho_A$ after a local quantum quench in a CFT with
(a) Euclidean (b) mixed (partial Euclidean and partial Lorentz), and (c) complex spacetime metrics.
The location of the branch cuts (blue lines) corresponds to the subsystem $A=[a,b]$.
There are two slits along the imaginary time direction in each configuration, which are used to prepare the initial state in \eqref{InitialState_local}.
}
\label{Local}
\end{figure}
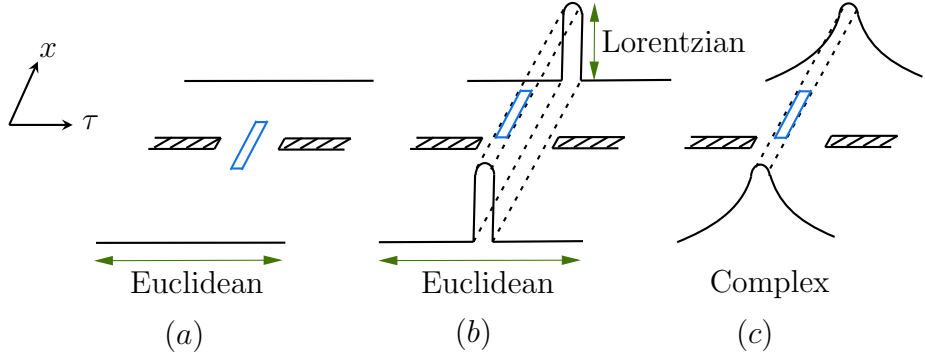

We consider two CFTs defined on $(-\infty,0)$ and $(0,+\infty)$, with the same conformal boundary condition imposed at the two ends respectively. Then at $t=0$, we join the two ends of the CFTs at $x=0$ suddenly and let the system evolve in time under 
the uniform CFT Hamiltonian $H_{\text{CFT}}$ that is defined over $(-\infty,+\infty)$.
Note that the initial state itself is not translationally invariant:
\be
\label{InitialState_local}
|\psi_0\rangle=e^{-\lambda H_{\text{CFT}}}\left(|G_L\rangle\otimes |G_R\rangle\right),\quad \lambda\in \mathbb R,
\ee 
where $|G_L\rangle$ and $|G_R\rangle$ denote the ground states of the two decoupled CFTs on the left and right sides respectively,
and the factor $e^{-\lambda H_{\text{CFT}}}$ is introduced as a regularization.
The path integral of the above initial state $|\psi_0\rangle$ is then straightforward: one can introduce a slit that goes from $-i\infty$ to $-i\lambda$ 
in the imaginary time direction, and then evolve the state from $-i\lambda$ to $0$ using a CFT Hamiltonian.
Then, we can consider the time evolution of this initial state as well as the corresponding (reduced) density matrix under different spacetime metrics, as shown in Fig.\ref{Local}.

Here we are interested in the complex time evolution in \eqref{WF_Quench}. The time-dependent density matrix we consider here is
\be
\label{rho_local}
\rho(t)=e^{-\left(\lambda+\epsilon t+it\right) H_{\text{CFT}}}(|G_L\rangle\otimes |G_R\rangle)(\langle G_L|\otimes \langle G_R|) e^{-\left(\lambda+\epsilon t -i t\right) H_{\text{CFT}}}.
\ee
To study this density matrix and related physical quantities, one can first consider the two-time density matrix in Euclidean space as follows
\be
\label{rho_tau1tau2_local}
\rho(\tau_1,\tau_2)=e^{-\left(\lambda+\epsilon \tau_1+\tau_2\right) H_{\text{CFT}}}(|G_L\rangle\otimes |G_R\rangle)(\langle G_L|\otimes \langle G_R|) e^{-\left(\lambda+\epsilon \tau_1-\tau_2\right) H_{\text{CFT}}},
\ee
based on which one can evaluate $\rho_A(\tau_1,\tau_2)=\text{Tr}_{\bar A}\rho(\tau_1,\tau_2)$. 
Then in the final step, one can obtain the reduced density matrix $\rho_{A}(t)$ by taking an analytical continuation $\tau_1\to t$ and $\tau_2\to it$.

\subsection{Entanglement Hamiltonian and entanglement spectrum evolution}

To study the entanglement Hamiltonian $K_A(t)$ for $A=[0,+\infty)$, let us start from
the path integral of the two-time reduced density matrix $\rho_A(\tau_1,\tau_2)$ in the Euclidean space as follows:
\be
\begin{tikzpicture}[x=0.75pt,y=0.75pt,yscale=-0.8,xscale=0.8]
%uncomment if require: \path (0,235); %set diagram left start at 0, and has height of 235

%Straight Lines [id:da6513516693134163] 
\draw    (105,78) -- (105.3,16.71) ;
%Straight Lines [id:da3973171567437922] 
\draw    (95.3,77.71) -- (105.3,67.71) ;
%Straight Lines [id:da6785772395277093] 
\draw    (95,67) -- (105,57) ;
%Straight Lines [id:da9711843788227461] 
\draw    (95.3,56.71) -- (105.3,46.71) ;
%Straight Lines [id:da8999930357491253] 
\draw    (95,48) -- (105,38) ;
%Straight Lines [id:da8478501864882666] 
\draw    (95.3,37.71) -- (105.3,27.71) ;
%Straight Lines [id:da6714568002918281] 
\draw    (95,27) -- (105,17) ;
%Shape: Arc [id:dp33132549052073923] 
\draw  [draw opacity=0] (105.78,104.57) .. controls (104.52,106.52) and (102.34,107.82) .. (99.85,107.86) .. controls (95.88,107.92) and (92.6,104.76) .. (92.54,100.78) .. controls (92.48,96.81) and (95.65,93.54) .. (99.62,93.47) .. controls (102.33,93.43) and (104.71,94.89) .. (105.97,97.08) -- (99.73,100.67) -- cycle ; \draw   (105.78,104.57) .. controls (104.52,106.52) and (102.34,107.82) .. (99.85,107.86) .. controls (95.88,107.92) and (92.6,104.76) .. (92.54,100.78) .. controls (92.48,96.81) and (95.65,93.54) .. (99.62,93.47) .. controls (102.33,93.43) and (104.71,94.89) .. (105.97,97.08) ;  
%Straight Lines [id:da5567565992405998] 
\draw [color={rgb, 255:red, 74; green, 144; blue, 226 }  ,draw opacity=1 ][line width=1.5]    (105.97,97.08) -- (187.97,97.08) ;
%Straight Lines [id:da8599214410394936] 
\draw [color={rgb, 255:red, 74; green, 144; blue, 226 }  ,draw opacity=1 ][line width=1.5]    (105.78,104.57) -- (187.78,104.57) ;
%Straight Lines [id:da16248839938020443] 
\draw    (95.3,77.71) -- (95.6,16.43) ;
%Straight Lines [id:da29125650985538254] 
\draw [color={rgb, 255:red, 0; green, 0; blue, 0 }  ,draw opacity=1 ][line width=0.75]    (95.3,77.71) -- (105.3,77.71) ;
%Straight Lines [id:da9762482367538045] 
\draw    (95.29,141.43) -- (95.01,202.72) ;
%Straight Lines [id:da5815247396612884] 
\draw    (104.99,141.71) -- (94.99,151.72) ;
%Straight Lines [id:da5051547434061004] 
\draw    (105.29,152.43) -- (95.3,162.43) ;
%Straight Lines [id:da7585072828119681] 
\draw    (105,162.71) -- (95,172.72) ;
%Straight Lines [id:da7602718636537532] 
\draw    (105.3,171.43) -- (95.3,181.43) ;
%Straight Lines [id:da40299722551923733] 
\draw    (105,181.71) -- (95.01,191.72) ;
%Straight Lines [id:da41277825346757147] 
\draw    (105.31,192.43) -- (95.31,202.43) ;
%Straight Lines [id:da7604282258270554] 
\draw    (104.99,141.71) -- (104.71,203) ;
%Straight Lines [id:da31190279395371456] 
\draw [color={rgb, 255:red, 0; green, 0; blue, 0 }  ,draw opacity=1 ][line width=0.75]    (104.99,141.71) -- (94.99,141.72) ;

\small
\draw (48,77) node  {$i(\lambda+\epsilon\, \tau_1)$};
\draw (72,100) node  {$i\tau_2$};

\draw (42,141) node  {$-i(\lambda+\epsilon\, \tau_1)$};

\begin{scope}[xshift=-80pt,yshift=-30pt]
\draw [line width=0.5] [>=stealth,->]   (282,95) -- (282,75) ;
\draw [line width=0.5] [>=stealth,->]   (282,95) -- (300,95) ;

\draw (285,68) node  {$\tau$};
\draw (308,92) node  {$x$};
\end{scope}

\end{tikzpicture}
\label{local_PI}
\ee
There are two slits along the imaginary time direction, one is from $-i\infty$ to $-i(\lambda+\epsilon \tau_1)$, and the other is from 
$i(\lambda+\epsilon \tau_1)$ to $+i\infty$.
The branch cut corresponding to subsystem $A$ is along $C=\{i\tau_2+x, \, x\ge 0\}$.
A small disk of radius $\epsilon_0$ is removed at the entanglement point $z_0=0+i\tau_2$, with a conformal boundary condition imposed along the boundary.

Next, we consider the conformal mapping:
\be
\label{conformalMap_local}
w=f(z)=\log\frac{\sqrt{[(\lambda+\epsilon \tau_1)^2-\tau_2^2]\cdot[(\lambda+\epsilon \tau_1)^2+z^2]}-i\tau_2\,z-(\lambda+\epsilon \tau_1)^2}{(\lambda+\epsilon\tau_1)(z-i\tau_2)},
\ee
which maps $\rho_A(\tau_1,\tau_2)$ in \eqref{local_PI} to a $w$-cylinder (See, e.g., the right plot in Fig.\ref{Rho_A_Global1}).
The length of this cylinder in the $\Im w$ direction is $2\pi$, and the length in the $\Re w$ direction is denoted as $W(t)$.
Then the entanglement Hamiltonian corresponds to the generator of translation in $\Im w$ direction of this cylinder.
By following the same procedure in Sec.\ref{Sec:EH_global1}, one can obtain
\be
\label{EH_local}
K_A(t)=\int_0^\infty \frac{x\sqrt{(x-t)^2+(\lambda+\epsilon t)^2}}{\sqrt{(\lambda+\epsilon t)^2+t^2}} T(x-t) dx
+
\int_0^\infty \frac{x\sqrt{(x+t)^2+(\lambda+\epsilon t)^2}}{\sqrt{(\lambda+\epsilon t)^2+t^2}} \bar{T}(x+t) dx.
\ee

Now let us consider the entanglement temperature as defined in \eqref{EE_temperature}.
For $\epsilon=0$ and $t\gg \lambda$, one can find that the entanglement temperature $\beta_E^{-1}(x)$ is low everywhere except near $x=t$:
\be
\label{beta_local_real}
\beta_E(x=t,t)=2\pi \cdot \frac{\lambda t}{\sqrt{\lambda^2+t^2}} \simeq 2\pi \lambda.
\ee
That is, near $x=t$, the entanglement temperature $1/\beta_E(x)\simeq \frac{1}{2\pi \lambda}$ is very high, which means that this region is highly entangled with $\bar A=(-\infty,0)$.
In fact, if one studies the entanglement Hamiltonian for $\bar A$, one can find $1/\bar \beta_E(x=-t,t)\simeq \frac{1}{2\pi \lambda}$, which indicates there is a maximal entanglement between the regions near $x=-t$ and $x=t$. This agrees with the quasi-particle picture as studied in Ref.\cite{CC2007_local,2009CC,2013Nozaki,2015Wen_EN,2014Asplund}.

Now, let us consider the complex time evolution with $\epsilon>0$. For the entanglement Hamiltonian of $A=[0,+\infty)$, one can find that the entanglement temperature at 
$x=t$ becomes:
\be
\beta_E(x=t,t)=2\pi \frac{(\lambda+\epsilon t)t}{\sqrt{(\lambda+\epsilon t)^2+t^2}}.
\ee
In the long time limit $ t\gg \lambda/\epsilon$, one can find 
\be
\beta_E(x=t,t)\propto 2\pi t.
\ee
That is, as $t$ grows, the entanglement temperature $1/\beta_E(x)\propto \frac{1}{2\pi t}$ will decay to zero in time. 
In other words, the entanglement between the two regions near $x=-t$ and $x=t$ keeps decreasing in time due to the damping effect.

\bigskip

\textit{-- Time evolution of entanglement spectrum:}

\bigskip

Next, let us consider the time evolution of the entanglement spectrum. Similar to the analysis in Sec.\ref{Sec:Global1}, the spacing of entanglement spectrum is 
determined by \eqref{ES_general}. Based on the conformal map in \eqref{conformalMap_local}, one can find
\be
\label{W_local}
W(t)=\log \frac{2[ t^2+(\lambda+\epsilon t)^2]}{\epsilon_0(\lambda+\epsilon t)}
+\mathcal O(\epsilon_0).
\ee
In a real time evolution ($\epsilon=0$), one can find that $W(t)\simeq 2\log t$ for $t\gg \lambda$. 
In a complex time evolution ($\epsilon>0$), one can find that $W(t)\simeq \log t$ for $t\gg \lambda/\epsilon $.
Then from \eqref{ES_general}, the long time evolution of entanglement spectrum has the following scaling behavior:
\be
\label{Local_ES}
\left\{
\begin{split}
&\text{Spacing of entanglement spectrum}\sim \frac{1}{2\log t}, \quad \text{real time evolution }(\epsilon=0),\\
&\text{Spacing of entanglement spectrum}\sim \frac{1}{\log t}, \quad \text{complex time evolution }(\epsilon> 0).
\end{split}
\right.
\ee
Note there is a factor $2$ difference between the real and complex time evolutions.

\begin{figure}[h]
\center
\centering
\includegraphics[width=3.05in]{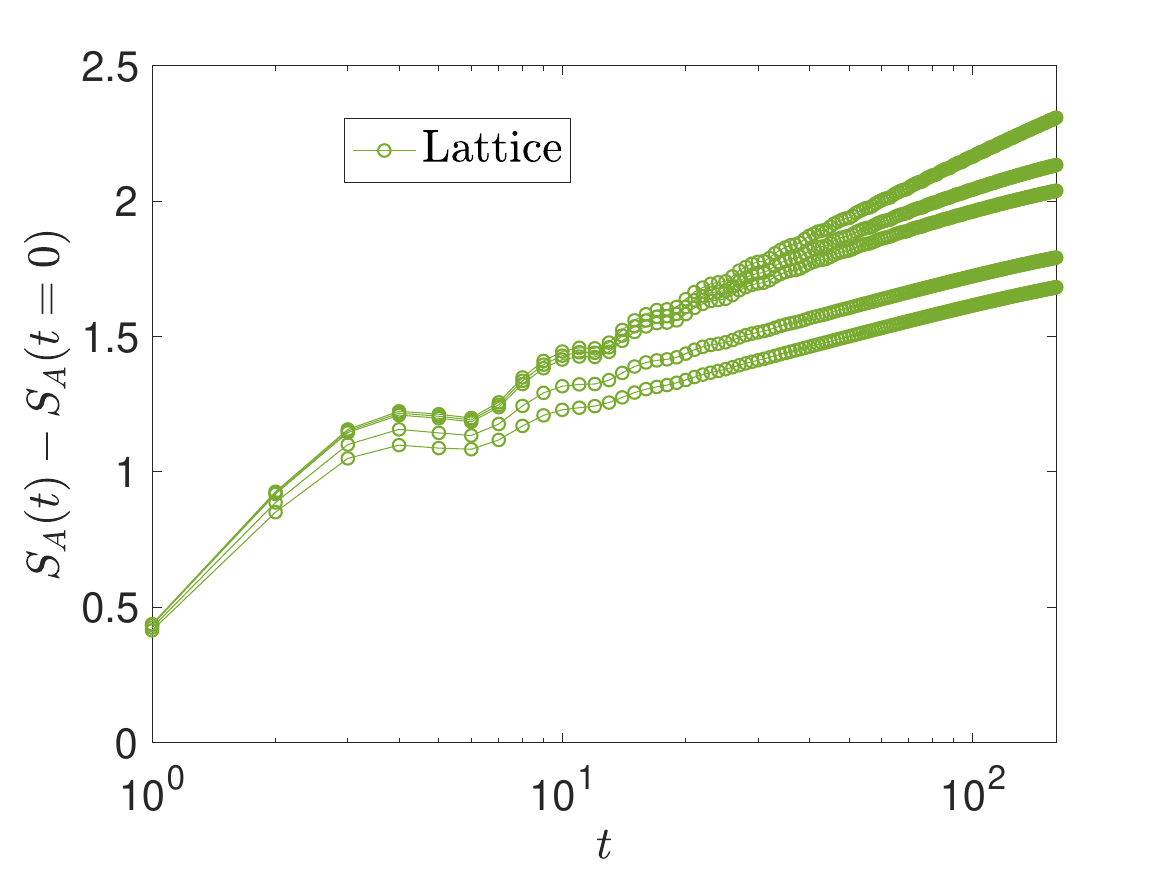}
\includegraphics[width=3.05in]{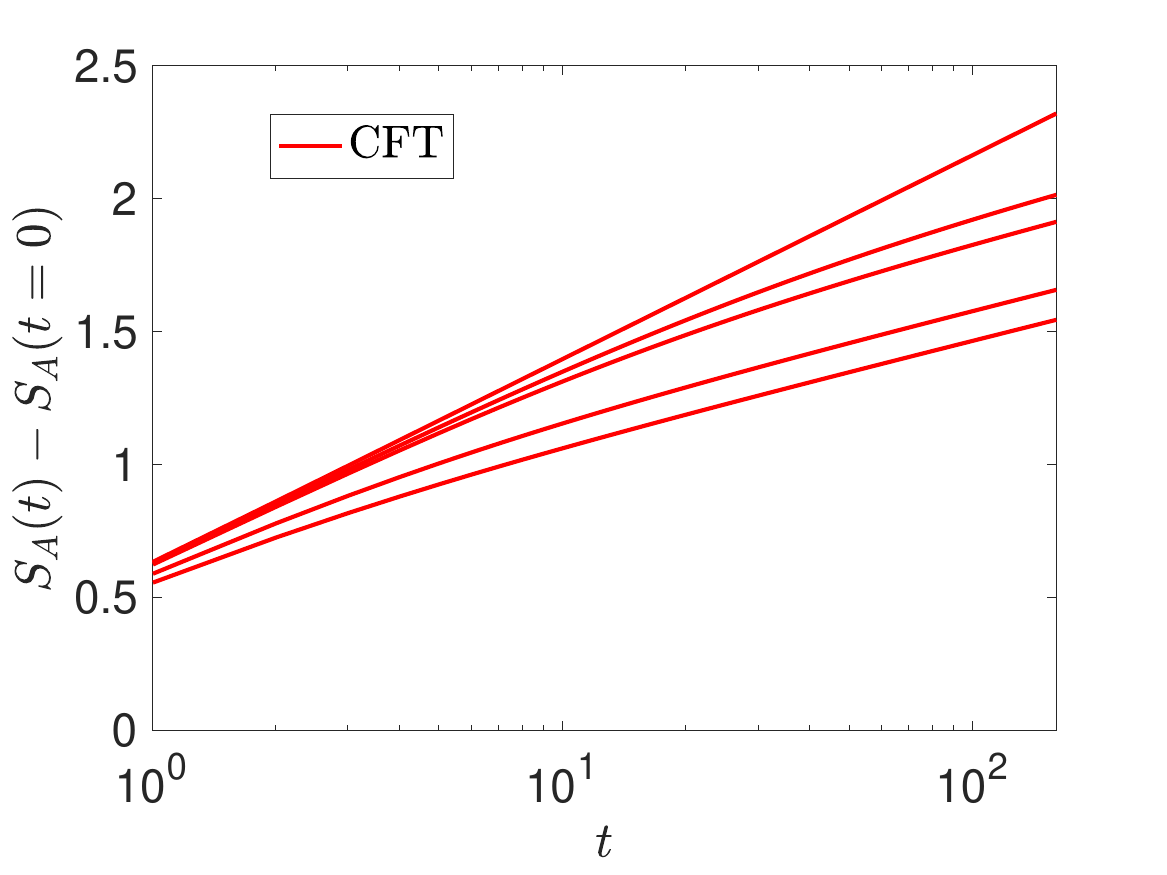}
\caption{
Comparison of complex time evolution of entanglement entropy $S_A(t)$ after a local quench
in a lattice system (left) and in CFT calculation (right).
The lattice system is defined on $[0, L]=[0, 800]$, and $A=[0,400]$.
The CFT result is plotted according to \eqref{EE_local} where we choose $\lambda=0.15$.
Note that $\epsilon_0$ is canceled out in $S_A(t)-S_A(t=0)$.
From top to bottom, we choose $\epsilon=0$, $0.005$, $0.01$, $0.05$, and $0.1$.
}
\label{LatticeEE_LocalQuench}
\end{figure}

\subsection{Entanglement entropy evolution}

The entanglement entropy evolution is related to $W(t)$ in \eqref{W_local} via the same formula in \eqref{SA_general}.
It is found that 
 \be
\label{EE_local}
S_A(t)=\frac{c}{6}\log \frac{2[ t^2+(\lambda+\epsilon t)^2]}{\epsilon_0(\lambda+\epsilon t)}+\log g_a+\log g_b
+\mathcal O(\epsilon_0).
\ee
A plot of $S_A(t)$ from both the lattice model calculation and the CFT calculation can be found in Fig.\ref{LatticeEE_LocalQuench}.
It is apparent that the complex time ($\epsilon>0$) will suppress the entanglement entropy evolution in time.
In particular, one can find the following simple scaling behavior in the long time limit:

\be
\label{SA:Local_longtime}
\left\{
\begin{split}
&S_A(t)\simeq \frac{c}{3}\log t, \quad t\gg \lambda \text{ in real time evolution }(\epsilon=0),\\
&S_A(t)\simeq \frac{c}{6}\log t, \quad t\gg \lambda/\epsilon  \text{ in complex time evolution }(\epsilon> 0).
\end{split}
\right.
\ee
Similar to the entanglement spectrum evolution in \eqref{Local_ES}, here we have a factor $2$ difference in the real and complex time evolutions of $S_A(t)$.

\subsection{Energy density evolution}

In real time, the energy density evolution after a local quench  has been studied Ref.\cite{2014Asplund,2013Ugajin}.
Since the initial state in \eqref{InitialState_local} is not translation invariant, one will observe a flow of energy density emitted from $x=0$ where we join the two CFTs, after the quench. Here we are interested in the complex time evolution.

\medskip

\begin{figure}[t]
\centering
\begin{tikzpicture}

    \node[inner sep=0pt] (russell) at (-65pt,-20pt)
    {\includegraphics[width=.37\textwidth]{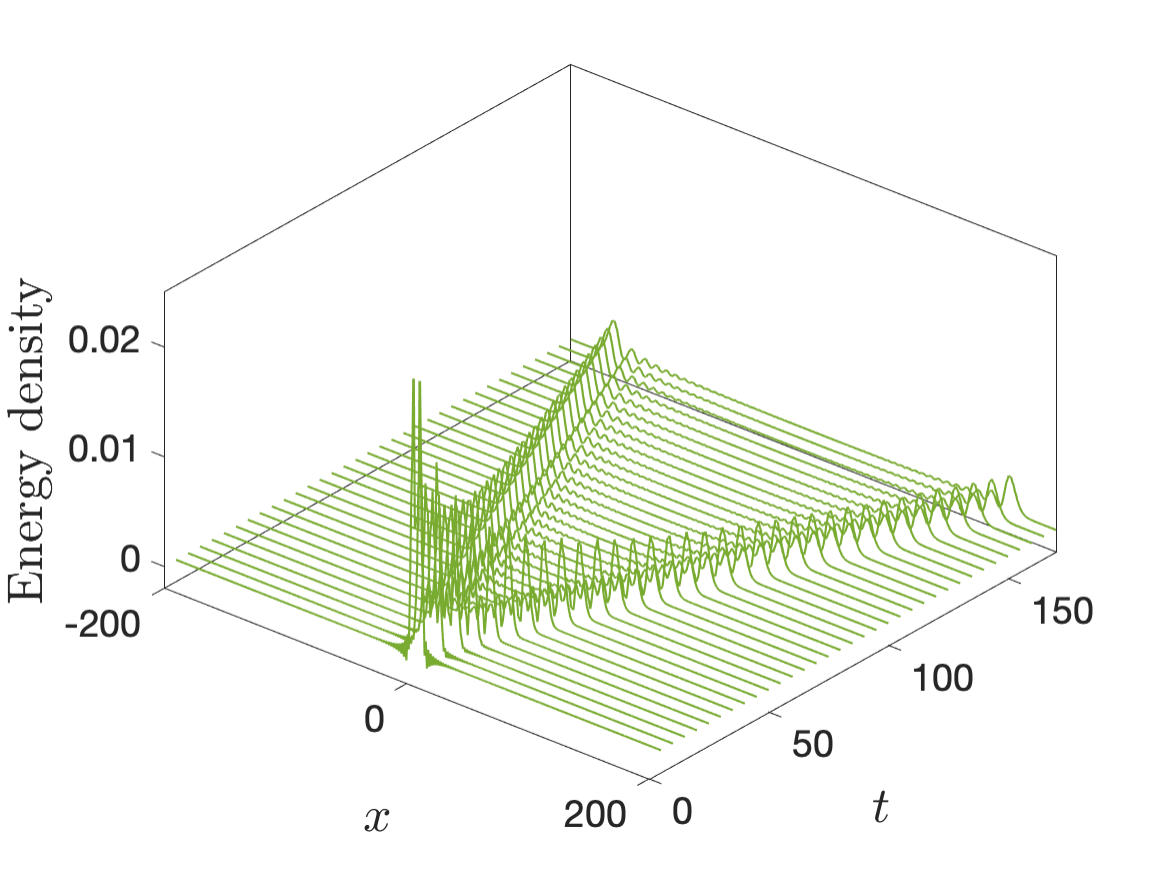}};
        \node[inner sep=0pt] (russell) at (100pt,-20pt)
    {\includegraphics[width=.37\textwidth]{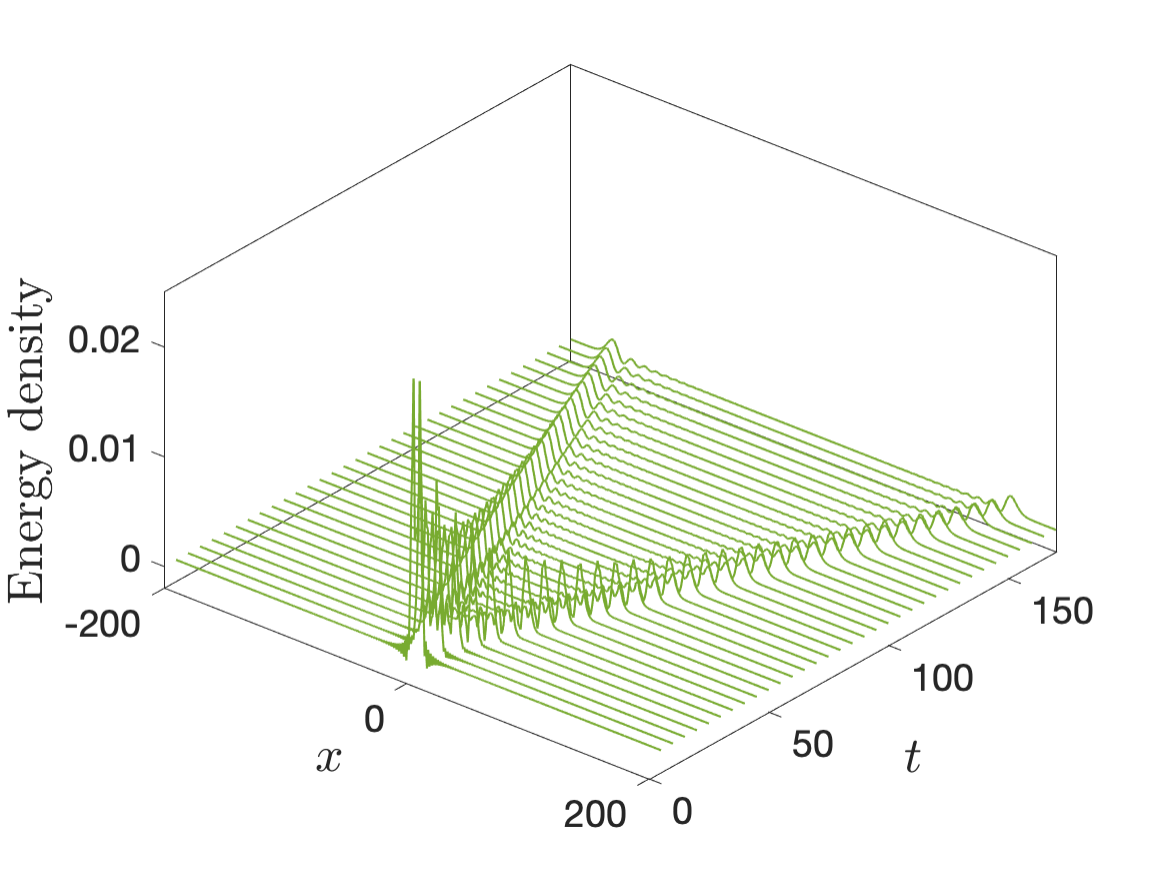}};
            \node[inner sep=0pt] (russell) at (265pt,-20pt)
    {\includegraphics[width=.37\textwidth]{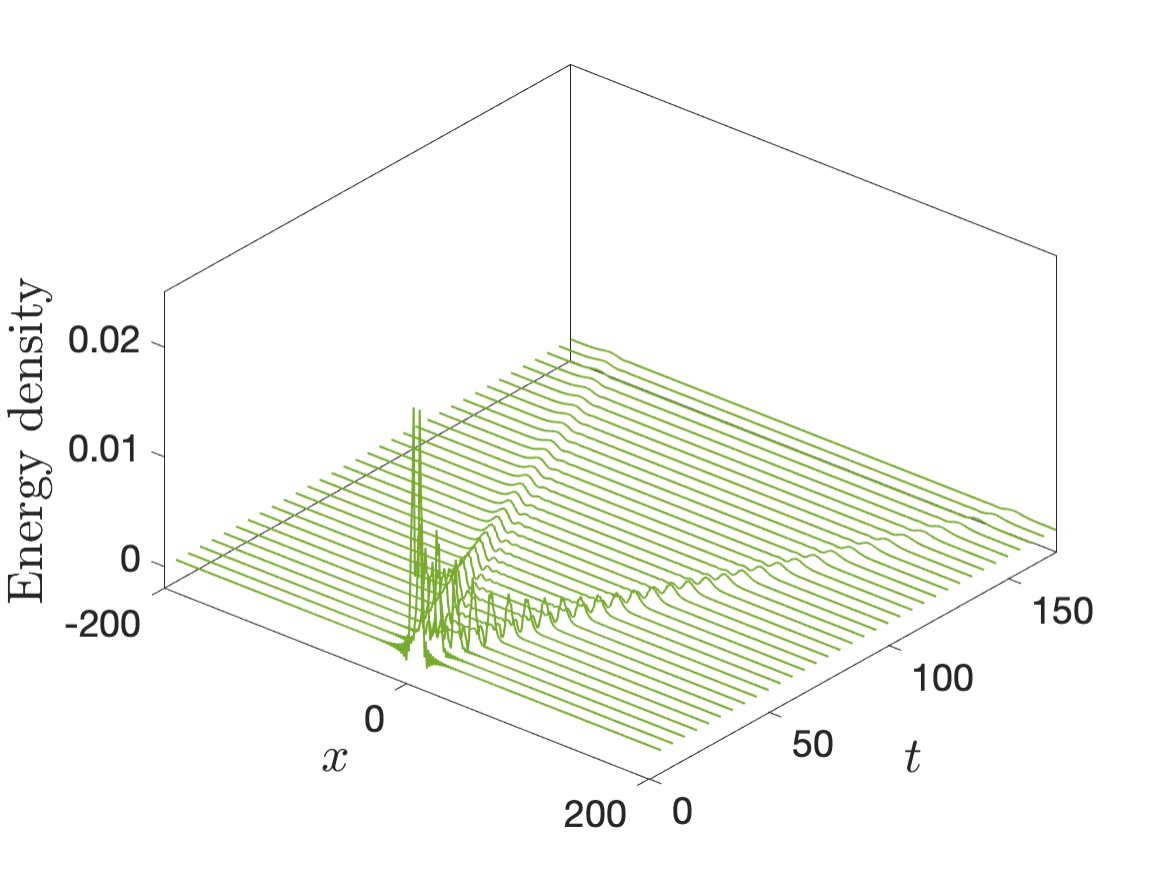}};
    \node[inner sep=0pt] (russell) at (-65pt,-150pt)
    {\includegraphics[width=.37\textwidth]{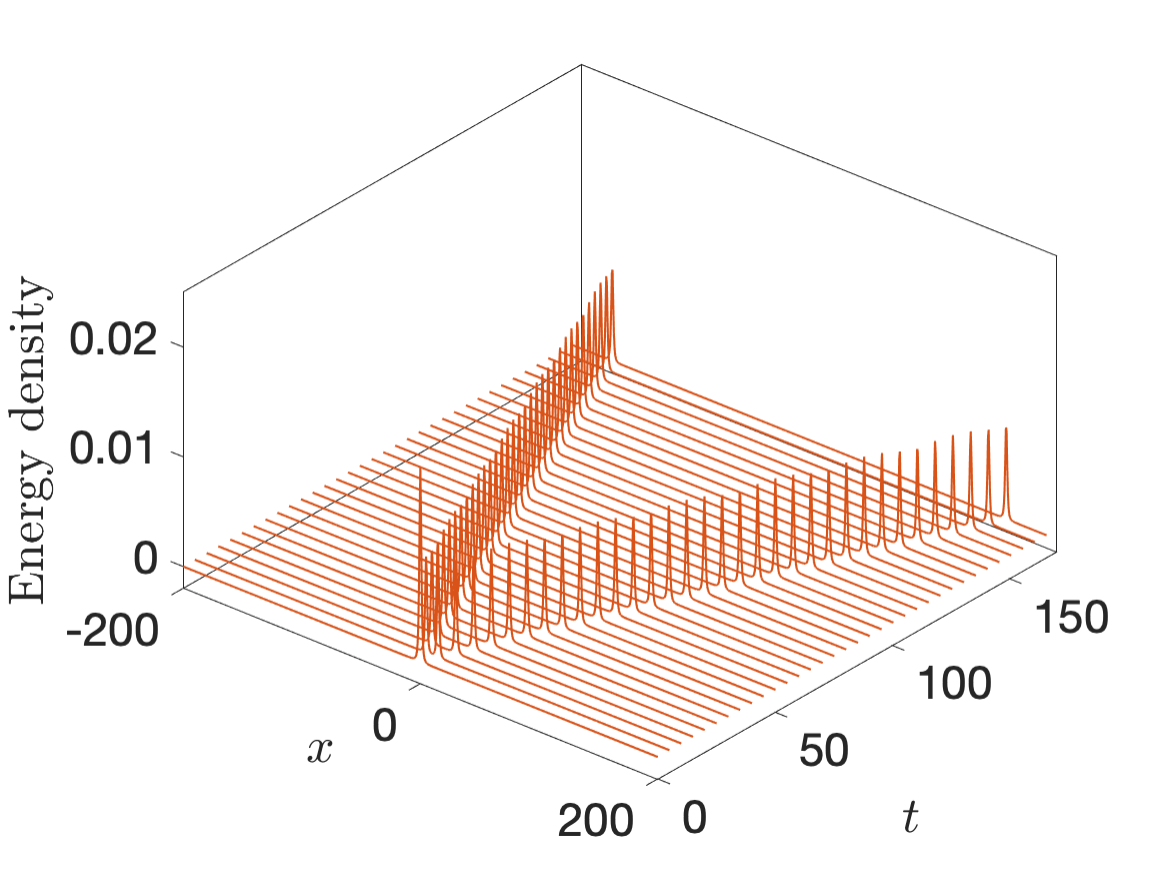}};    
    \node[inner sep=0pt] (russell) at (100pt,-150pt)
    {\includegraphics[width=.37\textwidth]{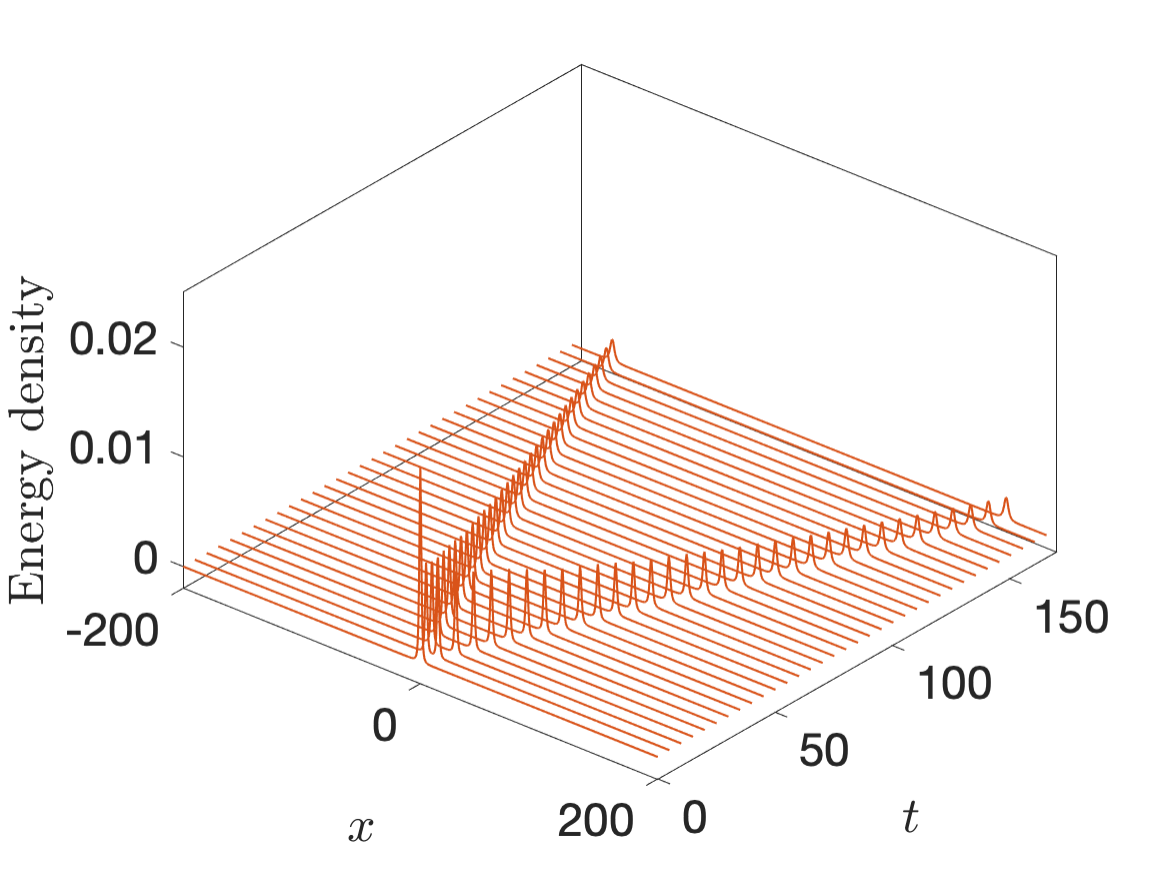}};    
        \node[inner sep=0pt] (russell) at (265pt,-150pt)
    {\includegraphics[width=.37\textwidth]{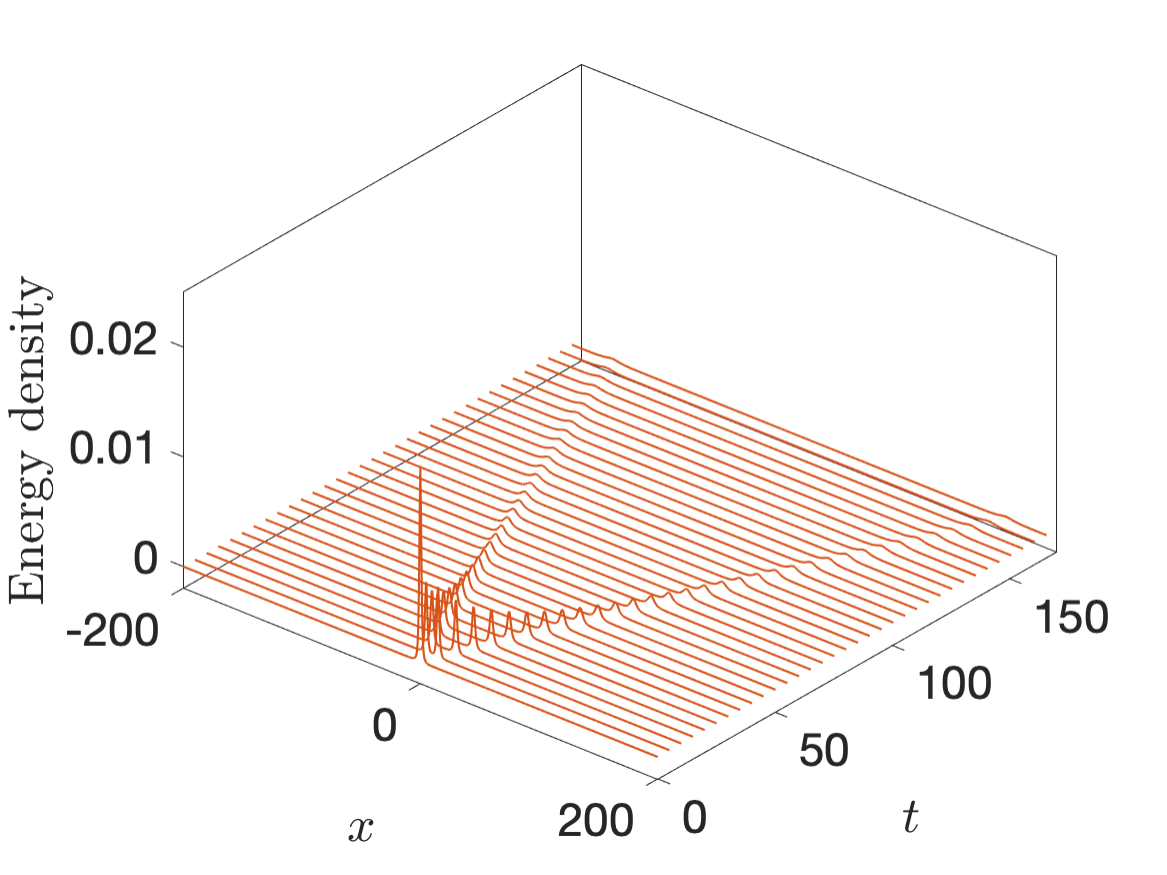}};

             \node at (-30pt, 8pt){\small$\epsilon=0$, lattice};
            \node at (125pt, 8pt){\small$\epsilon=0.01$, lattice};
            \node at (290pt, 8pt){\small$\epsilon=0.05$, lattice};
             \node at (-30pt, -128pt){\small$\epsilon=0$, CFT};         
	    \node at (125pt, -128pt){\small$\epsilon=0.01$, CFT};    
	    \node at (290pt, -128pt){\small$\epsilon=0.05$, CFT};               
                        
    \end{tikzpicture}
\caption{Complex-time evolution of energy density  $\langle T_{00}(x,t)\rangle$ after a local quantum quench in the free-fermion lattice (top) and CFT (bottom) calculations,
with $\epsilon=0$, $0.01$, and $0.05$ respectively.
In the CFT calculation, we take $\lambda=1.5$ in \eqref{EnergyDensity_Local}.}
\label{LatticeEnergy_LocalQuench}
\end{figure}
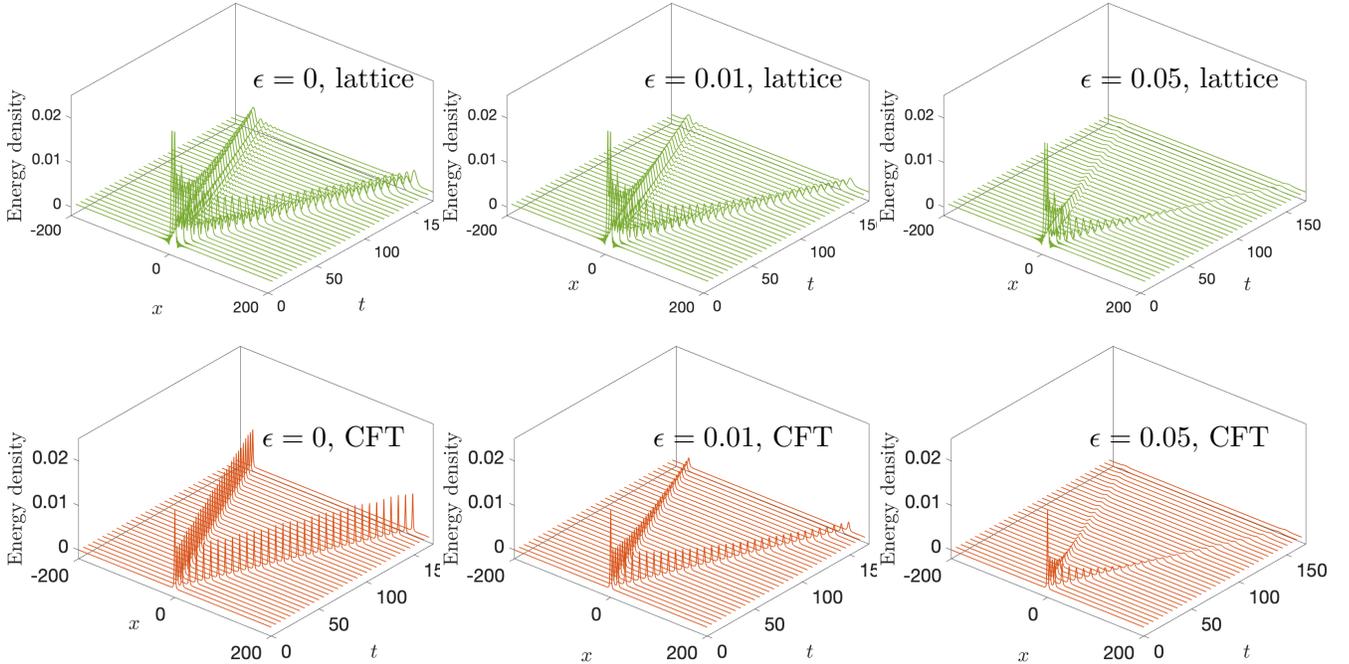

We consider the one point function $\langle T(z)\rangle$ by inserting the holomorphic stress-energy tensor operator at $z=x+i\tau_2$ in 
 the configuration in \eqref{local_PI} (with the branch cuts removed). By using the conformal mapping 
\be
w=f(z)=\frac{1}{\lambda+\epsilon \tau_1}\left(z+\sqrt{z^2+(\lambda+\epsilon \tau_1)^2}\right),
\ee
one can map \eqref{local_PI} to a right-half-plane (RHP). Since $\langle T(w)\rangle_{\text{RHP}}=0$, one can obtain 
\be
\langle T(x,t)\rangle=-\frac{c}{12}\{w,z\}=\frac{c}{8}\cdot \frac{(\lambda+\epsilon t)^2}{\left[(x-t)^2+(\lambda+\epsilon t)^2\right]^2},
\ee
where in the last step we have taken the analytical continuation $\tau_1\to t$ and $\tau_2\to it$. Similarly, one can obtain 
\be
\langle \bar T(x,t)\rangle=\frac{c}{8}\cdot \frac{(\lambda+\epsilon t)^2}{\left[(x+t)^2+(\lambda+\epsilon t)^2\right]^2}.
\ee
Therefore, the time evolution of energy density $\langle T_{00}(x,t)\rangle=\frac{1}{2\pi}
\left[\langle T(x,t)\rangle+\langle \bar T(x,t)\rangle\right]$ is
\be
\label{EnergyDensity_Local}
\langle T_{00}(x,t)\rangle
=
\frac{c}{16\pi}\left(
\frac{(\lambda+\epsilon t)^2}{[(x-t)^2+(\lambda+\epsilon t)^2]^2}+\frac{(\lambda+\epsilon t)^2}{[(x+t)^2+(\lambda+\epsilon t)^2]^2}
\right).
\ee
For $\epsilon=0$, it reproduces the real-time evolution result\cite{2014Asplund}.
In particular, at $x=\pm t$, one can observe two peaks in the energy density with 
\be
\langle T_{00}(x=\pm t,t)\rangle\simeq \frac{c}{16\pi} \cdot\frac{1}{\lambda^2}.
\ee
This is in comparison with the regions $|x\pm t|\gg \lambda$, where $T_{00}(x,t)\simeq 0$.
This feature of propagating energy density peaks can be observed in both the CFT and lattice calculations, as shown in Fig.\ref{LatticeEnergy_LocalQuench}.
It has been shown that these two peaks, which are emitted from $x=0$, are strongly entangled with each other \cite{2013Nozaki,2015Wen_EN,2014Asplund}.
See also a related discussion on the entanglement temperature near \eqref{beta_local_real}.

Now, by considering a complex-time evolution, one can still observe two peaks in the energy density evolution. 
The difference is that those propagating excitations die out in time. More explicitly, the peaks at $x=\pm t$ are suppressed in time as:
\be
\label{Energy_local_Quench}
\langle T_{00}(x=\pm t,t)\rangle\simeq
\frac{c}{16\pi} \cdot\frac{1}{(\lambda+\epsilon\, t)^2},
\ee
which decays to zero as a function of $1/t^2$ in the long time limit. A sample plot of $\langle T_{00}(x,t)\rangle$ for both the lattice model calculations and the CFT calculations can be found in Fig.\ref{LatticeEnergy_LocalQuench}, where one can see clearly the energy density peaks emitted from $x=0$ gradually die out in time for $\epsilon>0$.

\section{Periodically driven critical systems with complex metrics: 
Competition between driving and damping
}
\label{Sec:FloquetAllowableM}

\begin{figure}[h]
\centering

\begin{tikzpicture}[baseline={(current bounding box.center)}][x=0.75pt,y=0.75pt,yscale=-0.8,xscale=0.8]

       \begin{scope}[xshift=-20pt,yshift=-5pt]            
\node[inner sep=0pt] (russell) at (175pt,25pt)
        {\includegraphics[width=1.45in,height=1.9in]{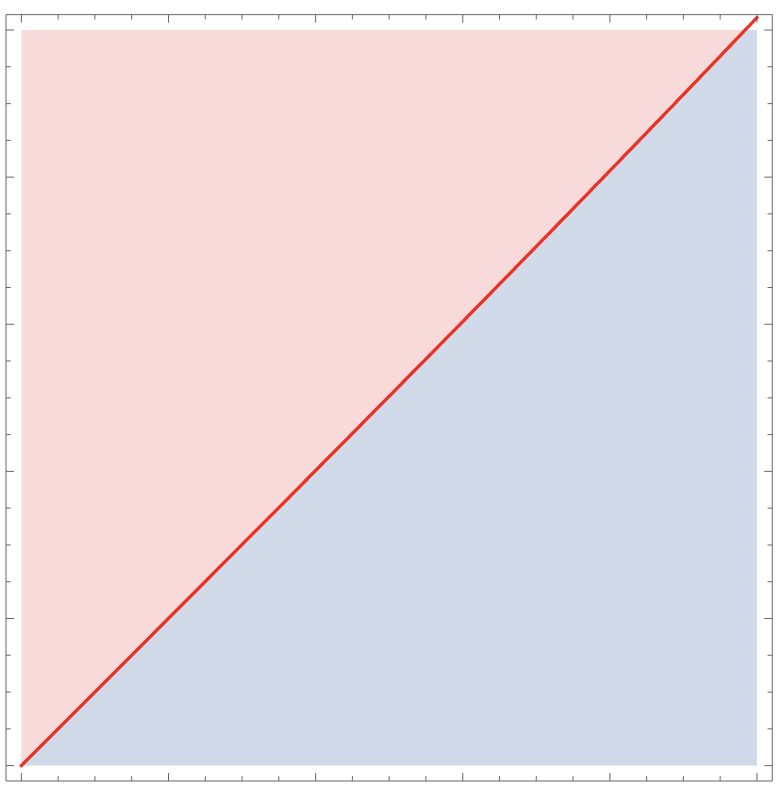} };
    
              \node at (118pt, -40pt){ \footnotesize $0$};    
		  \node at (113pt, 24pt){ \footnotesize $0.05$};    
		  \node at (115pt, 89pt){ \footnotesize $0.1$};        
    
              \node at (125pt, -48pt){ \footnotesize $0$};    
              \node at (175pt, -48pt){ \footnotesize $0.05$};    
              \node at (225pt, -48pt){ \footnotesize $0.1$};    
              
            \node at (98pt, 50pt){ \small $T_1/l$};

               \node at (155pt, 40pt){\textcolor{red}{heating}};

\draw (180pt, -18pt) node [color={rgb, 255:red, 65; green, 117; blue, 5 }  ,opacity=1 ]  {non-heating};

        \node at (175pt, -65pt){\small $T_0/l$};
    \end{scope}

\node[inner sep=0pt] (russell) at (345pt,10pt)
    {\includegraphics[width=3.25in]{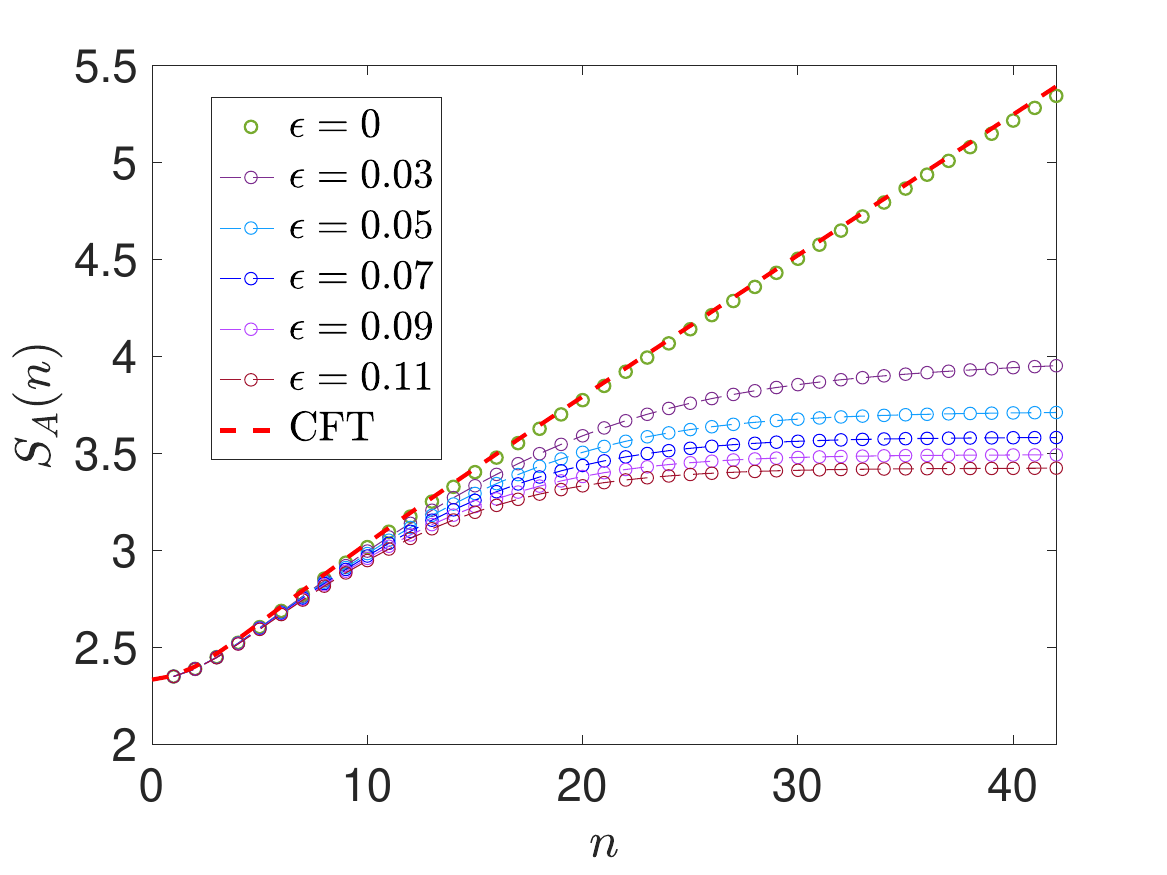} };

\end{tikzpicture}
\caption{
Left: Phase diagram of the Floquet CFT in a real time evolution in \eqref{TwoStep_Example}.
Right:
Complex time evolution of entanglement entropy $S_A(n)$ during a periodic driving, where
$n$ is the driving period in \eqref{psi_n} and the driving Hamiltonians are chosen in \eqref{2_Hamiltonian}.
We choose $L=2l=400$ with periodical boundary conditions, and the subsystem is $A=[0,\,L/2]$.
The driving parameters are chosen as $T_0/l=1/50$ and $T_1/l=1/25$ in \eqref{psi_n}, which is in the heating phase of a Floquet CFT if $\epsilon=0$.
Dashed red line corresponds to the CFT result in the real time evolution, which can be obtained based on the approach in \cite{wen2018floquet}.
For $\epsilon>0$, the entanglement entropy will reach a steady value, which decreases as one increases $\epsilon$.
}
\label{Floquet}
\end{figure}

\begin{figure}[h]
\center
\centering
\includegraphics[width=3.25in]{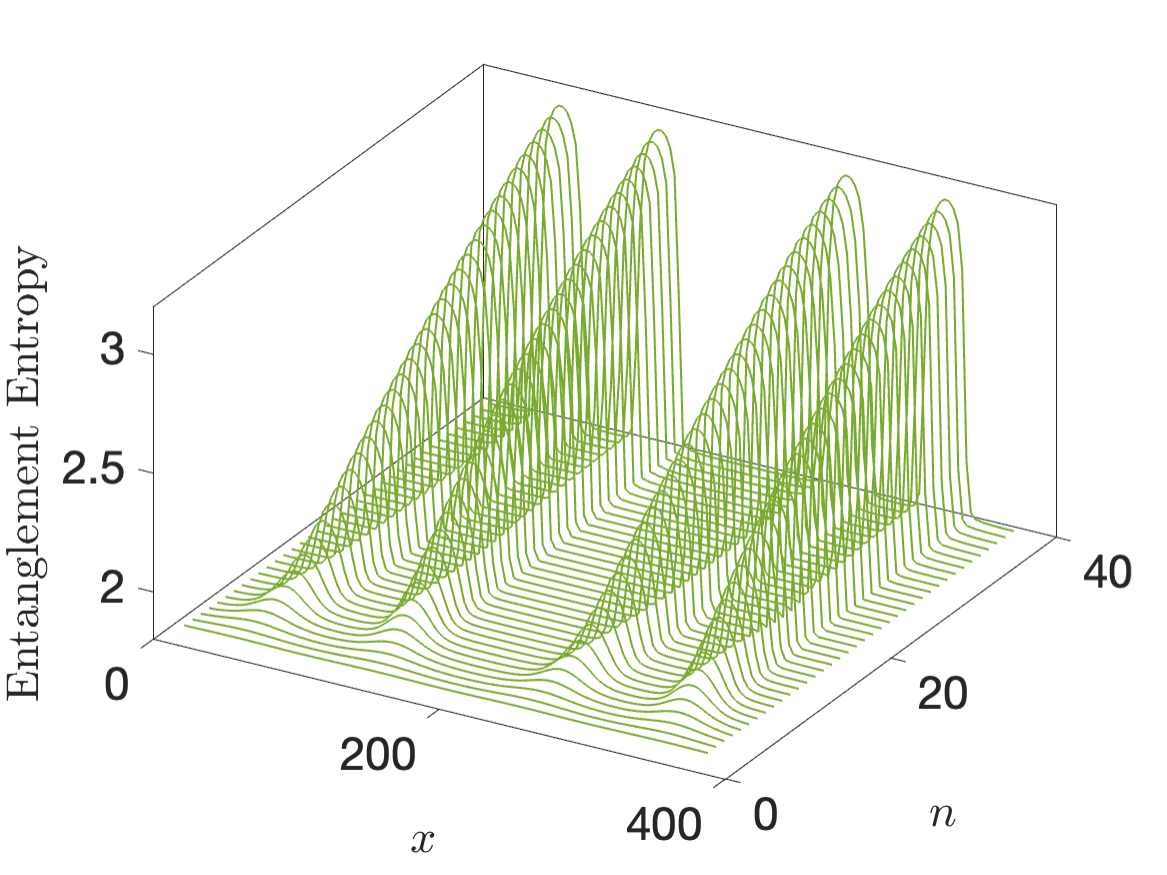}
\includegraphics[width=3.25in]{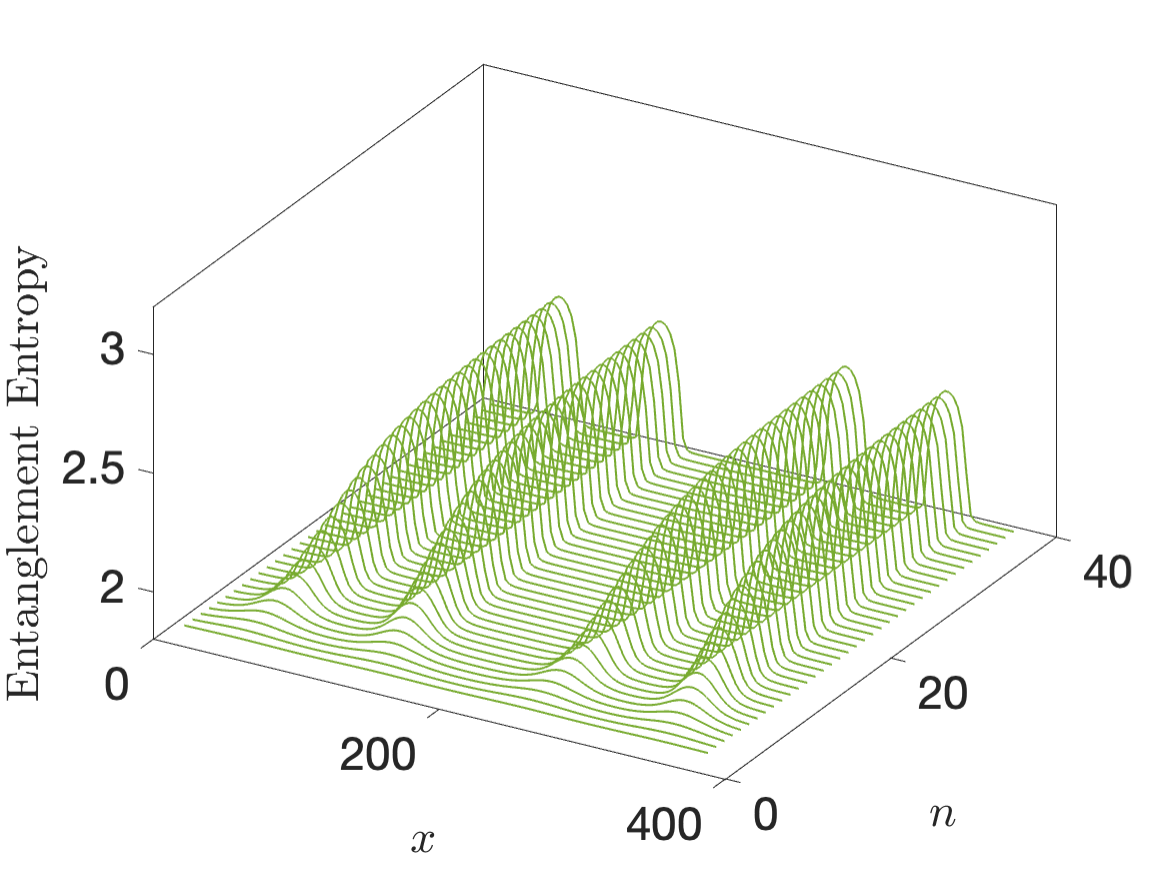}
\caption{Complex time evolution of entanglement entropy density $S_A(x,n)$ for subsystem $A=[x-l_A/2, x+l_A/2]$ during a periodic driving.
We choose $L=400$ with periodical boundary conditions, and the length of the subsystem is $l_A=30$.
The driving parameters are the same as those in Fig.\ref{Floquet}. 
In a real time evolution with $\epsilon=0$ (left), the entanglement entropy density
 forms peaks in the real space, and these peaks grow linearly in time.
In a complex time evolution with $\epsilon=0.11$ (right), the peaks in $S_A(x,n)$ form at the same locations, grow first and then saturate at a steady value.
}
\label{Floquet_EE_3d}
\end{figure}

In the previous sections on quantum quenches, we have seen that complex time evolutions give rise to a damping effect in the non-equilibrium dynamics.
Note that in a quantum quench the excitations are injected \textit{only} at $t=0$, and then there is no further driving.
Then as time evolves, the system will gradually decay to the ground state due to the damping effect.

\begin{figure}[h]
\center
\centering
\includegraphics[width=3.25in]{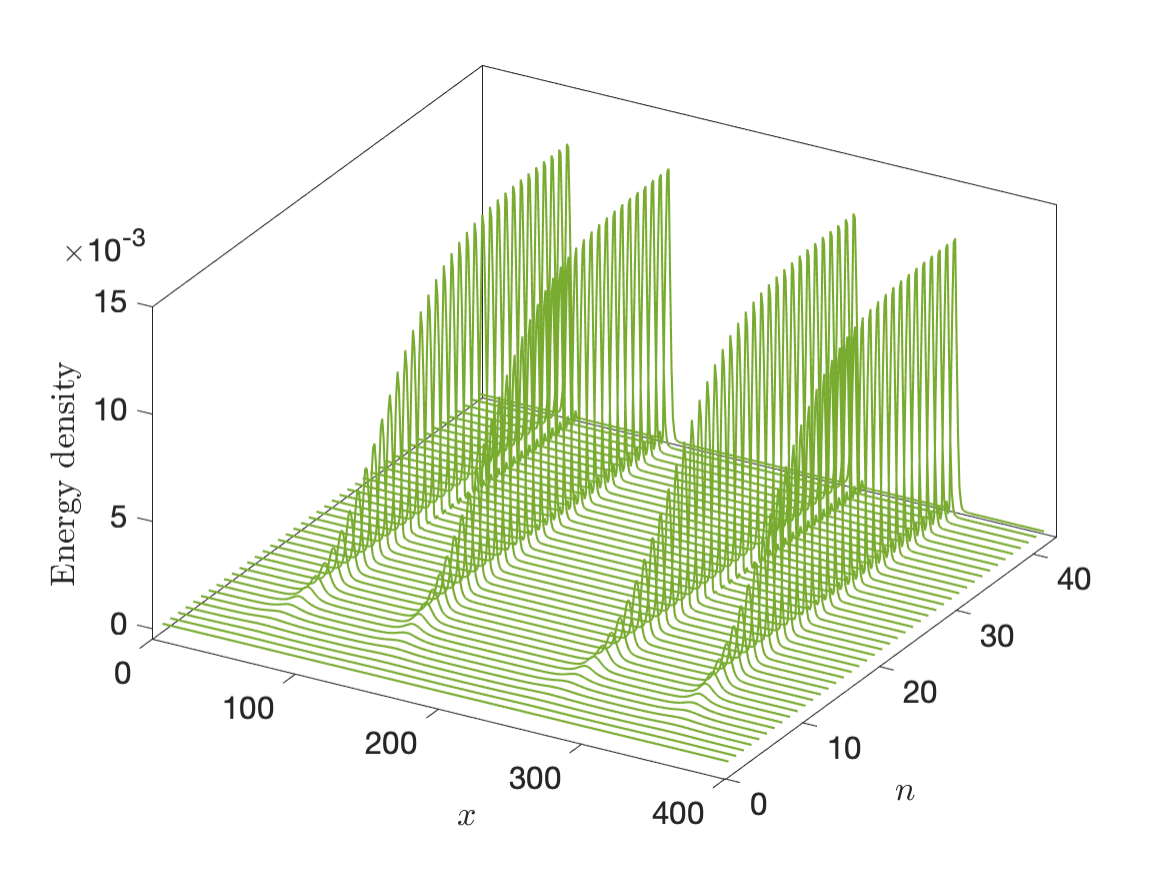}
\includegraphics[width=3.25in]{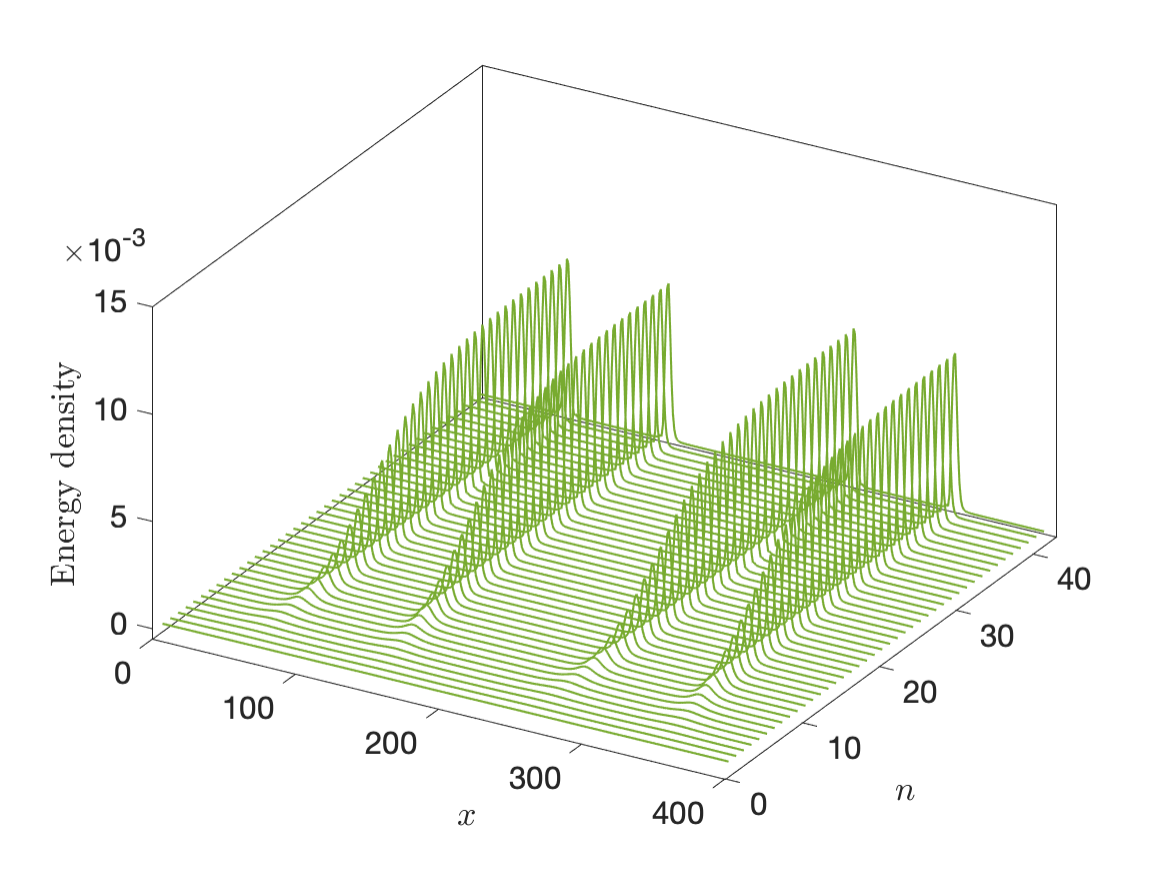}
\caption{
Complex time evolution of energy density in a Floquet CFT with 
$L=400$. The driving parameters are the same as those in Fig.\ref{Floquet}, with 
 $\epsilon=0.07$ (left) and $\epsilon=0.11$ (right).
A larger $\epsilon$ leads to a smaller steady value of  the energy density at the hot spots. 
}
\label{Floquet_energy}
\end{figure}

In this section, we are interested in the case of time-dependent drivings, and in particular the competition between driving and damping. 
The setup we consider here is the so called Floquet CFT, the real time evolution of which is exactly solvable\cite{wen2018floquet}. 
See some recent progress along this direction in quantum field theory
\cite{wen2018floquet,wen2020periodically,Fan_2020,Wen_2018,Lapierre:2019rwj,fan2020General, lapierre2020geometric,han2020classification,2020Lapierre,2020Andersen,2021Ageev,2021Das,RandomCFT2021,
2022Das_OTOC,2022Bermond,2022Choo,2022Cooling,2023_OBC,2023_StatePrepare,2024Ryu,2023Das,2023_Nozaki_Scrambling, 2024_Krylov,2024_Guo}
and the holographic dual\cite{2021Nozaki,2023Caputa,2023_Geometry, 2023_Briding,2024_Ge,2024Nozaki2,2024Mezei,2023_Brane,2024_Modular}. 
The basic idea is to drive the system with time-dependent Hamiltonians
\be
H(t)=\int_0^L f(x,t) \, T_{00}(x) \, dx,
\ee
where $T_{00}(x)$ is the energy density in a conformal field theory, and $f(x,t)$ is a real-value function. For a fixed time $t$, the effect of $f(x,t)$ is to deform the Hamiltonian density in space. The effect of such time dependent driving is to perform a conformal transformation on operators, which is the underlying reason why this setup is exactly solvable.
One interesting feature in this setup is that there can be different phases, including the heating and non-heating phases with a phase transition, 
depending on the driving parameters\cite{wen2018floquet}. In particular, in the heating phase, the total energy of the system grows exponentially in time, and 
the entanglement entropy grows linearly in time. 
The energy and entanglement growths also exhibit interesting spatial features:
The absorbed energy during the driving is mainly accumulated at certain hot spots (See, e.g., Fig.\ref{Floquet_energy} in a later discussion), and the entanglement entropy is mainly contributed by the entanglement between excitations localized at the neighboring hot spots \cite{Fan_2020}.

Now let us consider a concrete and minimal setup of Floquet CFT. We start from the ground state $|G\rangle$ of a uniform CFT Hamiltonian $H_0$, and 
evolve the state $|\psi_0\rangle=|G\rangle$ with periodically changing Hamiltonians as follows:
\be
\label{TwoStep_Example}
\begin{tikzpicture}[baseline={(current bounding box.center)}][x=0.75pt,y=0.75pt,yscale=-0.4,xscale=0.4]

\node at (-110pt, 10pt){$
\left\{
\begin{split}
H_0=&\int_0^LT_{00}(x) dx,\\
H_1=&\int_0^L \cos\left(\frac{2\pi x}{l}\right) T_{00}(x)\, dx
\end{split}
\right.
$};

\draw [thick](0pt,20pt)--(20pt,20pt);
\draw [thick](20pt,20pt)--(20pt,0pt);
\draw [thick](20pt,0pt)--(40pt,0pt);

\draw [thick](40pt,0pt)--(40pt,20pt);

% cycle 2:
\draw [thick](40pt,20pt)--(60pt,20pt);
\draw [thick](60pt,20pt)--(60pt,0pt);
\draw [thick](60pt,0pt)--(80pt,0pt);

\draw [thick](80pt,0pt)--(80pt,20pt);

% cycle 3:
\draw [thick](80pt,20pt)--(100pt,20pt);
\draw [thick](100pt,20pt)--(100pt,0pt);
\draw [thick](100pt,0pt)--(120pt,0pt);

\draw [thick](120pt,20pt)--(120pt,0pt);

% cycle 4:
\draw [thick](120pt,20pt)--(140pt,20pt);

\node at (150pt, 10pt){$\cdots$};

\node at (8pt,28pt){$H_1$};
\node at (30pt, 7pt){$H_0$};

   \draw  [-stealth](50pt, -10pt)--(110pt, -10pt);
   \node at (80pt, -15pt){time};

\end{tikzpicture}
\ee
That is, we evolve the system with Hamiltonians $H_0$  and $H_1$ for time $T_0$  and $T_1$ respectively in a periodic way (See also Eq.\eqref{psi_n_real_intro}).
The corresponding non-equilibrium phase diagram is shown in Fig.\ref{Floquet}.

Now, by introducing a complex time evolution in the heating phase, there will be a competition between the driving and damping. 
Although the real time evolution is exactly solvable,
unfortunately we don't know how to analytically solve the \textit{complex} time evolution in a Floquet CFT with allowable complex spacetime metrics\cite{Segal_2021}.
Therefore, in the following discussions, we will rely on numerical calculations based on a free-fermion lattice model to study the complex time evolution.
As a remark, in the real time evolution, it has been shown that such free fermion model calculations agree with the CFT calculations in a remarkable way\cite{wen2020periodically,RandomCFT2021}.

\begin{figure}[h]
\center
\centering
\includegraphics[width=3.25in]{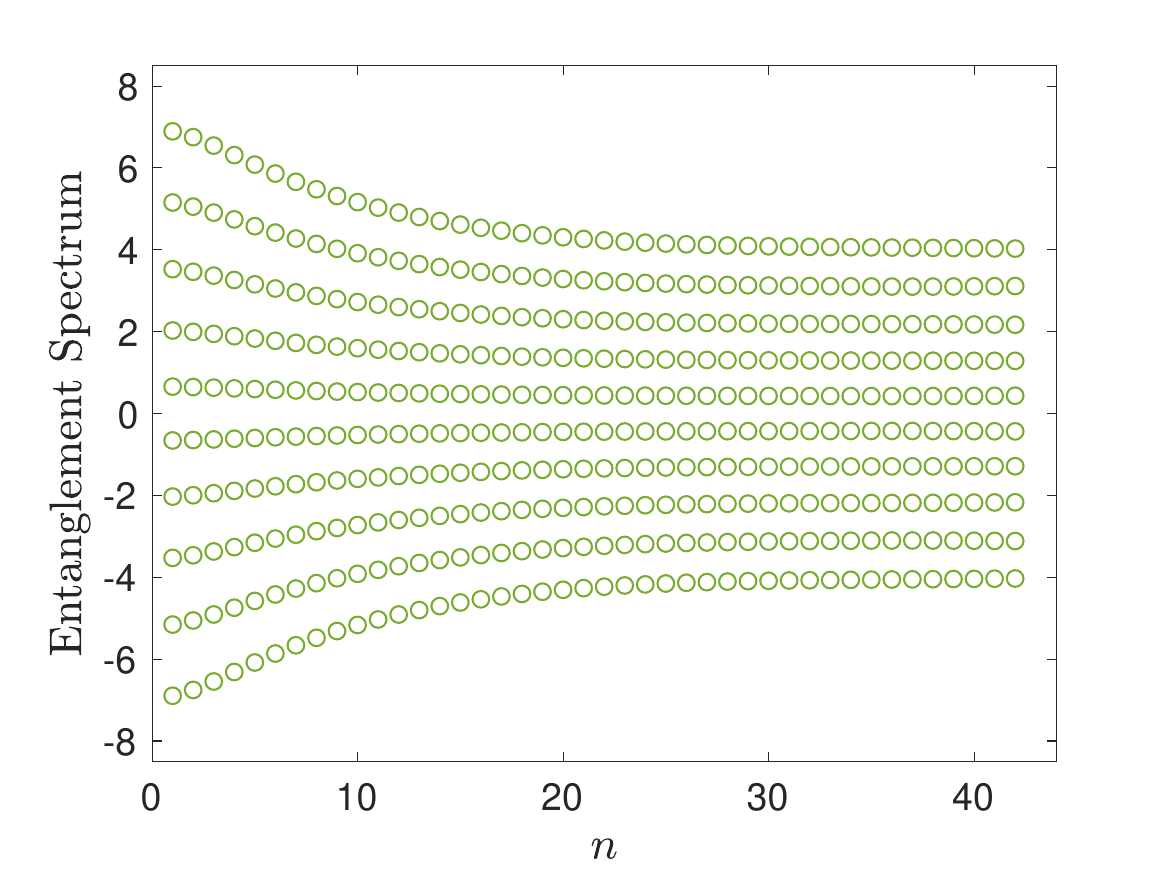}
\includegraphics[width=3.25in]{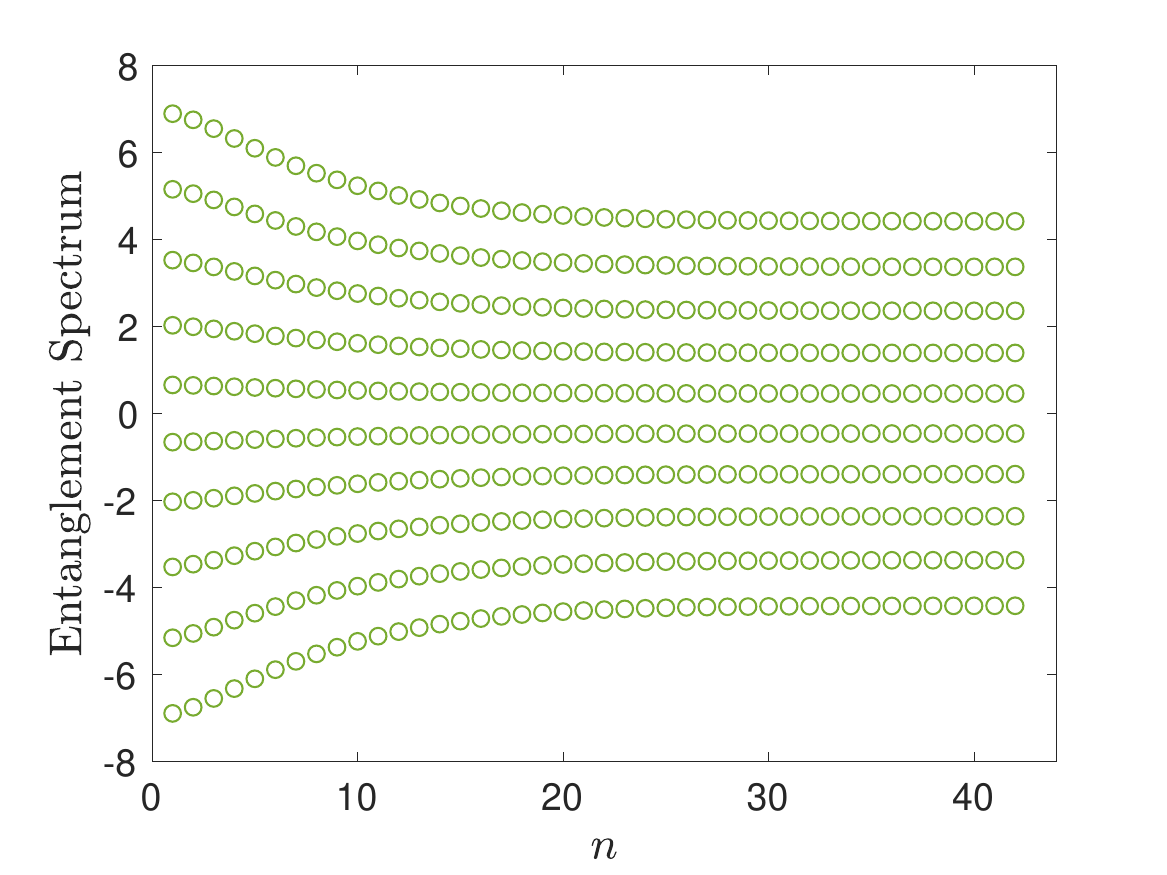}
\caption{Complex time evolution of entanglement spectrum for subsystem $A=[0,L/2]$.
The  parameters are the same as those in Fig.\ref{Floquet}, with $\epsilon=0.05$ (left) and $\epsilon=0.11$ (right).
As $\epsilon$ increases, the steady value of spacing in the entanglement spectrum increases, which
results in a smaller entanglement entropy in Fig.\ref{Floquet}.
}
\label{Floquet_ES}
\end{figure}

To be concrete, we consider the complex time evolution with the following two driving Hamiltonians on a lattice:
\be
\label{2_Hamiltonian}
\left\{
\begin{split}
H_0=&-\frac{1}{2}\sum_{i=1}^L c_i^\dag c_{i+1}+h.c.,\\
H_1=&-\frac{1}{2}\sum_{i=1}^L f_j \, c_i^\dag c_{i+1}+h.c., 
\end{split}
\right.
\ee
where $c_i$  and $c_i^\dag$ are fermionic annihilation and creation operators that 
satisfy the anticommutation relations $\{c_i, c_j\}=\{c_i^\dag, c_j^\dag\}=0$, and $\{c_i,c_j^\dag\}=\delta_{ij}$.
Here we choose $f_j$ as the discrete version of $f(x)=\cos\left(\frac{2\pi x}{l}\right)$ in \eqref{TwoStep_Example}, with $L=2l$. That is, $H_0$ is a free fermion lattice with uniform hopping, 
while the Hamiltonian density in $H_1$ is deformed by $f_j$. 
The initial state is chosen as the ground state $|G\rangle$ of $H_0$ with a half filling. 
This setup of driving in a real time evolution is exactly the same as that considered in \cite{2022Cooling}.
Now, the complex time evolution we consider is
\be
\label{psi_n}
|\psi(n)\rangle=\left(e^{-i(1-i\epsilon)H_0 T_0}e^{-iH_1 T_1}\right)^n |G\rangle,
\ee
where we introduce the complex time only during the driving by $H_0$. The motivation is that the time-dependent driving can lead to a heating phase\cite{wen2018floquet}, 
where the system keeps absorbing energy and evolves to a highly excited state of $H_0$ in time.
At the same time, the factor $e^{-\epsilon H_0 T_0}$ in the driving has a damping effect which tends to evolve the system to the ground state $|G\rangle$ of $H_0$. 
When both driving and damping are present, it is expected that the system will reach a steady state.

\subsection{Entanglement entropy and entanglement spectrum evolution}

To see the competition between driving and damping, we first study the complex time evolution of entanglement entropy. As shown in Fig.\ref{Floquet},
by first setting $\epsilon=0$, we tune the driving parameters $T_0$ and $T_1$ in \eqref{psi_n} to the heating phase in a Floquet CFT\cite{wen2018floquet}, 
where the entanglement entropy of a subsystem grows linearly in time. By turning on the complex time with $\epsilon>0$ in \eqref{psi_n}, one can find that the entanglement 
entropy will reach a steady state value. As one increases $\epsilon$, the steady value decreases -- this is as expected, since a larger $\epsilon$ corresponds to a 
larger damping rate to the ground state.
Next, to see the spatial feature of the entanglement, we check the entanglement entropy density $S_A(x,n)$ for a small subsystem $A=[x-l_A/2, x+l_A/2]$.
As shown in Fig.\ref{Floquet_EE_3d}, in a real time evolution ($\epsilon=0$), the entanglement entropy density $S_A(x,n)$ form peaks in the real space, and the peaks grow linearly in time \cite{Fan_2020}. By turning on the complex time evolution with $\epsilon>0$, one can still observe the peaks of entanglement entropy density at the same locations.
These peaks grow in time at the beginning, and then 
 saturate at a steady value which depends on the concrete choice of $\epsilon$.

\medskip
We further check the complex time evolution of the entanglement spectrum in the subsystem $A=[0,L/2]$.
In a real time evolution, since the entanglement entropy grows linearly in time, the spacing of entanglement spectrum will keep decreasing in time.
In a complex time evolution, however, the spacing of entanglement spectrum reaches a steady value, as seen in Fig.\ref{Floquet_ES}.
This steady value will increase as we increase $\epsilon$, which results in a decreasing entanglement entropy.
This is consistent with the result in the time evolution of entanglement entropy in Fig.\ref{Floquet}.

\subsection{Energy density evolution}

In the end, we check the time evolution of energy density. In a real time evolution in the heating phase, it is known that the total energy will grow exponentially in time, and the absorbed energy is mainly accumulated at certain hot spots in the real space\cite{Fan_2020,Lapierre:2019rwj,2022Cooling}. In a complex time evolution, we find that the absorbed energy is still accumulated at such hot spots (see Fig.\ref{Floquet_energy}), but will reach a steady value, which is similar to the feature of the entanglement entropy/spectrum evolution. In addition, as we increase $\epsilon$, the steady value of the energy density at the hot spots will decrease, as expected.

\medskip

In short, based on a numerical study of the entanglement entropy and energy density evolution on a free-fermion lattice model in complex spacetime metrics, one can find that the competition between the driving and damping leads to a steady state, where the patterns of entanglement and energy density inherit from those in a Floquet CFT in a real time evolution. It is an interesting future problem to give an analytical study of the features observed in the above numerical calculations.

\section{Discussion and Conclusion}

\begin{figure}[h]
\center
\centering
\includegraphics[width=3.25in]{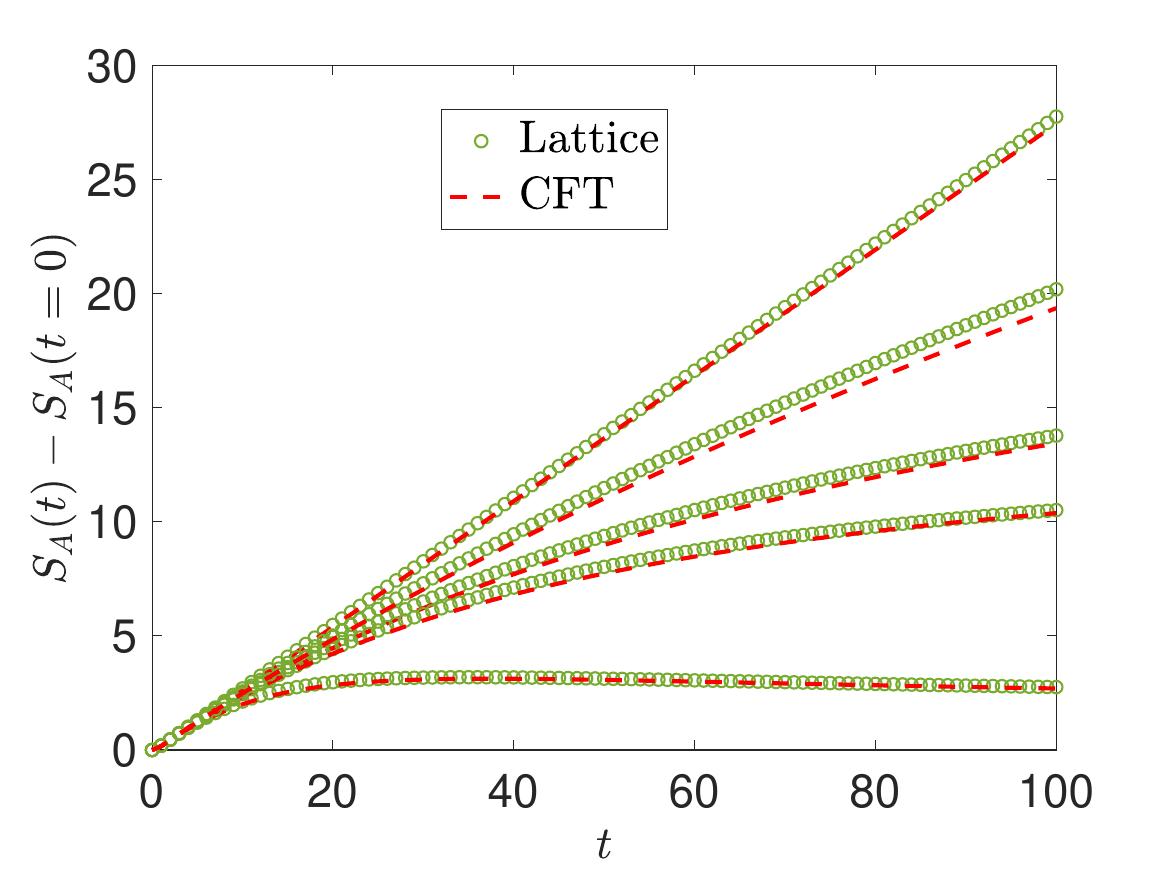}
\caption{
Comparison of complex-time evolution of 
 entanglement entropy $S_A(t)$ after a global quench according to \eqref{WF_general}
in a free-fermion lattice system and CFT calculations.
The lattice system is defined on $[0, L]=[0, 800]$, and the subsystem is $A=[0,400]$.
We choose $\epsilon=0.01$, and $\alpha=1.5$, $1.1$, $1$, and $0.8$ from bottom to top. 
For the top curve, it is a real time evolution with $\epsilon=0$. 
The CFT result is plotted according to \eqref{SA_general} and \eqref{W:Global1} by replacing $\epsilon t$ with $\epsilon t^\alpha$. The fitting parameters are 
$\beta=3.75$ and $\epsilon_0=0.1$. }
\label{LatticeEE_global_general}
\end{figure}

The method in this work can be generalized to other interesting cases. For example, one can consider the following complex time evolution
\footnote{We thank Shinsei Ryu for asking this question.}
\be
\label{WF_general}
|\psi(t)\rangle=e^{-i H_{\text{CFT}}\, (t-i\epsilon t^\alpha )}\, |\psi_0\rangle, \quad \epsilon \ge 0, \, \alpha\in \mathbb R.
\ee
In comparison with \eqref{WF_Quench}, here we have replaced $\epsilon t$ with $\epsilon t^\alpha$. The entanglement Hamiltonian and other quantities can be studied in 
the same way as what we did in the main text for each setup.
For example, for the global quantum quench as considered in Sec.\ref{Sec:Global1}, the time evolution of entanglement entropy can be obtained from \eqref{SA_general} and \eqref{W:Global1} by simply replacing $\epsilon t$ with $\epsilon t^\alpha$. See Fig.\ref{LatticeEE_global_general} for a comparison of lattice and CFT calculations for this case.

\bigskip

Now let us briefly conclude this work and mention some future problems.

\medskip

In this work, we have studied the effect of allowable complex spacetime metrics (as recently proposed in \cite{Segal_2021} and \cite{Witten2021_ComplexMetric}) on global/local quantum quench dynamics as well as time-dependent drivings in $(1+1)$ dimensional conformal field theories.
For quantum quenches, the time evolution of various physical quantities can be analytically solved at an arbitrary time.
The non-equilibrium dynamics in a complex time evolution shows universal features that are qualitatively different from those in a real time evolution.
See, e.g., \eqref{SA_global1_intro}, \eqref{SA_global2_intro}, and \eqref{SA_local_intro} for a comparison of the entanglement entropy evolution.
Physically, this qualitative difference is caused by the damping effect introduced by a complex time.
We further investigate the competition between the damping effect and external driving, by studying the recently proposed Floquet CFT \cite{wen2018floquet} in a complex spacetime metric. This Floquet CFT setup, although analytically solvable in the real time evolution, cannot be solved analytically in the allowable complex spacetime metrics to our knowledge. By performing a numerical study on a lattice model, it is found the competition between driving and damping can lead to a steady state with interesting patterns for the entanglement and energy density distribution.

\medskip

In the end, we want to point out several interesting future problems:

The complex time evolution considered in this work is generated by CFT Hamiltonians. 
For many interesting cases in non-unitary dynamics (See, e.g., some recent works \cite{2023Altaman2,2023Altman,2023Jian,2023Jian2,2023Alicea,2024Alicea,2022Granet,2020Chen,2023Schiro,2021_Tang,2022_Milekhin,2023_Silva}),  the corresponding Hamiltonians in the 
non-unitary time evolutions are not exactly the CFT Hamiltonians.
It will interesting to generalize our discussion in this work to the following Hamiltonians with a complex time evolution:
$
H_{\text{CFT}}\rightarrow H_{\text{CFT}} +\sum_i \lambda_i \int \Phi_i(x) \, dx,
$
where $\{\Phi_i\}$ could be irrelevant, marginal or relevant operators depending on the concrete physical problems, and $\{\lambda_i\}$ are real numbers.
A good understanding of this generalization may bring some insights to those problems in open quantum systems.

\medskip

Another interesting future problem is to give an analytical study of the time-dependent driven CFTs with allowable complex spacetime metrics.
From the numerical study on a free-fermion lattice model in Sec.\ref{Sec:FloquetAllowableM}, we have seen that the competition between driving and damping
results in a steady state with interesting entanglement features. It will be interesting to understand this steady state for a general CFT in an analytical way.

\bigskip
\textit{Note added:} During the preparation of this draft, I was aware of Ref.\cite{2024Nozaki}, which studied 
 the non-unitary time evolution in $2d$ CFTs in a different setup. Aside from the setup, the spacetime metrics considered therein are partially 
 Lorentz and partially Euclidean, while in this work we consider complex metrics.

\medskip
\textit{Note added 2:} We thank the authors of Ref.\cite{2024Su} for letting us know that they did a similar analysis on the complex time evolution of entanglement entropy  
after a global quantum quench, which has a partial overlap with our results in Sec.\ref{Sec:Semi_EE}.

\section{Acknowledgement}

The author thanks for the interesting discussions with Po-Yao Chang, Birgit Kaufmann, Ching Hua Lee,
Shinsei Ryu, Qicheng Tang, and Ashvin Vishwanath. 
This work is in part supported by the Simons Collaboration on Ultra-Quantum Matter, which is a grant from the Simons Foundation (618615, 651440).  
This work is also supported by a startup at Georgia Tech.

\appendix

\section{Details in the free-fermion lattice with a complex time evolution}
\label{Appendix:FreeFermion}

The complex time evolution of a free fermion lattice model can be studied by a straightforward generalization of the real time evolution.
See, e.g., a detailed study in \cite{2021_imaginary}.
Thanks to Wick’s theorem, to study the entanglement entropy/spectrum in a free fermion lattice, 
 it is enough to know the time dependent two-point functions \cite{Peschel2002}.

We consider a free-fermion Hamiltonian of the general form 
$H=\sum_{i,j=1}^L H_{ij} \, c_i^\dag c_j$,
where $L$ is the total number of lattice sites. Here $c_i$  and $c_i^\dag$ are fermionic annihilation and creation operators that 
satisfy the anticommutation relations $\{c_i, c_j\}=\{c_i^\dag, c_j^\dag\}=0$, and $\{c_i,c_j^\dag\}=\delta_{ij}$.
Note that although we are interested in the complex time evolution, the Hamiltonian we consider here is always hermitian.
The Hamiltonian can be diagonalized by a unitary matrix with $c_i=\sum_{j}U_{ij} \gamma_j$ 
such that $H=\sum_{i=1}^L E_i \gamma_i^\dag \gamma_i$.
Then the ground state with the lowest $N$ energy levels filled is
\be
\label{GroundState_G}
|G\rangle=\prod_{i=1}^N \gamma_i^\dag |\text{vac}\rangle=\prod_{i=1}^N \Big (\sum_{j=1}^L c_j^\dag U_{ji}\Big)|\text{vac}\rangle.
\ee
For later convenience, we denote $U^f$ as the first $N$ columns of the unitary matrix $U$. That is, $U^f$ is an $L$ by $N$ matrix.
One can rewrite the ground state as $|G\rangle=\prod_{i=1}^N \Big (\sum_{j=1}^L c_j^\dag U^f_{ji}\Big)|\text{vac}\rangle$.

Now we consider a quantum quench after a complex time evolution:
$
|\psi_1(t)\rangle=e^{-i (1-i\epsilon) H_1 t}|G\rangle
$.
Let us denote this state as $|\psi_1(t)\rangle=e^{-z_1 H_1 }|G\rangle$, where we have defined $z_1:=i(1-i\epsilon)t$.
By considering the general formula
\be
e^{\sum_{i,j}P_{ij}c_i^\dag c_j} \, c_l^\dag\, e^{-\sum_{i,j}P_{ij}c_i^\dag c_ j}=\sum_m (e^P)_{ml} c_m^\dag,
\ee
one can find that 
\be
\label{Eq:Psi1}
|\psi_1(t)\rangle=e^{-z_1H_1}|G\rangle= \prod_{i=1}^N \Big (\sum_{j=1}^L c_j^\dag [e^{-z_1H_1} U^f]_{ji}\Big)|\text{vac}\rangle 
=:
\prod_{i=1}^N \Big (\sum_{j=1}^L c_j^\dag\, \tilde U_{ji}\Big)|\text{vac}\rangle. 
\ee
Note that $\tilde U=e^{-z_1 H_1}U^f$ is an $L$ by $N$ matrix.
The first thing to notice is that even for a normalized initial state $\langle G|G \rangle=1$, we no longer have $\langle \psi_1(t)|\psi_1(t)\rangle=1$
after a non-unitary time evolution. Instead, we have 
\be\label{Psi1_normalize}
\langle \psi_1(t)|\psi_1(t)\rangle=\det [\tilde U^\dag \tilde U].
\ee
Next, based on \eqref{Eq:Psi1}, it is straightforward to check that
\be\label{2point_0}
\langle \psi_1(t)|c_m^\dag c_n|\psi_1(t)\rangle=\left[\tilde U \, \text{adj} (\tilde U^\dag \tilde U)\tilde U^\dag \right ]_{nm}
\ee
where $\text{adj} (M)$ denotes the adjugate of matrix  $M$.
Then, based on \eqref{Psi1_normalize} and \eqref{2point_0}, one can find the 2-point function after normalizing the state is
\be
\frac{\langle \psi_1(t)|c_m^\dag c_n|\psi_1(t)\rangle}{\langle \psi_1(t)|\psi_1(t)\rangle}=\left[\tilde U  (\tilde U^\dag \tilde U)^{-1} \tilde U^\dag \right ]_{nm},
\ee
where $M^{-1}$ is the inverse of $M$ with $M^{-1}=\text{adj}(M)/\det (M)$.
Based on this correlation matrix, one can follow the procedure in \cite{Peschel2002} to further obtain 
the entanglement spectrum and entanglement entropy.

The above discussion can be straightforwardly generalized to a time-dependent driving case, 
e.g., the periodically driven case in Sec.\ref{Sec:FloquetAllowableM},
by simply re-defining the matrix 
$\tilde U$ in \eqref{Eq:Psi1} as:
\be
\tilde{U}=\left(e^{-z_nH_n} \cdots e^{-z_1 H_1} \right)U^f,
\ee
where it is reminded that $U^f$ is the first $N$ columns of the unitary matrix $U$ in \eqref{GroundState_G}.

\medskip
Next, let us briefly introduce the lattice model we used to compare with the CFT calculation in the main text. 
In the global quench, the initial state $|\psi_0\rangle$ is prepared as the ground state of the following gapped Hamiltonian:
\be
\label{gap_Hamiltonian}
H_0=-\frac{1}{2}\sum_i c_i^\dag c_{i+1}+h.c.+m\sum_i (-1)^i c_i^\dag c_i.
\ee
Here the mass term $m\in \mathbb R$ determines the size of gap in the energy spectrum.
Then at $t=0$, we evolve the initial state as $|\psi_1(t)\rangle = e^{-i H_1 (1-i\epsilon)t}|\psi_0\rangle$, 
where $H_1$ is the critical Hamiltonian by setting $m=0$ in $H_0$, i.e.,
\be
\label{H1_lattice}
H_1=-\frac{1}{2}\sum_i c_i^\dag c_{i+1}+h.c.
\ee

In the local quench, the initial state $|\psi_0\rangle$ is considered as the ground state of two decoupled fermion chains as follows
\be
\label{H0_local}
H_0=-\frac{1}{2}\sum_{i=0}^{L/2-1} c_i^\dag c_{i+1} -\frac{1}{2}\sum_{i=L/2}^{L-1} c_i^\dag c_{i+1}+h.c.
\ee
Then at $t=0$, we evolve the initial state with the Hamiltonian $H_1$ in \eqref{H1_lattice}, i.e., we connect the two ends of
free fermion chains in \eqref{H0_local}.
The time dependent state is $|\psi_1(t)\rangle = e^{-i H_1 (1-i\epsilon)t}|\psi_0\rangle$.

\bibliography{NonUnitary1Ref.bib}

\end{document}